%% file: arXiv-Carlos2.tex
\documentclass[11pt, a4paper]{article}
\usepackage[textwidth = 7in, textheight=25cm,nohead]{geometry}
\usepackage{ulem}
\usepackage{arydshln}
\usepackage{graphicx, amssymb, amsmath, amsthm, amstext, booktabs, float, multirow, rotating,bm}
\usepackage{caption}
\usepackage{subcaption}
\usepackage{stackengine}
\usepackage{amssymb}
\usepackage{mathtools,bm}
\usepackage[mathscr]{euscript}

\usepackage{bbding}
\usepackage{pifont}

\usepackage{authblk}

\usepackage{type1cm}        
\usepackage{makeidx}         
\usepackage{graphicx}        
\usepackage{multicol}        

\usepackage{newtxtext}       %

\usepackage[
backend=bibtex,
style=numeric,
]{biblatex}
\usepackage{multirow}
\usepackage{caption}
\usepackage{subcaption}

\usepackage{mathdots}

\usepackage{tikz}
\usepackage{setspace}
\usepackage{enumerate}
\usepackage{multirow}
\usepackage{graphicx,nicefrac}

\usepackage[
backend=bibtex,
style=numeric,
]{biblatex}
\numberwithin{equation}{section} 
\numberwithin{figure}{section} 
\numberwithin{table}{section} 
\newcommand*\diff{\mathop{}\!\mathrm{d}} 
\newcommand{\Rd}{\mathbb{R}^d}

\newcommand{\Zd}{\mathbb{Z}^d}

\newcommand{\Abs}[1]{\left|#1\right|}
\newcommand{\Norm}[1]{\left\|#1\right\|}

\newcommand{\psiq}{\psi^{(q)}}
\newcommand{\psiqjz}{\psi^{(q)}_{j,z}}

\newcommand{\betahatqjz}{\hat{\beta}^{(q)} _{j,z}}
\newcommand{\bdelta}{\boldsymbol\delta}
\newcommand{\Bhat}{Bhattacharyya }

\def\fs{\kern 0.33em}
\def\ppn{\vskip 6pt \noindent }

\def\R{\ensuremath{\mathbb{R}}}
\def\N{\ensuremath{\mathbb{N}}}
\def\Z{\ensuremath{\mathbb{Z}}}

\def\E{\ensuremath{\mathbb{E}}}

\newcommand{{\Xs}}{{\cal X}}
\newcommand{{\Ys}}{{\cal Y}}
\newcommand{{\Ls}}{{\cal L}}
\newcommand{{\Ss}}{{\cal S}}
\newcommand{{\Ms}}{{\cal M}}
\newcommand{{\Gs}}{{\cal G}}
\newcommand{{\Hs}}{{\cal H}}
\newcommand{{\Ns}}{{\cal N}}
\newcommand{{\Is}}{{\cal I}}
\newcommand{{\As}}{{\cal A}}
\newcommand{{\Bs}}{{\cal B}}
\newcommand{{\Cs}}{{\cal C}}
\newcommand{{\Rs}}{{\cal R}}
\newcommand{{\Us}}{{\cal U}}
\newcommand{{\Es}}{{\cal E}}
\newcommand{{\Fs}}{{\cal F}}
\newcommand{{\pp}}{{\mathbf p}}
\newcommand{{\Ps}}{{\cal P}}
\newcommand{{\KK}}{{\mathbf K}}
\newcommand{{\HH}}{{\mathbf H}}
\newcommand{{\II}}{{\mathbf I}}
\newcommand{{\yy}}{{\mathbf y}}
\newcommand{{\ab}}{{\mathbf a}}

\newcommand{{\toL}}{{\ \overset{\mathcal{L}}{\longrightarrow}\ }}

\newtheorem{assumption}{Assumption}
\newtheorem{theorem}{Theorem}

\begin{document}

\setlength{\belowdisplayskip}{5pt} \setlength{\belowdisplayshortskip}{3pt}
\setlength{\abovedisplayskip}{5pt} \setlength{\abovedisplayshortskip}{0pt}

\title{Hellinger-\Bhat cross-validation for shape-preserving multivariate wavelet thresholding}
\author[1]{\textsc{Carlos Aya Moreno}\thanks{{\tt carlosayam@gmail.com}.}}
\author[1]{\textsc{Gery Geenens}\thanks{Corresponding author: {\tt ggeenens@unsw.edu.au}.}}
\author[1]{\textsc{Spiridon Penev}\thanks{{\tt s.penev@unsw.edu.au}.}}
\affil[1]{School of Mathematics and Statistics, UNSW Sydney, Australia}

\date{\today}
\maketitle
\thispagestyle{empty} 


\begin{abstract} The benefits of the wavelet approach for density estimation are well established in the literature, especially when the density to estimate is irregular or heterogeneous in smoothness. However, wavelet density estimates are typically not {\it bona fide} densities. In \cite{Aya18} a `shape-preserving' wavelet density estimator was introduced, including as main step the estimation of the square-root of the density. A natural concept involving square-root of densities is the Hellinger distance - or equivalently, the \Bhat affinity coefficient. In this paper, we deliver a fully data-driven version of the above `shape-preserving' wavelet density estimator, where all user-defined parameters, such as resolution level or thresholding specifications, are selected by optimising an original leave-one-out version of the Hellinger-\Bhat criterion. The theoretical optimality of the proposed procedure is established, while simulations show the strong practical performance of the estimator. Within that framework, we also propose a novel but natural `jackknife thresholding' scheme, which proves superior to other, more classical thresholding options.
	
\end{abstract}

\ppn {\bf Keywords:} 

\ppn {\bf AMS Classification:} 

\section{Introduction} \label{sec:intro}

Let $\{\varphi_{j_0,z},\psi^{(q)}_{j,z}; j=j_0,\ldots,\infty, z \in \Z^d, q \in Q_d\}$ be an orthonormal wavelet basis for $L_2(\R^d)$ generated from a certain basic level $j_0 \in \N$ by a `father' $\varphi: \R^d \to \R$ and `mothers' $\psi^{(q)}: \R^d \to \R$, $q \in Q_d = \{1,\ldots,2^d-1\}$, as $\varphi_{j_0,z}\left(x\right) = 2^{d\,j_0 /2}\varphi(2^{j_0} x - z)$ and $\psiqjz(x) = 2^{d\,j /2} \psiq(2^j x - z)$; see details in \cite{Meyer92}. Any $d$-variate function $f \in L_2(\R^d)$ admits the expansion
\begin{equation} f(x) = \sum_{z \in \Z^d} a_{j_0,z}\varphi_{j_0,z}(x) + \sum_{j=j_0}^\infty \sum_{z \in \Z^d} \sum_{q \in Q_d} b^{(q)}_{j,z}  \psi^{(q)}_{j,z}(x), \label{eqn:multwavexp} \end{equation}
with $a_{j_0,z} = \int_{\R^d} \varphi_{j_0,z}(x) f(x)\diff x$ and $b^{(q)}_{j,z} = \int_{\R^d} \psi_{j,z}^{(q)}(x) f(x)\diff x$. For $f$ a probability density, we note that $a_{j_0,z}=\E(\varphi_{j_0,z}(X))$ and $b^{(q)}_{j,z} = \E(\psi_{j,z}^{(q)}(X))$ for $X$ a random variable whose distribution $F$ has density $f$, which suggests empirical estimation of these coefficients by simple averages $\widehat{a}_{j_0,z}$ and $\widehat{b}^{(q)}_{j,z}$ upon observation of a random sample $\Xs = \{X_1,\ldots,X_n\}$ drawn from $F$. This yields the classical wavelet density estimator of $f$:
\begin{equation} \hat{f}_J(x) = \sum_{z \in \Z} \widehat{a}_{j_0,z} \varphi_{j_0,z}(x) + \sum_{j=j_0}^J \sum_{z \in \Z} \sum_{q \in Q_d} \widehat{b}^{(q)}_{j,z} \psi^{(q)}_{j,z}(x), \label{eqn:classest}\end{equation}
after truncation of (\ref{eqn:multwavexp}) at `resolution' $J \geq j_0$. Such wavelet-based estimators have proved very efficient for reconstructing possibly irregular density functions, due to the localised nature of the basis functions in $\{\varphi_{j_0,z},\psi^{(q)}_{j,z}\}$ \cite{Hardle12,Meyer92}.

\medskip

An important downside of (\ref{eqn:classest}), though, is that it is in general not `shape-preserving', meaning that it is not necessarily non-negative everywhere, and does not integrate to one -- in other words, $\hat{f}_J$ is not a {\it bona fide} density. The problem was addressed in \cite{Aya18}, where a shape-preserving version of the wavelet density estimator was introduced. The idea revolves around estimating the square root of $f$, say $g = \sqrt{f}$, first. As $\int g^2 = \int f = 1$, $g$ obviously belongs to $L_2(\R^d)$, and we can write as in (\ref{eqn:multwavexp}):
\begin{equation} g(x) = \sum_{z \in \Z^d} \alpha_{j_0,z}\varphi_{j_0,z}(x) + \sum_{j=j_0}^\infty \sum_{z \in \Z^d} \sum_{q \in Q_d} \beta^{(q)}_{j,z}  \psi^{(q)}_{j,z}(x), \label{eqn:gexp} \end{equation}
where $\alpha_{j,z} = \int_{\R^d} \varphi_{j,z}(x) \sqrt{f}(x)\diff x$ and $\beta^{(q)}_{j,z} = \int_{\R^d} \psi_{j,z}^{(q)}(x) \sqrt{f}(x)\diff x$.

\medskip

The challenge is that, as opposed to $a_{j_0,z}$ and $b^{(q)}_{j,z}$, $\alpha_{j_0,z}$ and $\beta^{(q)}_{j,z}$ are not simple expectations, easily estimated by their sample counterparts. Indeed we are left with an inconvenient expression involving $\sqrt{f}$. Now, define $X_{(1);i}$ the closest observation to $X_i$ in the sample $\Xs$; $R_{i} = \|X_{(1);i}-X_i \|$ the Euclidean distance between $X_i$ and $X_{(1);i}$; and $V_{i}$ the volume of the ball of radius $R_{i}$ centred at $X_i$. In \cite{Aya18} it was proved that, for any square-integrable function $\phi: \R^d \to \R$, 
\begin{equation} S_n \doteq \frac{2}{\sqrt{\pi}} \frac{1}{\sqrt{n}} \sum_{i=1}^n \phi(X_i) \sqrt{V_i}  \label{eqn:Sn1} \end{equation}
is a consistent estimator of $\int \phi(x) \sqrt{f}(x)\diff x$ under standard assumptions. Therefore, 
\begin{align} \hat{\alpha}_{j_0,z} &=  \frac{2}{\sqrt{\pi}}\,\frac{1}{\sqrt{n}}\sum _{i=1}^n \varphi _{j_0,z}\left(X_i\right) \sqrt{V_{i}}, \qquad  z \in \Zd \label{eqn:coefficients1} \\
	\betahatqjz &=  \frac{2}{\sqrt{\pi}} \,\frac{1}{\sqrt{n}}\sum _{i=1}^n \psiqjz\left(X_i\right) \sqrt{V_{i}}, \qquad j \in \N;z \in \Zd;  q \in Q \label{eqn:coefficients2}
\end{align}
are consistent estimators of $\alpha_{j_0,z}$ and $\beta^{(q)}_{j,z}$ \cite[Proposition 1]{Aya18}. This suggests, for $g$, the primary estimator
\begin{equation}
	\mathring{g}_{J}(x)=\sum _{z \in \Zd} \hat{\alpha }_{j_0,z} \varphi_{j_0,z}(x) + \sum_{j=j_0}^{J} \sum_{z \in \Z^d} \sum_{q \in Q_d} \betahatqjz \, \psiqjz(x), \label{eqn:ringgJ}
\end{equation}
akin to (\ref{eqn:classest}). Due to the orthonormal nature of the wavelet basis $\{\varphi_{j_0,z},\psi^{(q)}_{j,z}\}$, it is easy to see that 
\begin{equation}\label{eqn:normgJ}
	\Norm{\mathring{g}_J}^2 = \sum_{z \in \Z^d} \left(\hat{\alpha}_{j_0,z}\right)^2 + \sum_{j=j_0}^{J} \sum_{z \in \Z^d} \sum_{q \in Q_d} \left(\hat{\beta}^{(q)} _{j,z}\right)^2.
\end{equation}
Therefore, the normalised estimator
\begin{equation}
	\hat{g}_J \doteq \frac{\mathring{g}_J}{\Norm{\mathring{g}_J}}  \label{eqn:hatgJ}
\end{equation}
is such that $\hat{g}_J^2$ is always a proper density (non-negative and integrating to 1), hence a `shape-preserving' wavelet estimator for $f$. Theorem 2 in \cite{Aya18} established the uniform $L_2$-consistency of this estimator for $f$ over Sobolev balls of order $m$, provided that $2^J \sim n^{1/(2m+d)}$. Simulation studies suggested promising practical performance, too.

\medskip

Two questions were not investigated in \cite{Aya18}, though. The first one is how to select the resolution $J$ in practice? As pointed out in \cite{Vannucci97}, this is indeed one of the major problems in orthogonal series density estimation. In fact, the resolution plays a similar role to the bandwidth in kernel density estimation, and so a correct choice of this quantity can alleviate difficulties caused by over-smoothing, when too small, or over-fitting, when too large.

\medskip 

The second one relates to the fact that wavelet-based approaches are known to exhaust their full potential only if conjugated with an appropriate thresholding scheme for the coefficients. Indeed, thresholding, i.e., the suppression and/or shrinkage of some of the estimated wavelet coefficients, allows the underlying multiresolution analysis (MRA) to come fully into play, which translates into better theoretical performance in general Besov spaces \cite{Donoho98,DJKP96}. Such thresholding was not investigated in \cite{Aya18}. 

\medskip 

This paper fills in those two gaps: first, we propose a novel data-driven rule for selecting the right resolution $J$ in practice; and secondly, we explore a thresholding scheme for the shape-preserving wavelet estimator. Interestingly, both problems are tackled by making use of the same, novel criterion, which we call the Hellinger-\Bhat cross-validation. Indeed, it is an empirical `leave-one-out-cross-validation' type of the Hellinger distance (or equivalently the \Bhat affinity coefficient) between $f$ and its estimator.  

\medskip 

\section{Data-driven selection of the resolution level}

\subsection{The empirical Bhattacharyya coefficient} \label{subsec:empBhat}

Following common statistical practice, a data-driven way of selecting $J$ would be to pick the value which minimises an empirically estimated divergence between the `true' density $f$ and its estimator, here $\hat{g}^2_J$. Among the numerous measures available for quantifying the distance between two probability densities, the Hellinger distance seems the most natural here, given that it involves the square-roots of the said densities \cite{Basu11}. Specifically, the (squared) Hellinger distance between $f$ and $\hat{g}^2_J$ is:
\begin{equation} \mathcal{H}^2(J) = \frac{1}{2}\int \left(|\hat{g}_J|(x) - \sqrt{f}(x) \right)^2 \, \diff x. \label{eqn:Hell1} \end{equation}
(Note that $\hat{g}_J$ may take on negative values at some locations, so $\sqrt{\hat{g}^2_J}$ is really $|\hat{g}_J|$.) 
Expanding the square, we obtain
\begin{equation} \Hs^2(J) = 1 - \int |\hat{g}_J|(x) \sqrt{f}(x)\, \diff x \doteq 1- \Bs(J), \label{eqn:bhat} \end{equation}
owing to $\int \hat{g}^2_J = \int f = 1$. The integral $\Bs(J)$ is the `\Bhat affinity coefficient' \cite{Bhattacharyya46} between the estimator $\hat{g}_J^2$ and $f$. Minimising $\Hs^2(J)$ is obviously equivalent to maximising $\Bs(J)$ -- so we are looking for making $f$ and $\hat{g}^2_J$ resemble each other as much as possible in the \Bhat sense.

\medskip 

As $\Bs(J)$ involves the unknown $f$, it needs to be estimated. Here we propose to combine (\ref{eqn:Sn1}), allowing consistent estimation of quantities of type $\int \phi \sqrt{f}$, and ideas reminiscent of classical cross-validation for selecting a smoothing parameter in nonparametric density estimation \cite{Bowman84,Rudemo82,Stone84}. Specifically, we estimate $\Bs(J)$ by 
\begin{equation} \widehat{\Bs}(J) \doteq  \frac{2}{\sqrt{\pi}} \frac{1}{\sqrt{n}} \sum_{i=1}^n|\hat{g}^{(-i)}_J|(X_i) \sqrt{V_i}, \label{eqn:BhatJ} \end{equation} 
where, like in Section \ref{sec:intro}, $V_{i}$ is the volume of the ball centred at $X_i$ whose radius is the distance from $X_i$ to its closest neighbour in the sample. As usual in such a framework, $\hat{g}^{(-i)}_J$ is the `leave-one-out' estimator, here the estimator (\ref{eqn:hatgJ}) computed on the subsample $\Xs^{(-i)} = \Xs \backslash \{X_i\}$. Note that obtaining such `leave-one-out' estimator does not require re-computing entirely (\ref{eqn:ringgJ})-(\ref{eqn:hatgJ}) $n$ times. Indeed, removing $X_i$  from the sample just removes the $i$th term in (\ref{eqn:coefficients1})-(\ref{eqn:coefficients2}) and only amends $\sqrt{V_{i'}}$ in terms $i' \neq i$ for those $X_{i'}$ whose nearest-neighbour happens to be $X_i$ -- in other words, a very limited fraction of the observations of the sample. The estimator $\hat{g}^{(-i)}_J$ can, therefore, be computed very efficiently by careful amendment of a small number of terms from the initial estimator $\hat{g}_J$ and by computing all $2^{nd}$-nearest neighbour distances in advance, a facility available in most implementations.

\medskip

It appears from \cite[Lemma 1]{Aya18} that $\widehat{\Bs}(J)$ is a consistent estimator of $\Bs(J)$. We call it the {\it empirical Bhattacharyya coefficient}, and we choose for resolution $J$ in (\ref{eqn:ringgJ})-(\ref{eqn:hatgJ}) its maximiser:
\begin{equation} \widehat{J} \doteq \arg \max_{J \in \{j_0,j_0+1,\ldots\}} \widehat{\Bs}(J). \label{eqn:Jhat}\end{equation}
The final estimator of $f$ is thus $\hat{g}_{\widehat{J}}$, which is always a {\it bona fide} density, by definition.

\medskip 

Note that, in the above procedure, the resolution level is selected {\it after} normalisation, that is, assuming that the estimator is automatically normalised via (\ref{eqn:hatgJ}). Another, slightly different approach would be to select $J$ {\it before} the normalisation step. In other words, we would look for selecting the value of $J$ which makes the unnormalised estimator (\ref{eqn:ringgJ}) as close to $\sqrt{f}$ as possible, and then, in a final step, we would make sure that the resulting estimate is a proper density by normalising it. One can write the analogue to (\ref{eqn:Hell1}) for the estimator $\mathring{g}_{J}$:
\begin{equation} \Hs_\circ^2(J) \doteq  \frac{1}{2}\int \left(|\mathring{g}_{J}|(x) - \sqrt{f}(x) \right)^2 \, \diff x. \label{eqn:H2circJ}\end{equation}
Note that, in spite of the slight abuse of notation, $\Hs_\circ^2$ is not any squared `Hellinger distance', given that $\mathring{g}^2_{J}$ is not a proper density, in general. Expanding the square, we obtain this time
\[\Hs_\circ^2(J) = \frac{1}{2}\int \mathring{g}^2_{J}(x)\,\diff x - \int |\mathring{g}_{J}|(x) \sqrt{f}(x)\,\diff x + \frac{1}{2}, \]
whose minimiser would be the maximiser of 
\begin{equation*} \Bs_\circ(J) \doteq \int |\mathring{g}_{J}|(x) \sqrt{f}(x)\,\diff x - \frac{1}{2}\int \mathring{g}^2_{J}(x)\,\diff x.   \end{equation*}
For the same reason as above, $\Bs_\circ$ is not a `\Bhat affinity coefficient' as such either --  which is quite obvious here, given the second term. Nevertheless, we can follow (\ref{eqn:BhatJ}) and estimate $\Bs_\circ(J)$ by 
\begin{equation} \widehat{\Bs}_\circ(J) \doteq  \frac{2}{\sqrt{\pi}} \frac{1}{\sqrt{n}} \sum_{i=1}^n|\mathring{g}^{(-i)}_J|(X_i) \sqrt{V_i} - \frac{1}{2}\int \mathring{g}^2_{J}(x)\,\diff x, \label{eqn:BcircJ} \end{equation} 
noting that the second term only involved the estimator and is, therefore, fully known to us. In fact, it is efficiently computed by \eqref{eqn:normgJ}. According to this criterion, the best resolution in (\ref{eqn:ringgJ}) would be
\begin{equation} \widehat{J}_\circ \doteq \arg \max_{J \in \{j_0,j_0+1,\ldots\}} \widehat{\Bs}_\circ(J). \label{eqn:Jhatcirc}  \end{equation}
The `best' unnormalised estimator would then be here $\mathring{g}_{\widehat{J}_\circ}$. The normalisation
\[ \hat{g}_{\widehat{J}_\circ} = \frac{\mathring{g}_{\widehat{J}_\circ}}{\|\mathring{g}_{\widehat{J}_\circ}\|},\]
makes $\hat{g}^2_{\widehat{J}_\circ}$ a shape-preserving estimator of $f$.

\subsection{Optimality}

Denote $J^* \doteq \arg \max_{J \in \{j_0,j_0+1,\ldots\}} \Bs(J)$ the true optimal resolution level in the sense of maximum \Bhat affinity coefficient/minimum Hellinger distance (\ref{eqn:bhat}), when these quantities do not need to be estimated. Likewise, let $J_\circ^* \doteq \arg \max_{J \in \{j_0,j_0+1,\ldots\}} \Bs_\circ(J)$. Below we prove that the empirically selected resolution levels $\widehat{J}$ and $\widehat{J}_\circ$ in (\ref{eqn:Jhat}) and (\ref{eqn:Jhatcirc}) are asymptotically equivalent to these `oracles' $J^*$ and $J^*_\circ$, hence asymptotically optimal in the Hellinger-\Bhat sense. 

\medskip
For simplicity of exposition, we re-write here the estimator (\ref{eqn:ringgJ}) as
\begin{equation} \mathring{g}_J(x) = \sum_{z \in \Z^d} \hat{\alpha}_{J+1,z}\varphi_{J+1,z}(x) \label{eqn:alphaonly}\end{equation}
with $\hat{\alpha}_{J+1,z} = \frac{2}{\sqrt{\pi}}\,\frac{1}{\sqrt{n}}\sum _{i=1}^n \varphi _{J+1,z}\left(X_i\right) \sqrt{V_{i}}$ like in (\ref{eqn:coefficients1}), as the MRA nature of the wavelet basis allows us to (see \cite[equation (13)]{Aya18}). Explicitly, this is 
\begin{align}
	\mathring{g}_J(x) &= \sum_{z \in \Z^d} \left( \frac{2}{\sqrt{\pi n}} \sum_{i=1}^n \varphi_{J+1,z}(X_i) \sqrt{V_{i}} \right)  \varphi_{J+1,z}(x) \nonumber \\
	&= \sum_{i=1}^n \left(\sum_{z \in \Z^d} \varphi_{J+1,z}(X_i) \varphi_{J+1,z}(x) \right)\frac{2}{\sqrt{\pi n}} \sqrt{V_{i}} \nonumber \\
	& \doteq \sum_{i=1}^n W_i \bdelta_{J}(x,X_i), \label{eqn:deltarootalpha}
\end{align}
where $\bdelta_{J}(x,X_i) = \sum_{z \in \Z^d} \varphi_{J+1,z}(X_i) \varphi_{J+1,z}(x)$ and $W_i = 2\sqrt{V_{i}/(\pi n)}$. It so appears that $\mathring{g}_J(x)$ is some sort of `weighted delta sequence estimator' for $\sqrt{f}$. 

\medskip 

`Delta sequence estimators' are well-studied density estimators, which write $\tilde{f}_\lambda(x) = n^{-1} \sum_{i=1}^n \bdelta_\lambda(x, X_i)$, where $\bdelta_\lambda: \Rd \times \Rd \rightarrow \R$ is such that, for any smooth function $\phi: \R^d \to \R$, $\lim_{\lambda \to 0} \int \bdelta_\lambda(x,y)\phi(y) \diff y = \phi(x)$. The parameter $\lambda$ typically acts as a smoothing parameter. Traditional kernel and orthogonal series estimators are known to be part of that class, among many others. Their properties have been studied in \cite{Marron87,Walter79,Walter92}; in particular, \cite[Theorem 2 and Corollary 2]{Marron87} showed that choosing the value $\lambda$ via cross-validation in the general framework of delta sequences is, in a strong sense, asymptotically equivalent to minimising the true Integrated Squared Error (ISE) of the estimator.

\medskip

Similar results apply here, noting that the ISE of the estimator $\mathring{g}_J$ for $\sqrt{f}$ is nothing else but $2\Hs_\circ^2(J)$ in (\ref{eqn:H2circJ}). A major subtlety, though, is the presence of the (random) weight $W_i$ in (\ref{eqn:deltarootalpha}) as well as in the `cross-validation' criterion (\ref{eqn:BcircJ}). Setting $\lambda \doteq 2^{-J}$ and combining the above results of \cite{Marron87} with those of \cite{Evans08} about nearest-neighbour-type of statistics, allow us to state the following result. It requires the consistency of estimator (\ref{eqn:ringgJ}) for $\sqrt{f}$, which has been established in \cite{Aya18} under standard assumptions (their Assumption 1-4); as well as two technical assumptions in Appendix. A full proof can be found in \cite[Theorem 8]{Aya20}.

\begin{theorem}\label{thm:nonlinears}
	Let $\Xs$ be a random sample from a distribution $F$ admitting a density $f$. Under Assumptions 1-4 in \cite{Aya18}, and Assumptions \ref{ass:optionsbounded}-\ref{ass:valuebounded} in Appendix, the selected resolution level $\widehat{J}_\circ$ which maximises $\widehat{\Bs}_\circ$ in (\ref{eqn:Jhatcirc}) is asymptotically optimal, in the sense that
	\[\lim_{n \rightarrow \infty} \frac{\Bs_\circ(\widehat{J}_\circ)}{ \Bs_\circ(J_\circ^*)} = 1 \quad\text{ a.s.}.\]
	Likewise, the selected resolution level $\widehat{J}$ which maximises $\widehat{\Bs}$ in (\ref{eqn:Jhat}) is asymptotically optimal, in the sense that
	\[\lim_{n \rightarrow \infty} \frac{\Bs(\widehat{J})}{ \Bs(J^*)} = 1 \quad\text{ a.s.}.\]
\end{theorem}

\medskip 

We investigate the empirical performance of these two approaches to resolution selection in Section \ref{sec:simulationnonlinear}. Before that, we describe a thresholding scheme for the coeffcient in estimator (\ref{eqn:ringgJ}).

\section{Thresholding}

Several options exist for wavelet thresholding \cite[Chapter 11]{Hardle12}, and any of these could be implemented on (\ref{eqn:ringgJ}). For the sake of simplicity, though, the exposition below focuses only on hard thresholding, whereby the $\beta$-coefficients are retained in (\ref{eqn:ringgJ}) only if they are larger in magnitude than a given value. We define:
\begin{equation}
	\tilde{\beta}^{(q)}_{j z;\kappa}\dot{=}\left\{\begin{array}{ll}{\hat{\beta}^{(q)}_{j z},} & {\text { if }\big|\hat{\beta}^{(q)}_{j z}\big|>\kappa\, \gamma_j n^{-1 / 2}} \\ {0,} & \text {otherwise }\end{array}\right.,
	\label{def:hardthreshold}
\end{equation}
where $\kappa >0$ is a constant to be determined and $\gamma_j$ is known and makes the threshold potentially level-dependent. As is customary, we do not seek to threshold the $\alpha$-coefficients in (\ref{eqn:ringgJ}). 
Then we have the thresholded estimator:
\begin{equation}
	\mathring{g}_{J,\kappa}(x)=\sum _{z \in \Zd} \hat{\alpha }_{j_0,z} \varphi_{j_0,z}(x) + \sum_{j=j_0}^{J} \sum_{z \in \Z^d} \sum_{q \in Q_d} \tilde{\beta}^{(q)}_{j z;\kappa} \, \psiqjz(x), \label{eqn:ringgJthresh}
\end{equation}  
which relies on two free parameters: the resolution level $J$ again, and the cut-off point $\kappa$. We propose to act in two steps: first, pick the right resolution level $J$ like in the previous section, via (\ref{eqn:BcircJ})-(\ref{eqn:Jhatcirc}). Writing the estimator as (\ref{eqn:alphaonly}) makes it clear that this may be achieved without reference to $j_0$. We will therefore assume that $J$ has been fixed and that $j_0$ is an arbitrary number of levels below -- typically, we can take $J- j_0 = O(\log n)$ as in \cite{Hall01}. Thus, in what follows, we consider both $j_0$ and $J$ as fixed. We may also set $\gamma_j=\sqrt{j - j_0 + 1}$ for $j = j_0,\ldots,J$, in (\ref{def:hardthreshold}), as advocated by \cite{Delyon96} -- we comment further on this at the end of this section. 

\ppn Call
\begin{equation} \hat{\lambda}_{j,z,q}= \sqrt{n}\frac{|\hat{\beta}^{(q)}_{j z}|}{\gamma_j} \label{eqn:lambda} \end{equation}
and let $\hat{\lambda}_{j^{[1]},z^{[1]},q^{[1]}},\hat{\lambda}_{j^{[2]},z^{[2]},q^{[2]}},\ldots$ denote these ranked lambda values in descending ordering; i.e.,
\[ t > t' \iff \hat{\lambda}_{j^{[t]},z^{[t]},q^{[t]}} < \hat{\lambda}_{j^{[t']},z^{[t']},q^{[t']}}. \] 
With a slight abuse of notation, let $\hat{\beta}_{[t]}$ and $\psi_{[t]}(x)$ refer to the values of $j,q,z$ corresponding to $\hat{\lambda}_{j^{[t]},z^{[t]},q^{[t]}}$. This allows us to re-write (\ref{eqn:ringgJthresh}) as 
\begin{equation}
	\mathring{g}_{\tau}(x) \doteq \sum _{z \in \Zd} \hat{\alpha }_{j_0,z} \varphi _{j_0,z}(x) + \sum_{t=1}^{\tau} \hat{\beta}_{[t]} \, \psi_{[t]}(x) \label{def:ghato}
\end{equation}
where $\tau \doteq \tau(\kappa) = \max\{ t \in \N:\hat{\lambda}_{j^{[t]},z^{[t]},q^{[t]}} > \kappa \}$, following (\ref{def:hardthreshold}). Under this form, selecting $\kappa \in \R^+$ or equivalently $\tau \in \N$, becomes similar to the problem of selecting $J$ in the previous section, as it is about determining the truncation point in a orthogonal series expansion -- as mentioned, this truncation point is a smoothing parameter in delta sequence estimators \cite{Marron87,Walter79}. As a result, the Hellinger/\Bhat cross-validation approach described above may be taken here as well.

\ppn Specifically, similarly to (\ref{eqn:BcircJ}), we define 
\begin{equation} \widehat{\Bs}_\circ(\tau) \doteq  \frac{2}{\sqrt{\pi}} \frac{1}{\sqrt{n}} \sum_{i=1}^n|\mathring{g}_{\tau}^{(-i)}|(X_i) \sqrt{V_i} - \frac{1}{2}\int \mathring{g}^2_{\tau}(x)\,\diff x, \label{eqn:BcircTau} \end{equation} 
and we select the empirically optimal threshold $\tau$ as
\begin{equation} \widehat{\tau}_\circ = \arg \max_{\tau \in \N} \widehat{\Bs}_\circ(\tau). \label{eqn:tauhatcirc} \end{equation}
This `best' non-linear un-normalised estimator $\mathring{g}_{\widehat{\tau}_\circ}$ is finally normalised as in (\ref{eqn:hatgJ}):
\[\hat{g}_{\widehat{\tau}_\circ} = \frac{\mathring{g}_{\widehat{\tau}_\circ}}{\|\mathring{g}_{\widehat{\tau}_\circ}\|},\]
which is such that $\hat{g}^2_{\widehat{\tau}_\circ}$ is a shape-preserving non-linear wavelet estimator of $f$, always. Note that, for each candidate threshold $\tau$, we could also normalise (\ref{def:ghato}) directly and use this {\it bona fide} density estimate $\hat{g}_{{\tau}} \doteq \mathring{g}_{{\tau}}/\|\mathring{g}_{{\tau}}\|$ in the empirical \Bhat coefficient. We would have:
\begin{equation} \widehat{\Bs}(\tau) \doteq  \frac{2}{\sqrt{\pi}} \frac{1}{\sqrt{n}} \sum_{i=1}^n|\hat{g}_{\tau}^{(-i)}|(X_i) \sqrt{V_i}, \label{eqn:Bhattau} \end{equation}
defining the optimal threshold as 
\begin{equation} \widehat{\tau} = \arg \max_{\tau \in \N} \widehat{\Bs}(\tau). \label{eqn:tauhat} \end{equation}
These two approaches obviously parallel the criteria (\ref{eqn:Jhatcirc}) and (\ref{eqn:Jhat}) developed in Section \ref{subsec:empBhat}. Similarly to Theorem \ref{thm:nonlinears}, the optimality of these threshold selection rules can be established -- see \cite{Aya20}. 

\medskip 

Now, we propose a novel variation around \eqref{def:hardthreshold}. Ideally, the threshold for a given estimated coefficient $\hat{\beta}^{(q)}_{j,z}$ should depend on its standard error -- essentially, from a testing perspective, we set that coefficient to zero if we cannot show that $\beta^{(q)}_{j,z}$ is significantly different to zero, and such a testing procedure would obviously involve the standard error of the estimator. This motivates the level-dependent constant $\gamma_j$ in \eqref{def:hardthreshold}: the higher $j$, the narrower the support of the wavelets $\psiqjz$, hence the smaller the number of non-zero terms $\psiqjz(X_i)$ and the smaller the number of observations effectively used $\hat{\beta}^{(q)}_{j,z}$. Therefore, $\gamma_j n^{-1/2}$ in \eqref{def:hardthreshold} should be a rough proxy for the increasing standard error of $\hat{\beta}^{(q)}_{j,z}$. E.g, \cite{Delyon96} suggested $\gamma_j = \sqrt{j-j_0+1}$.

\medskip

In our framework, however, we have a seemingly better alternative. Indeed the leave-one-out estimator $\hat{g}^{(-i)}_{\tau}$ appearing in (\ref{def:ghato}) requires computing $\left(\hat{\beta}^{(q)}_{j,z}\right)^{(-i)}$ for each $i = 1, \ldots, n$ -- as explained in the lines below (\ref{eqn:BhatJ}), this can be done very efficiently. Now, these `leave-one-out' coefficients may be used to produce jackknife pseudo-values of $\hat{\beta}^{(q)}_{j,z}$ \cite{Efron81,Miller74,Tukey58}; viz.
\[\left(\hat{\beta}^{(q)}_{j,z}\right)_{i} \doteq n \hat{\beta}^{(q)}_{j,z} - (n-1)\left(\hat{\beta}^{(q)}_{j,z}\right)^{(-i)}.\]
If one calls $v^{2(q)}_{j,z}$ the variance of this collection of pseudo-values $\left\{\left(\hat{\beta}^{(q)}_{j,z}\right)_{i} \right\}_{i=1}^n$, then it can be shown \cite{Efron81} that $\widehat{\sigma}^{2(q)}_{\text{jack};j,z} \doteq v^{2(q)}_{j,z}/n$ is a consistent estimator of the variance of $\hat{\beta}^{(q)}_{j,z}$. This suggests the novel thresholding rule
\begin{equation} 
	\frac{\left|\hat{\beta}^{(q)}_{j z}\right|}{\widehat{\sigma}^{(q)}_{\text{jack};j,z}}>\kappa \label{eqn:lambda2}
\end{equation}
in \eqref{def:hardthreshold}. We have thus eliminated the need for a somewhat arbitrary choice for $\gamma_j$, a subject of theoretical controversy \cite{Delyon96,DonohoJohnstone96,DJKP95}, pushing further for a fully data-driven threshold design. From there, the empirical determination of $\kappa$ (and/or $\tau$) can be carried out exactly as above, just redefining $\hat{\lambda}_{j,z,q}$ appropriately in \eqref{eqn:lambda}. The simulations which we run in Section \ref{subsec:thresholdingsimulation} show that this approach is superior in most cases -- future theoretical work is required to analyse this `jackknife thresholding' further, though.

\section{Numerical experiments}\label{sec:simulationnonlinear}

In this section we investigate the practical performance of our proposed methodology. As described above, the final, non-linear estimator involves two steps: first, we determine the best overall resolution $J$, and second we implement the hard-thresholding scheme on the coefficients regarding that resolution as fixed. In the same manner, this section follows the same strategy: first, it presents results on the performance of the novel Hellinger-\Bhat approach for selecting the best resolution, and second, it investigates the various thresholding options and discuss them over a set of known probability density functions. Lastly, we conclude this section with a practical popular example with an unknown PDF, the Old Faithful geyser dataset, to illustrate a real application.

\subsection{Resolution level} \label{subsec:resolutionlevelsimulation}

\begin{figure}[t!]
	\begin{subfigure}[t]{0.5\textwidth}
		\includegraphics[width=\textwidth]{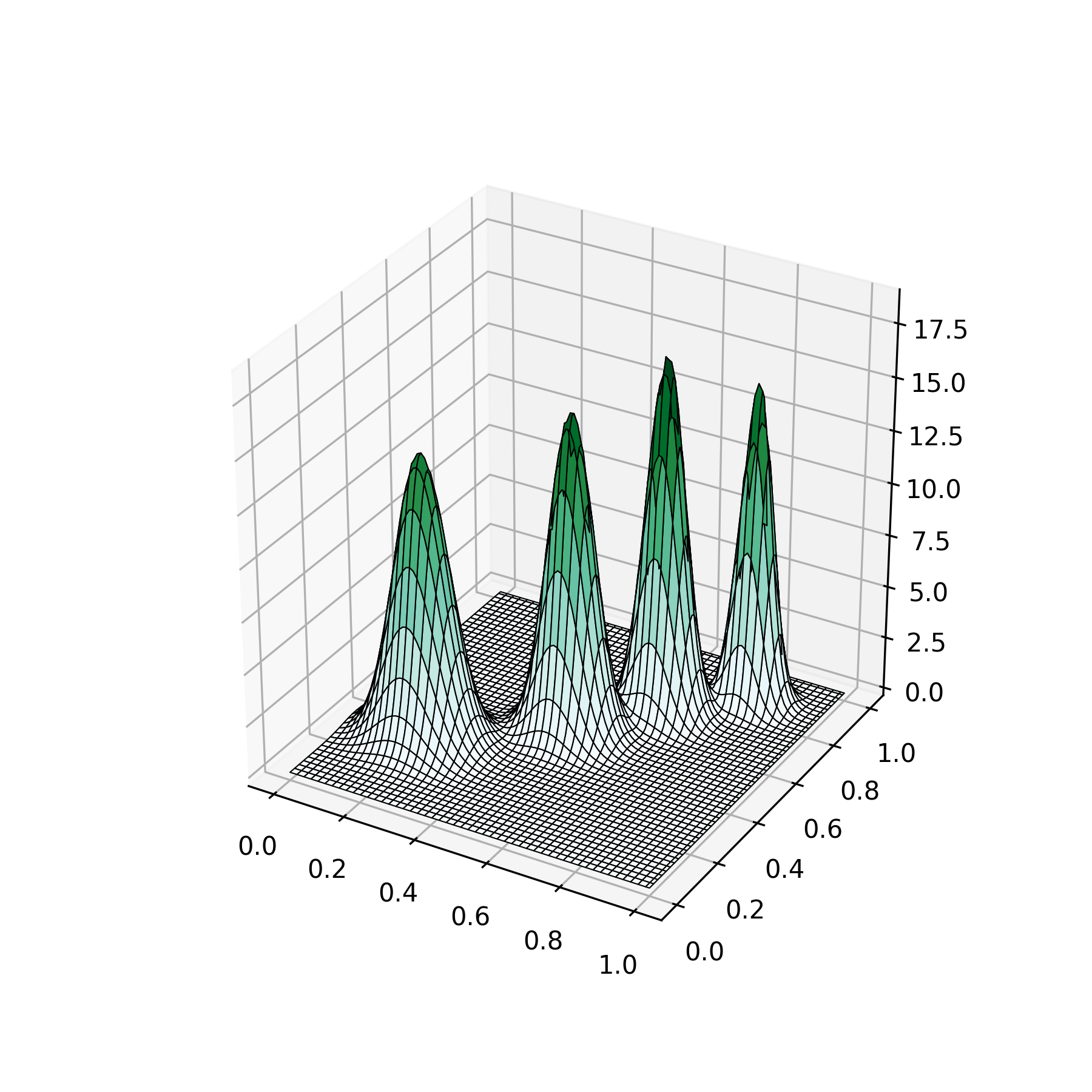}
		\caption{2D Comb}
	\end{subfigure}
	\begin{subfigure}[t]{0.5\textwidth}
		\includegraphics[width=\textwidth]{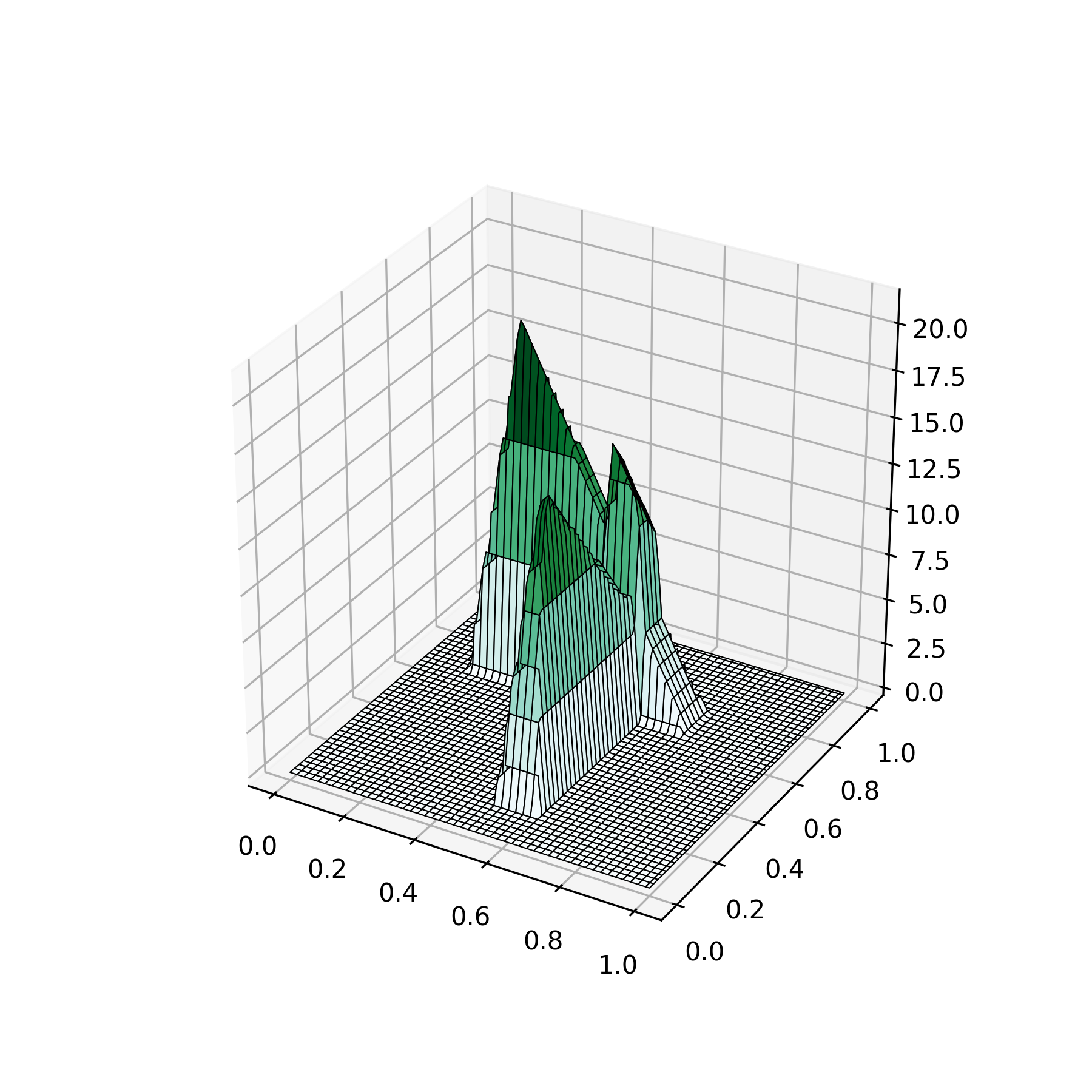}
		\caption{Pyramids mix}
	\end{subfigure}
	\begin{subfigure}[t]{0.5\textwidth}
		\includegraphics[width=\textwidth]{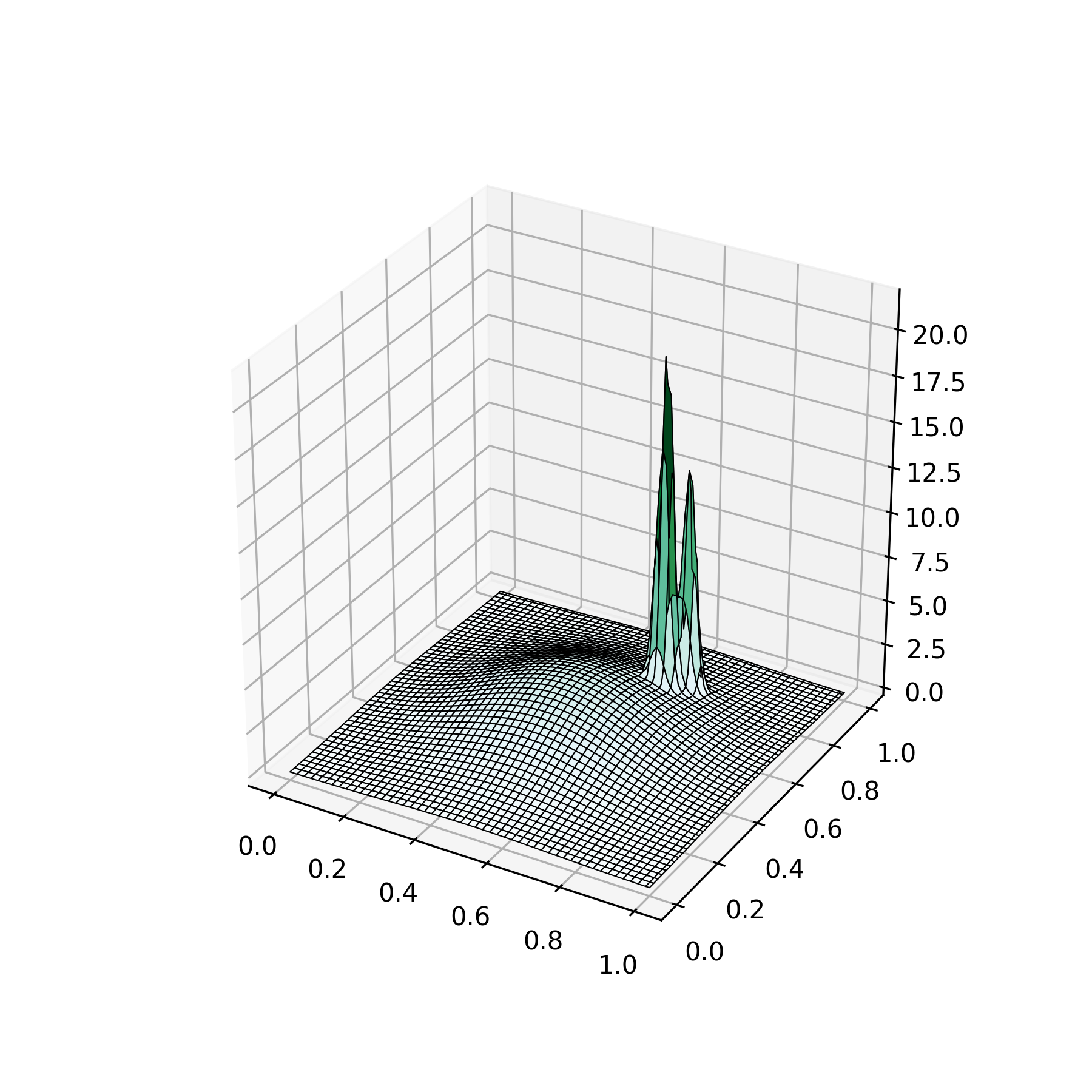}
		\caption{2D Gaussian mix 1}
	\end{subfigure}
	\begin{subfigure}[t]{0.5\textwidth}
		\includegraphics[width=0.9\textwidth]{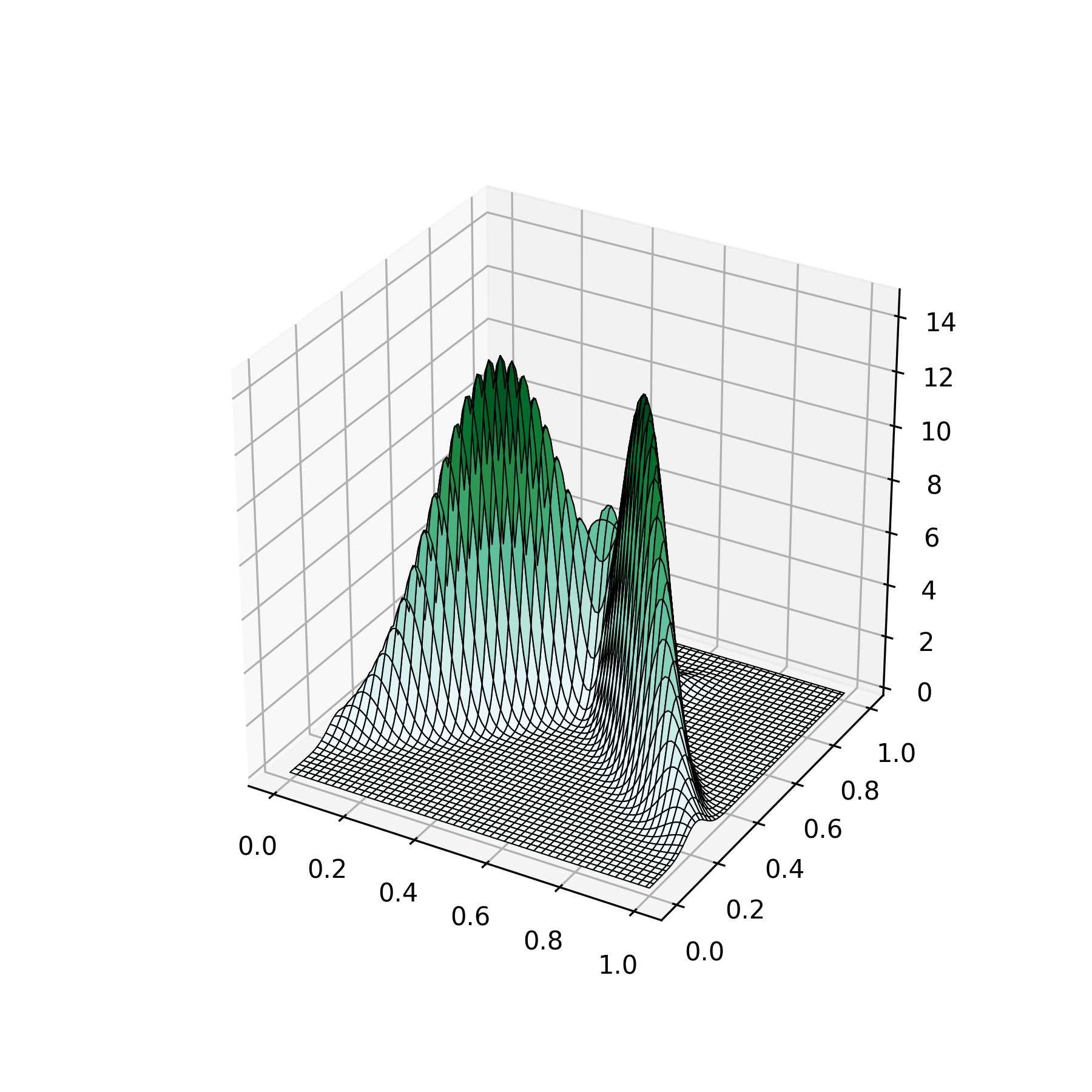}
		\caption{2D Gaussian mix 2}
	\end{subfigure}
	\caption{Densities used in `best $J$' simulation study.} \label{fig:truedensj}
\end{figure}

We proposed two Hellinger-\Bhat criteria (\ref{eqn:Jhat}) and (\ref{eqn:Jhatcirc}) for selecting $J$, depending on if the normalisation of the estimator is applied automatically at every step of the procedure or only one time at the end of it. Here these two options are compared in terms of their accuracy of estimation. For our numerical experiment, we have considered mixtures of different number of components, smoothness and covariance structures:  (a) a bivariate Gaussian comb (`2D comb'); (b) a mixture of two bivariate {\it pyramid} densities\footnote{These {\it pyramids} are constructed by taking the tensor product of two piecewise functions and ensuring it integrates to one.}, i.e. non smooth (`pyramids mix'); (c) a bivariate Gaussian mixture with very different spreads  (`2D Gaussian mix 1'); and (d) a bivariate mixture of Gaussian densities with very elongated covariances in different directions (`2D Gaussian mix 2'). These densities are shown in Figure \ref{fig:truedensj}; analytic expressions are found in \cite{Aya20}. 

\medskip 

For each of these densities and for each considered sample sizes $n$, 100 random samples were generated. From the true density, we can compute the true distances between estimators $\hat{g}^2_J$ or $\mathring{g}^2_J$ and $f$ as in (\ref{eqn:Hell1}) and (\ref{eqn:H2circJ}) and from this, identify the `oracle' best resolutions $J^*$ and $J^*_\circ$. For each sample, we obtain the practical resolutions $\widehat{J}$ and $\widehat{J}_\circ$ as per (\ref{eqn:Jhat}) and (\ref{eqn:Jhatcirc}). 

\medskip 

\begin{figure}[h]
	\begin{subfigure}[t]{0.49\textwidth}
		\includegraphics[width=\textwidth]{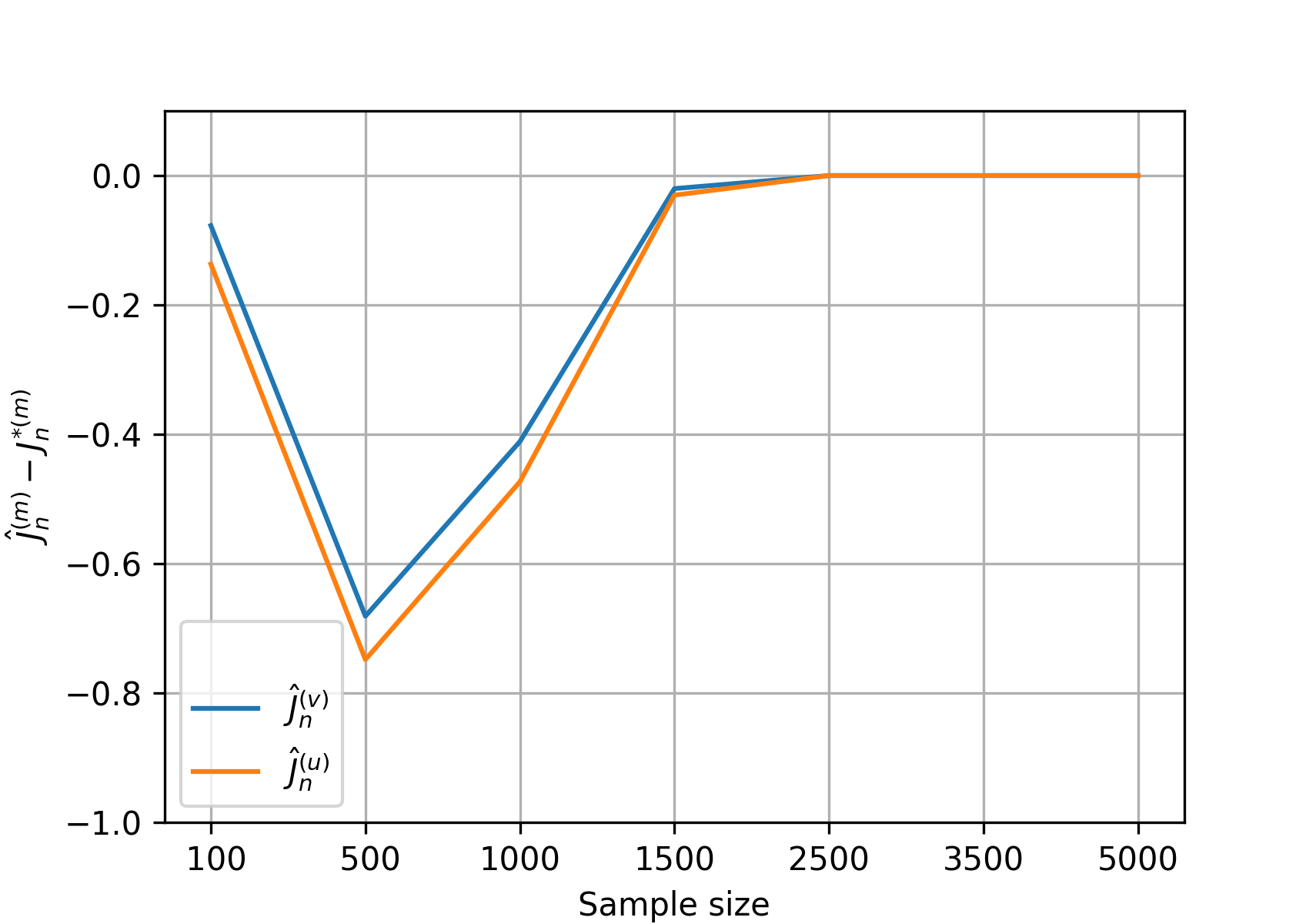}
		\caption{Average difference $\widehat{J}-J^*$ and $\widehat{J}_\circ-J_\circ^*$}
	\end{subfigure}
	\begin{subfigure}[t]{0.49\textwidth}
		\includegraphics[width=\textwidth]{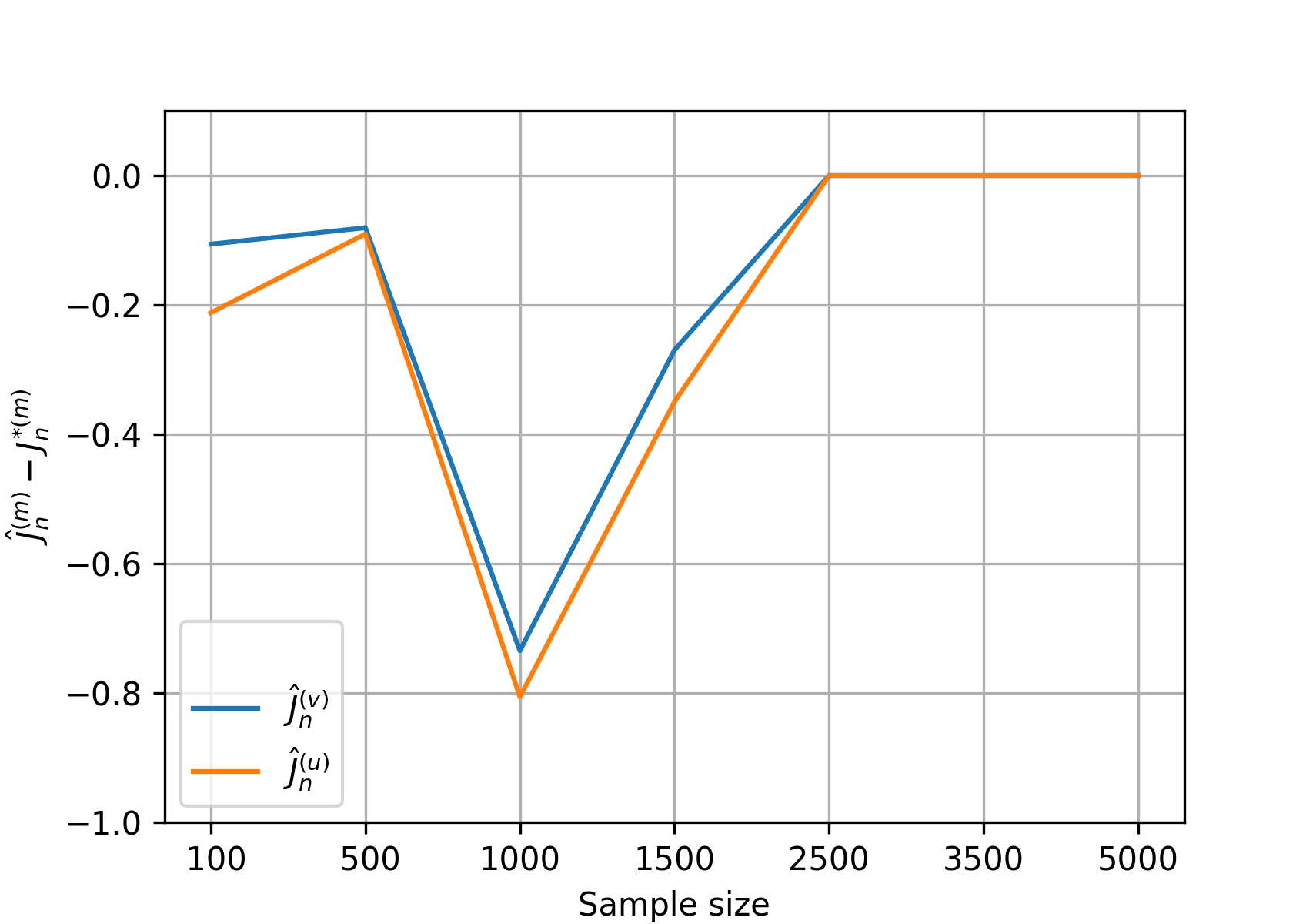}
		\caption{Average difference $\widehat{J}-J^*$ and $\widehat{J}_\circ-J_\circ^*$}
	\end{subfigure}
	\begin{subfigure}[t]{0.49\textwidth}
		\includegraphics[width=\textwidth]{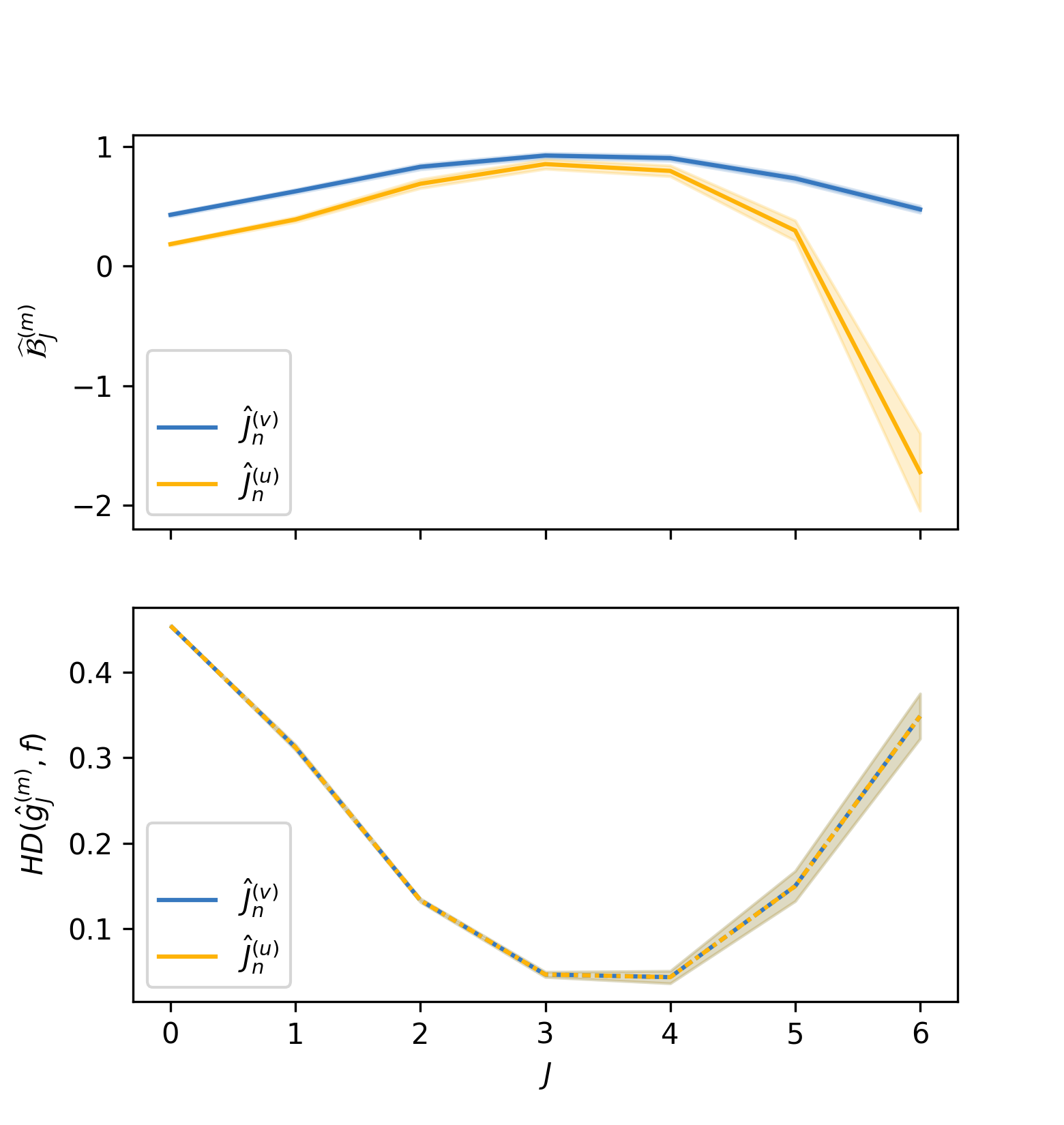}
		\caption{$\widehat{B}(J)$ and $\widehat{B}_\circ(J)$ (top) and $\Hs^2(J)$ (bottom); $n=500$}
	\end{subfigure}
	\begin{subfigure}[t]{0.49\textwidth}
		\includegraphics[width=\textwidth]{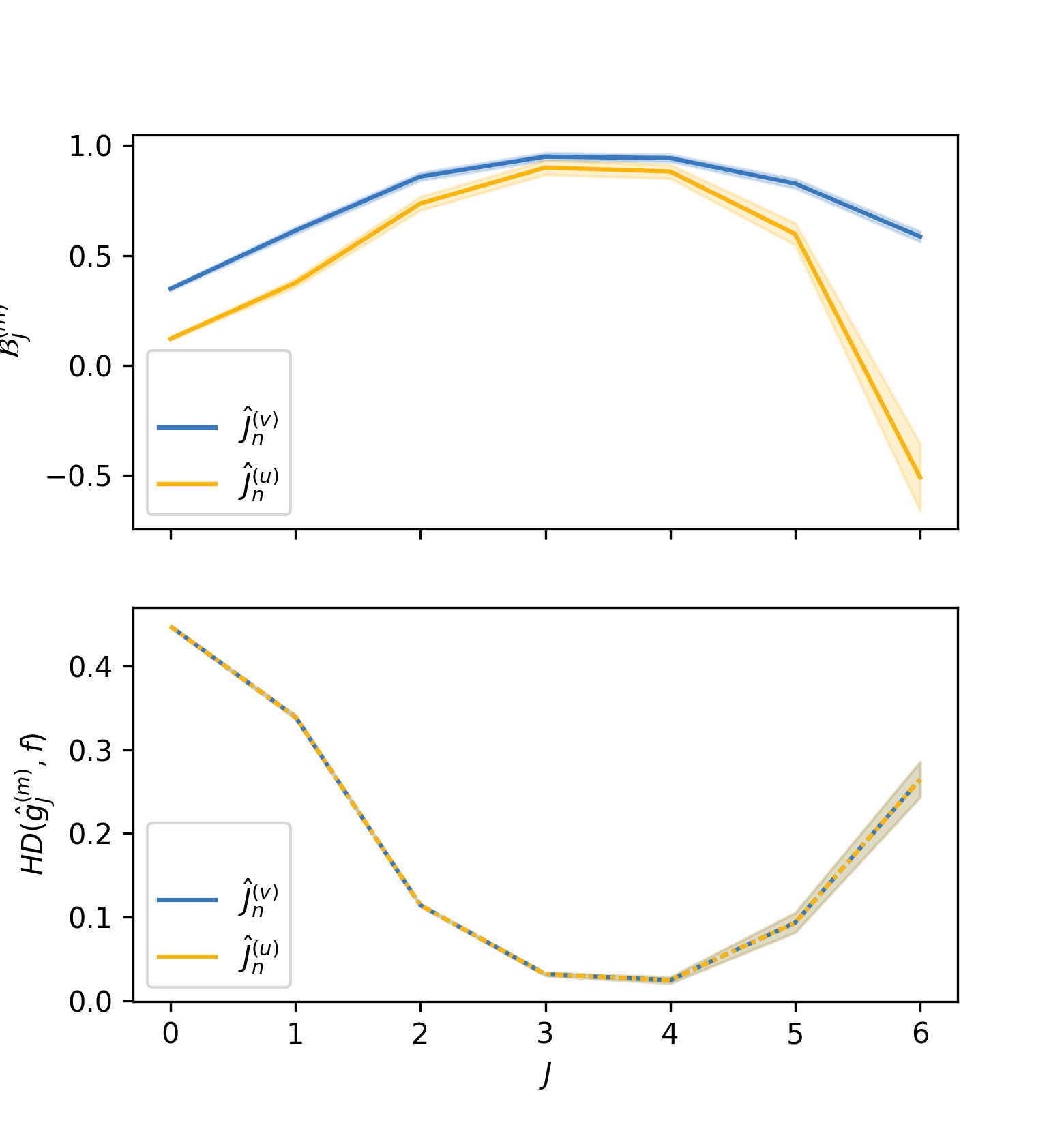}
		\caption{$\widehat{B}(J)$ and $\widehat{B}_\circ(J)$ (top) and $\Hs^2(J)$ (bottom); $n=1000$}
	\end{subfigure}
	\caption{Results for resolution selection rule; density \ref{fig:truedensj}(a) (Left: Daubechies 4, Right: Symlet 6).} \label{fig:bestjmix8}
\end{figure}

Figure \ref{fig:bestjmix8} (top row) show the average differences $\widehat{J}-J^*$ and $\widehat{J}_\circ-J_\circ^*$ for the `2D comb' (Figure \ref{fig:truedensj}(a)) as a function of the sample size $n$ for two choices of wavelet basis: Daubechies 4 (a) and Symlet 6 (b). In agreement with the theory (Theorem \ref{thm:nonlinears}), the empirical criterion always identifies the best `oracle' resolution level for $n$ `large'. For small samples (here: $n=100$), the `true' $J$ is expected/known to be low, i.e., one or two units above $j_0$ most certainly. The empirical \Bhat criterion recovers this rather easily. In between, when there exists a difference between $\widehat{J}$ and $J^*$ (or between  $\widehat{J}_\circ$ and $J_\circ^*$), it is of one unit only. Such error likely happens around a `critical' sample size $\tilde{n}$, for which the theoretically optimal $J^*$ switches from some $j^*$ for $n < \tilde{n}$ to $j^*+1$ for $n > \tilde{n}$ -- but this is not yet obvious empirically. The empirical procedure is, therefore, conservative, which explains why the average difference is always negative. The highest average difference occurs at $n=500$ and $n=1000$ for the Daubechies 4 and symlet 6 wavelet basis, respectively. This discrepancy is likely due to the different local geometry of these wavelet bases.

\medskip 

This is better understood in Figure \ref{fig:bestjmix8}(c), which shows the empirical \Bhat coefficients (\ref{eqn:Jhat})-(\ref{eqn:Jhatcirc}) and genuine Hellinger distances (\ref{eqn:Hell1}) as a function of $J$ for the Daubechies 4 wavelet at sample size $n=500$; and Figure \ref{fig:bestjmix8}(d) which shows the same for $n=1000$ and the Symlet 6 wavelet. In the first row, $\widehat{\Bs}(J)$, as genuine \Bhat affinity coefficient, is always in $[0,1]$, as it must be by definition. By contrast, $\widehat{\Bs}_\circ(J)$ does not necessarily satisfy this constraint -- which is not really a concern as we are only looking for the location of the maximum. In the area of the maximum, the two criteria behave similarly anyway: for $\mathring{g}^2_J$ to resemble $f$ as much as possible, it must behave like a density, even if not structurally forced to be so. The bottom row shows the average Hellinger distances (\ref{eqn:Hell1}) between the final normalised estimator and the true $f$ -- the two curves (one for the estimator based on $\widehat{J}$, the other one based on $\widehat{J}_\circ$) seem superimposed on each other as the difference is minimal (the final estimate is always normalised, regardless on how $J$ was selected). 

\medskip

For these sample sizes, which correspond to the highest differences in subfigures (a)-(b), the genuine Hellinger distance shows that $J=4$ would be the best resolution level here, but $J=3$ is a very close second. Indeed we are at that sample size where the theoretical best $J^*$ just switched from 3 to 4. Hence the empirical \Bhat criterion may sometimes pick $J=3$, as easily understood from the appearance of the $\widehat{\Bs}(J)$ and $\widehat{\Bs}_\circ(J)$ curves (almost flat between $J=3$ and $J=4$). This clarifies that the discrepancies between $\widehat{J}$ and $J^*$ (or $\widehat{J}_\circ$ and $J^*_\circ$) observed at some sample sizes do not flag a fail of the method; they just happen when $J^*$ and $\widehat{J}$ (or $\widehat{J}_\circ$ and $J^*_\circ$) lead to very similar (empirically undistinguishable) accuracy of estimation.

\medskip 

\begin{figure}[h]
	\begin{subfigure}[t]{0.5\textwidth}
		\includegraphics[width=\textwidth]{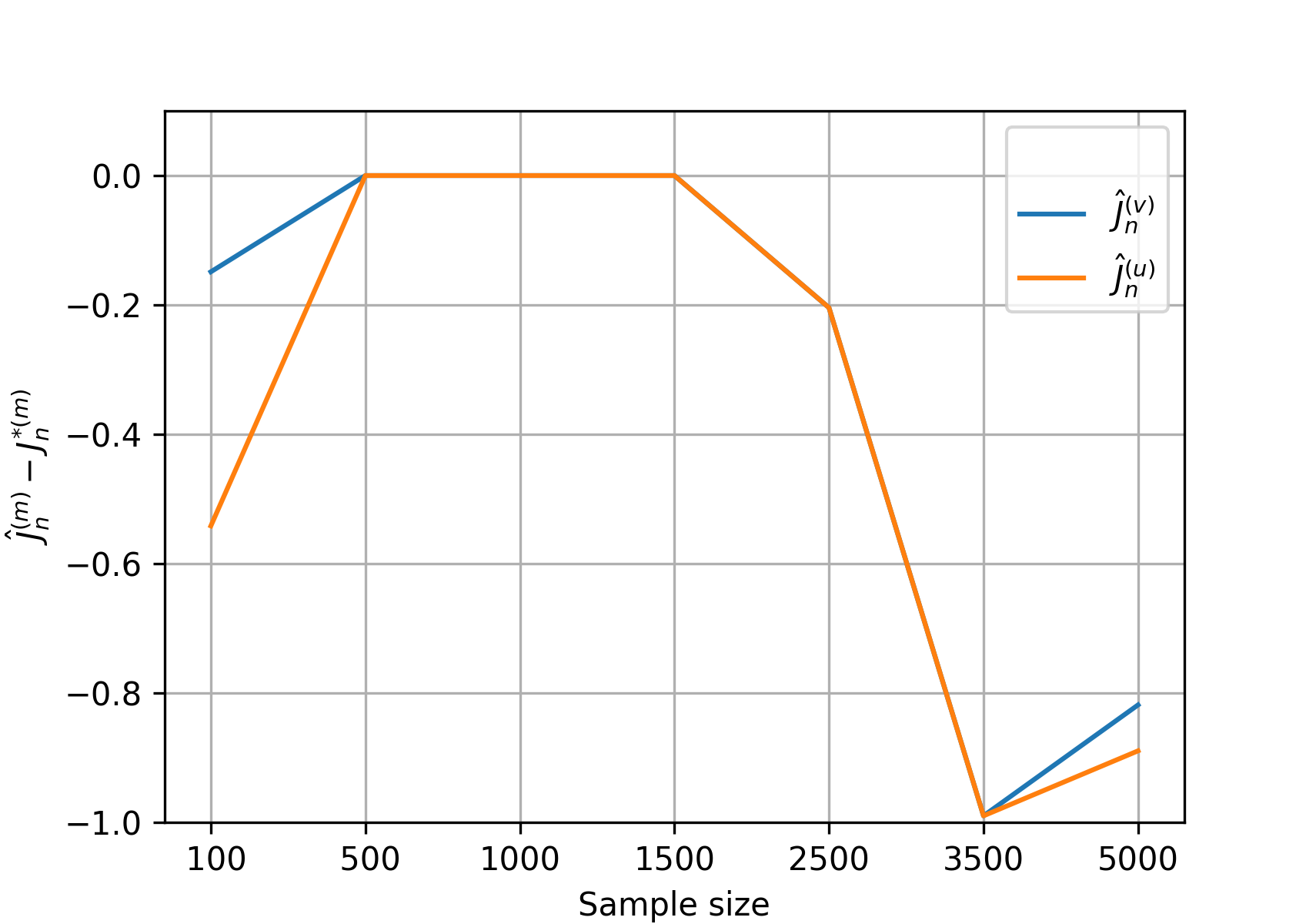}
		\caption{Average difference $\widehat{J}-J^*$ and $\widehat{J}_\circ-J_\circ^*$}
	\end{subfigure}
	\begin{subfigure}[t]{0.5\textwidth}
		\includegraphics[width=\textwidth]{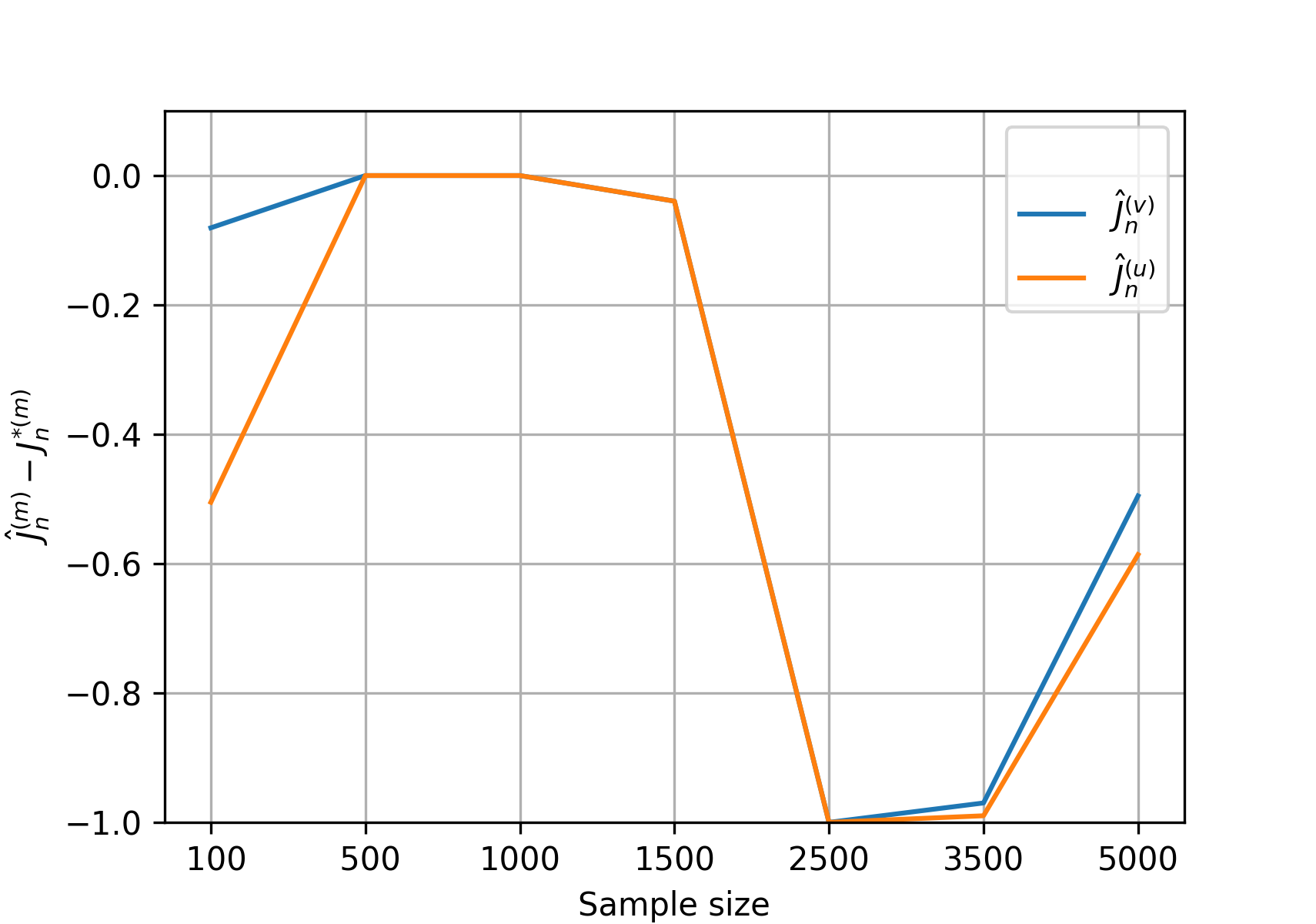}
		\caption{Average difference $\widehat{J}-J^*$ and $\widehat{J}_\circ-J_\circ^*$}
	\end{subfigure}
	\begin{subfigure}[t]{0.5\textwidth}
		\includegraphics[width=\textwidth]{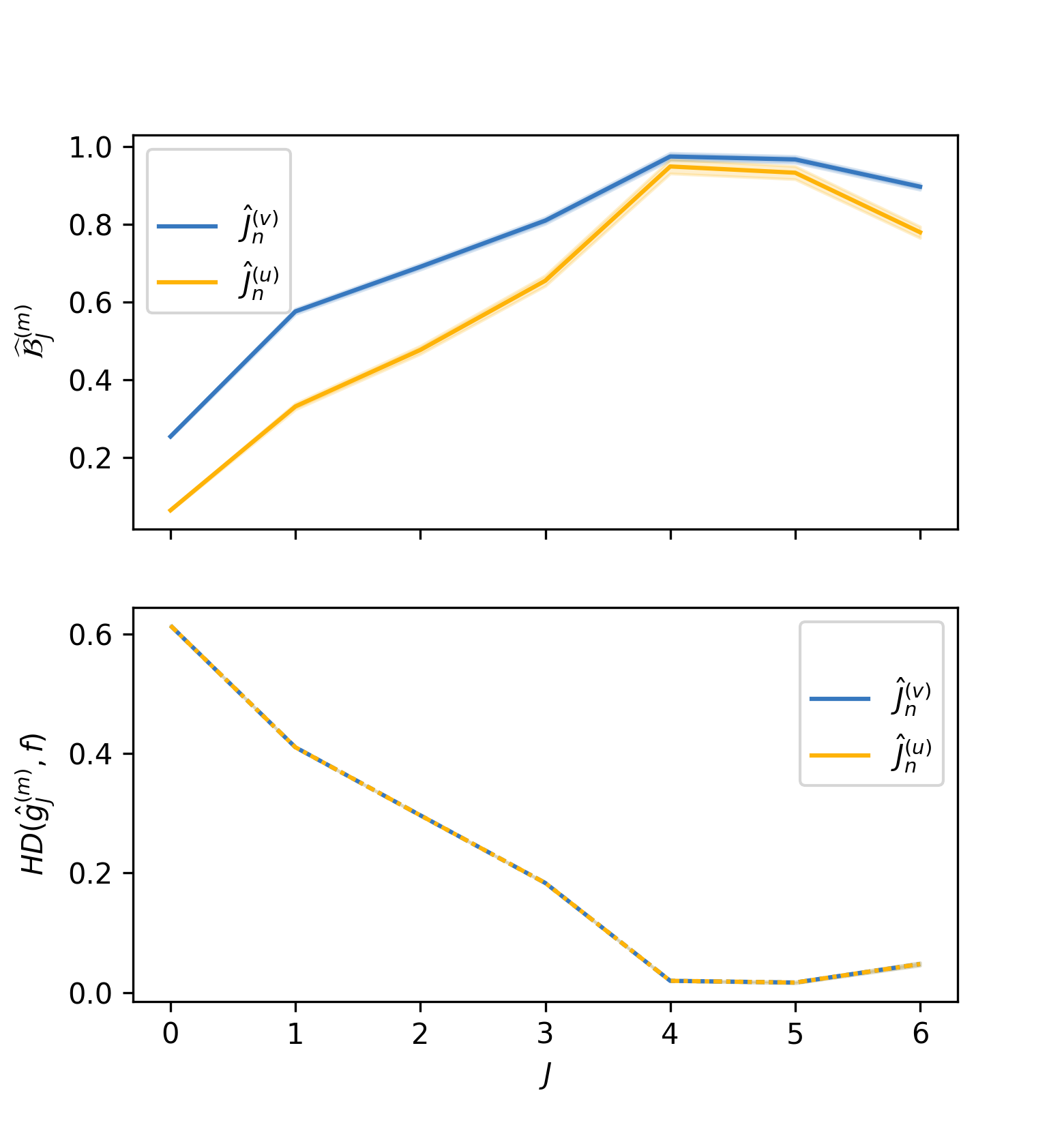}
		\caption{$\widehat{B}(J)$ and $\widehat{B}_\circ(J)$ (top) and $\Hs^2(J)$ (bottom); $n=3500$}
	\end{subfigure}
	\begin{subfigure}[t]{0.5\textwidth}
		\includegraphics[width=\textwidth]{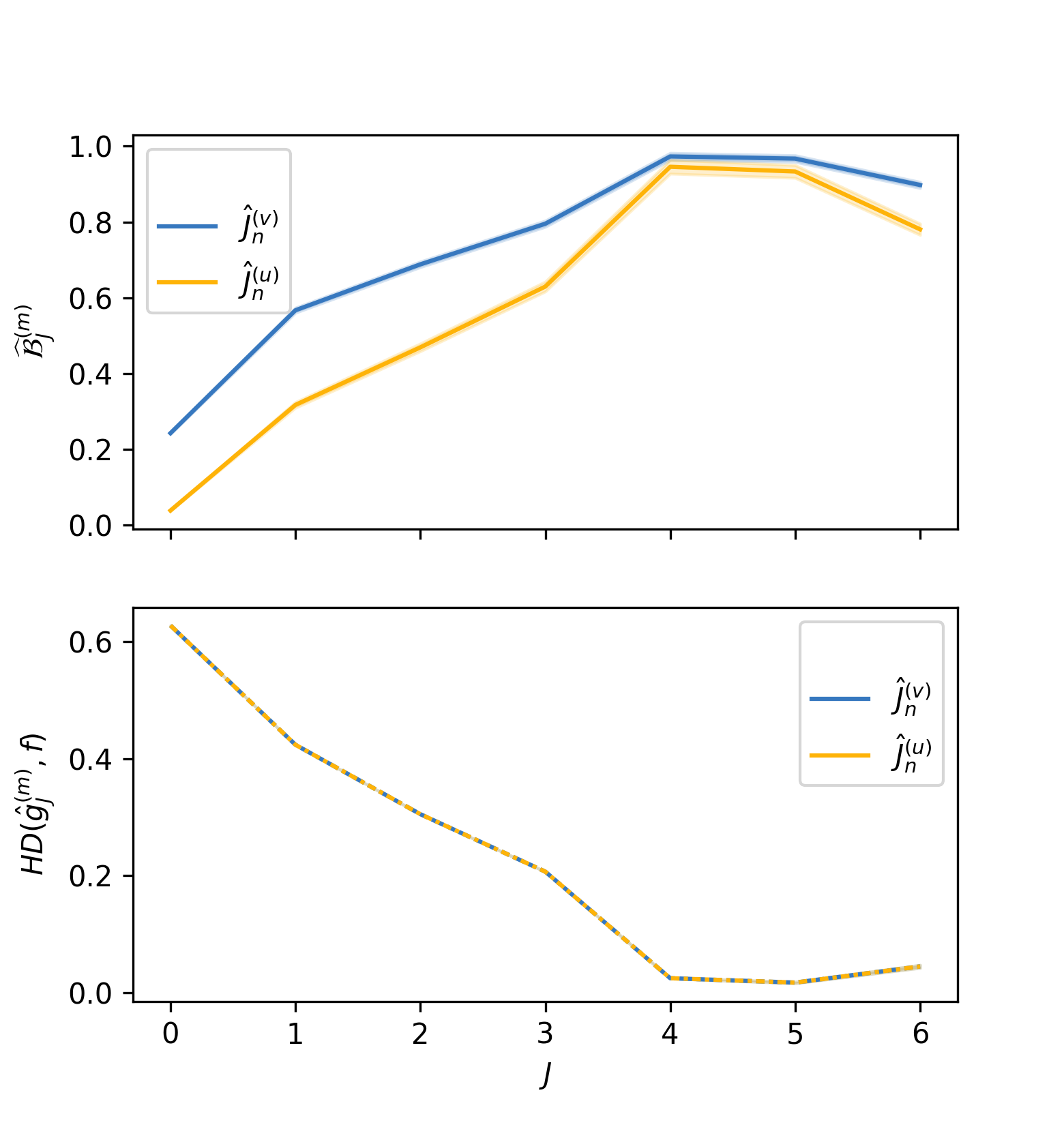}
		\caption{$\widehat{B}(J)$ and $\widehat{B}_\circ(J)$ (top) and $\Hs^2(J)$ (bottom); $n=3500$}
	\end{subfigure}
	\caption{Results for resolution selection rule; density \ref{fig:truedensj}(b) (Left: Symlet 6, Right: biorthogonal spline 2.8).} \label{fig:bestjtmx4}
\end{figure}

We note that this phenomenon may well occur at higher sample sizes, too -- and this is not in disagreement with Theorem \ref{thm:nonlinears}. For example, Figure \ref{fig:bestjtmx4} shows the same plots, this time for the `pyramids density' (Figure \ref{fig:truedensj}(b)). We see that, for $n=3500$, the Hellinger curve takes almost identical values at both $J=4$ or $J=5$, with only a slight advantage for $J=4$. The empirical \Bhat coefficient, on the other hand, shows a peak on the $J=4$ side. This explains the appearance of the curves shown in (a) and (b), with the same conclusion as above -- that it is mostly inconsequential.

\medskip

\begin{figure}[h]
	\begin{subfigure}[t]{0.5\textwidth}
		\includegraphics[width=\textwidth]{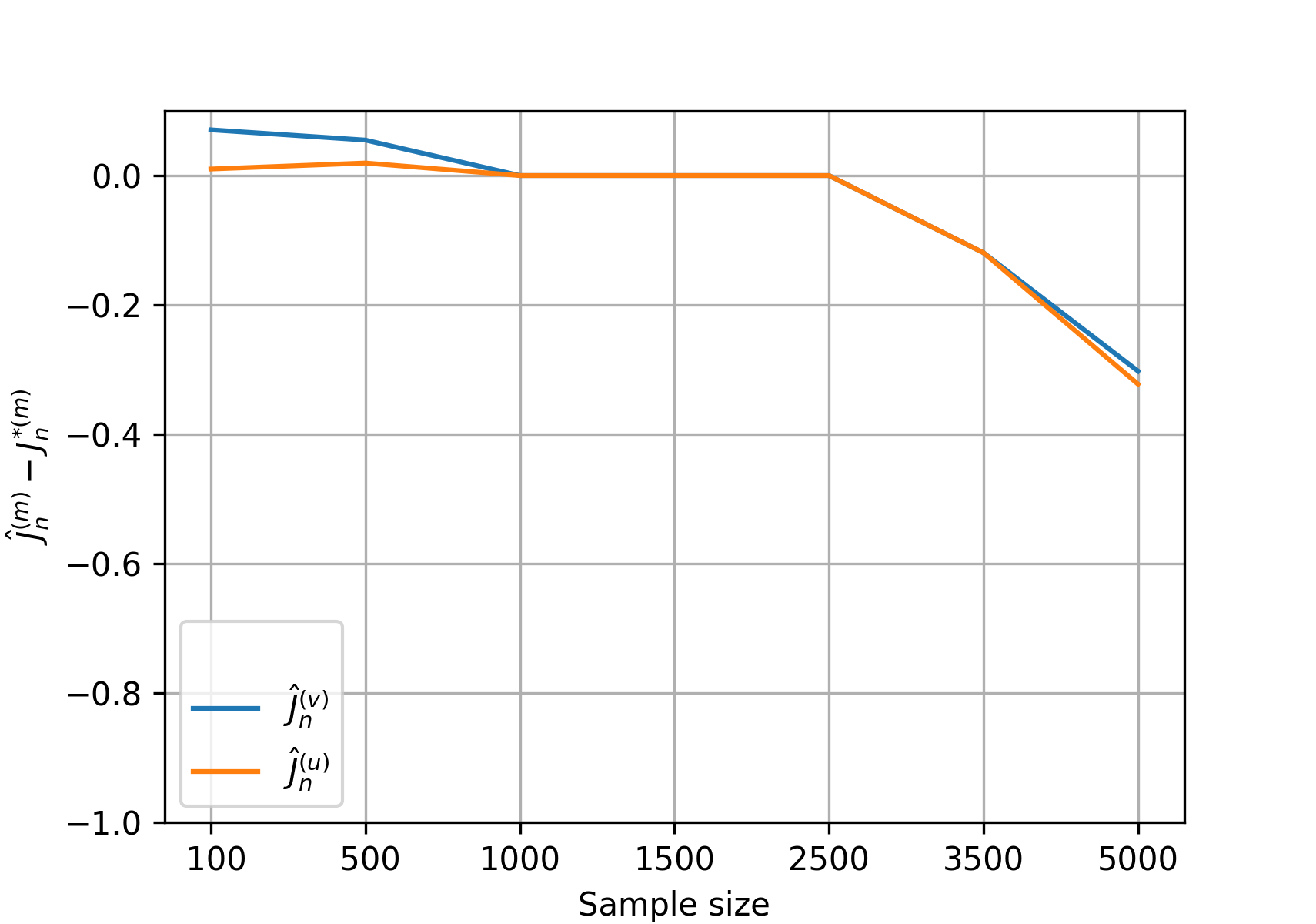}
		\caption{Average difference $\widehat{J}-J^*$ and $\widehat{J}_\circ-J_\circ^*$}
	\end{subfigure}
	\begin{subfigure}[t]{0.5\textwidth}
		\includegraphics[width=\textwidth]{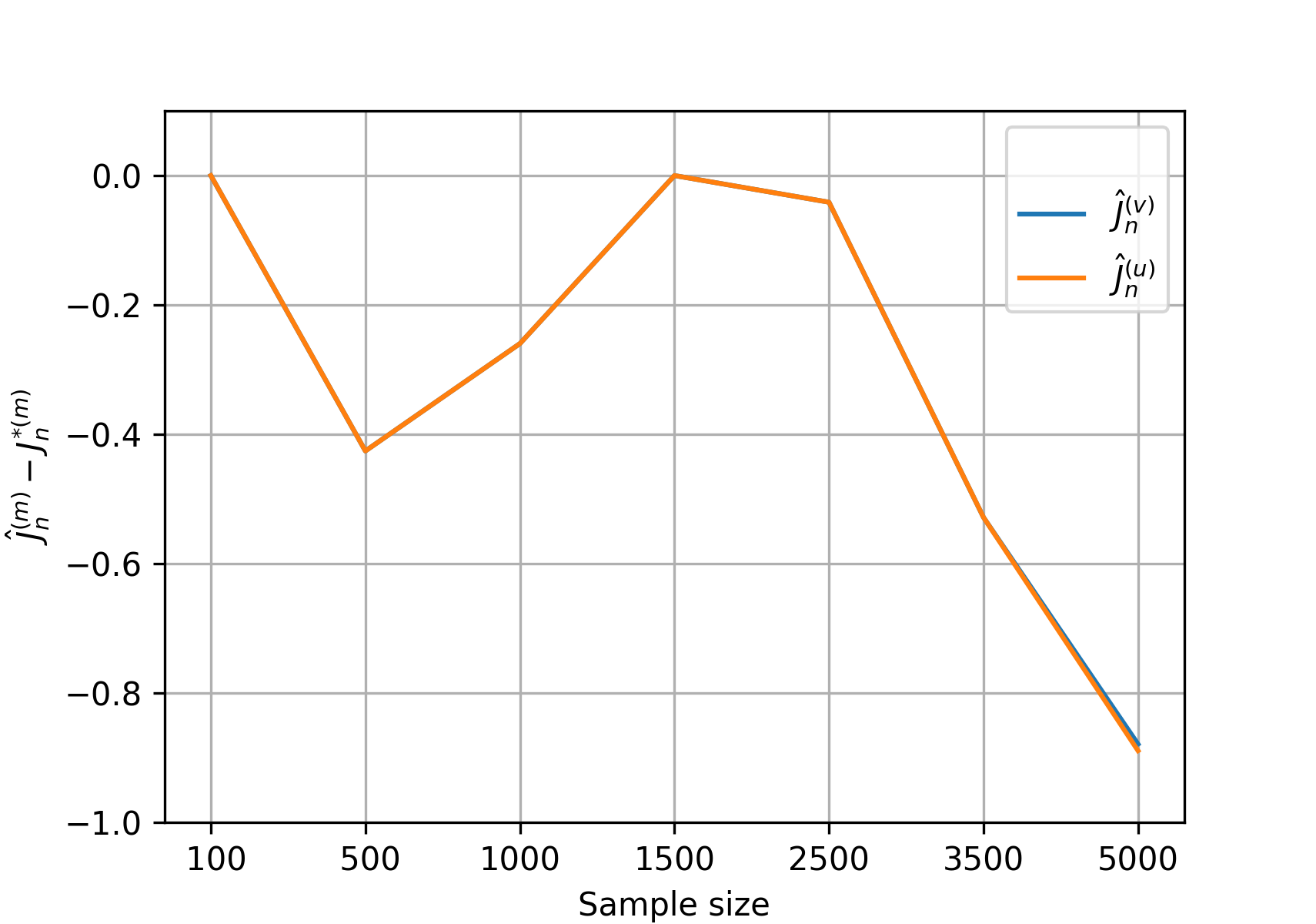}
		\caption{Average difference $\widehat{J}-J^*$ and $\widehat{J}_\circ-J_\circ^*$}
	\end{subfigure}
	\begin{subfigure}[t]{0.5\textwidth}
		\includegraphics[width=\textwidth]{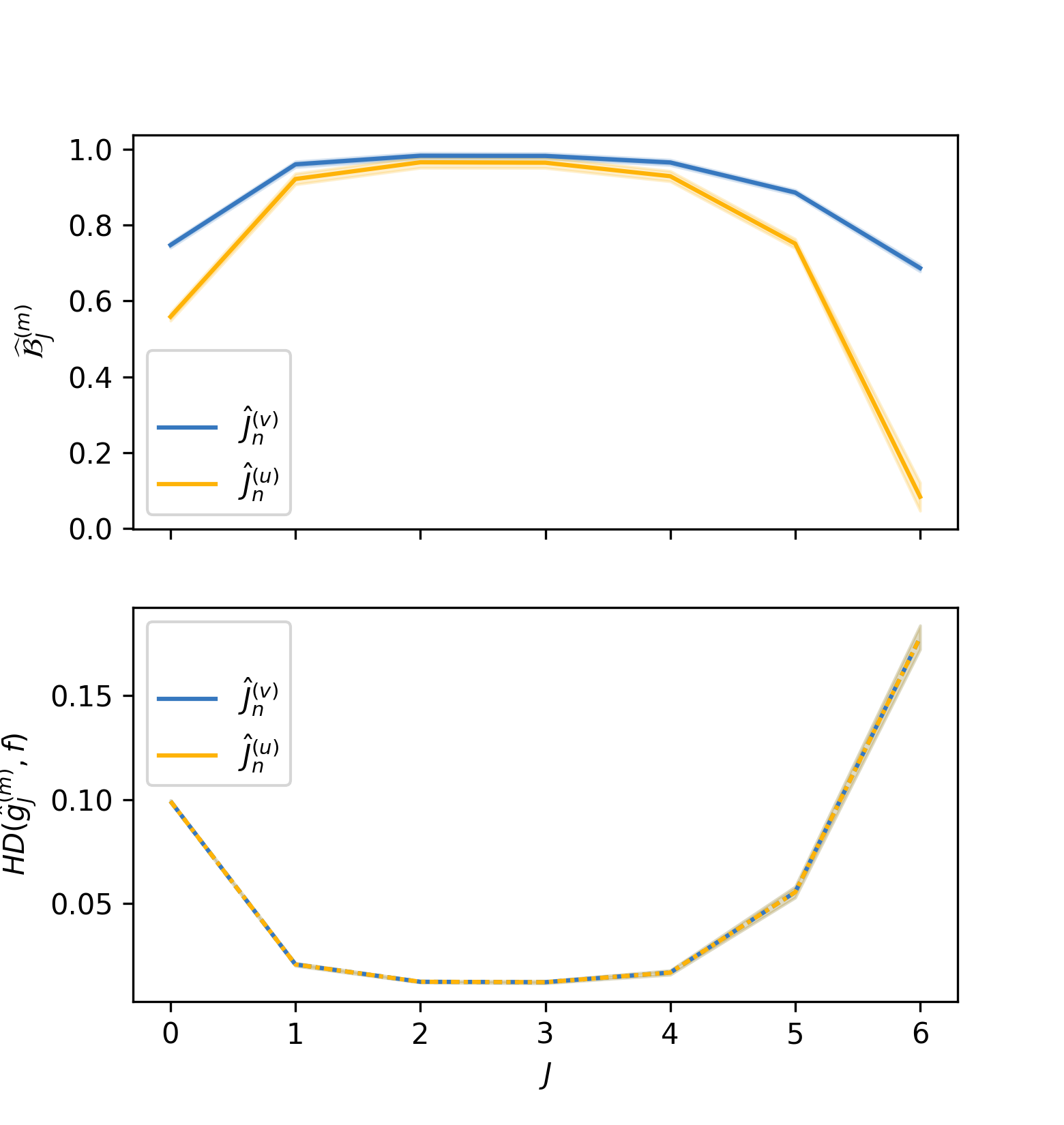}
		\caption{$\widehat{B}(J)$ and $\widehat{B}_\circ(J)$ (top) and $\Hs^2(J)$ (bottom); $n=5000$}
	\end{subfigure}
	\begin{subfigure}[t]{0.5\textwidth}
		\includegraphics[width=\textwidth]{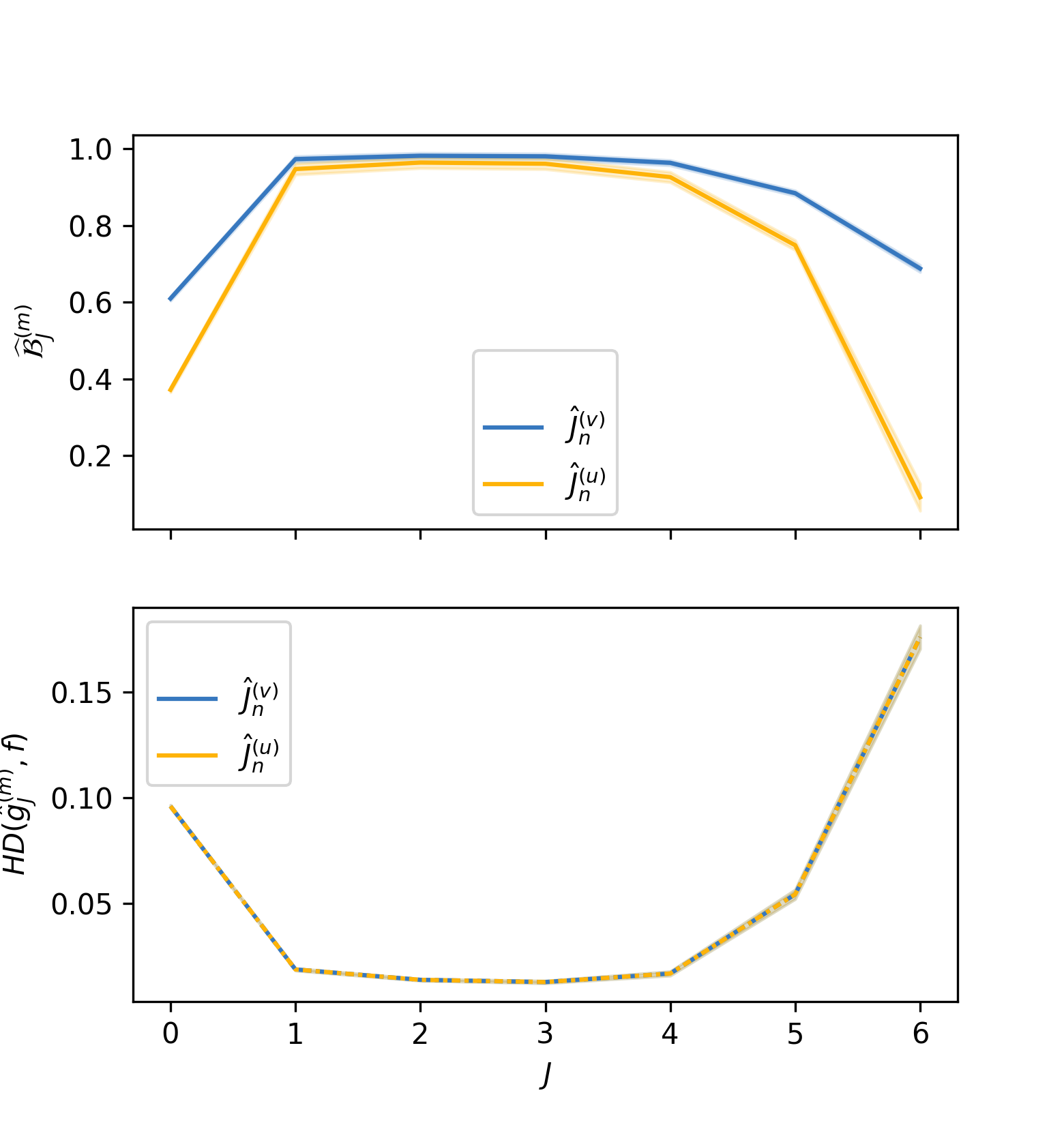}
		\caption{$\widehat{B}(J)$ and $\widehat{B}_\circ(J)$ (top) and $\Hs^2(J)$ (bottom); $n=5000$}
	\end{subfigure}
	\caption{Results for resolution selection rule; density \ref{fig:truedensj}(c) (Left: Daubechies 4, Right: Symlet 6).} \label{fig:bestjmix9}
\end{figure}

Density `2D Gaussian Mix 1' (Figure \ref{fig:truedensj}(c)) is a typical example of suitability of wavelet thresholding in density estimation, as it exhibits components with high locality. Without thresholding, we see in Figure \ref{fig:bestjmix9} (a) and (b), that the resolution level seems to be underpredicted for $n = 5000$. In this particular case, the corresponding curves (c) and (d) appear to have a plateau in the wide range $J=1,2,3,4$, which makes the selection of the best theoretical $J$ a very difficult one -- but again, consequences of a `wrong' choice are very limited in terms of accuracy of estimation. Thresholding is investigated empirically in the next section.

\medskip

\begin{figure}[h]
	\begin{subfigure}[t]{0.5\textwidth}
		\includegraphics[width=\textwidth]{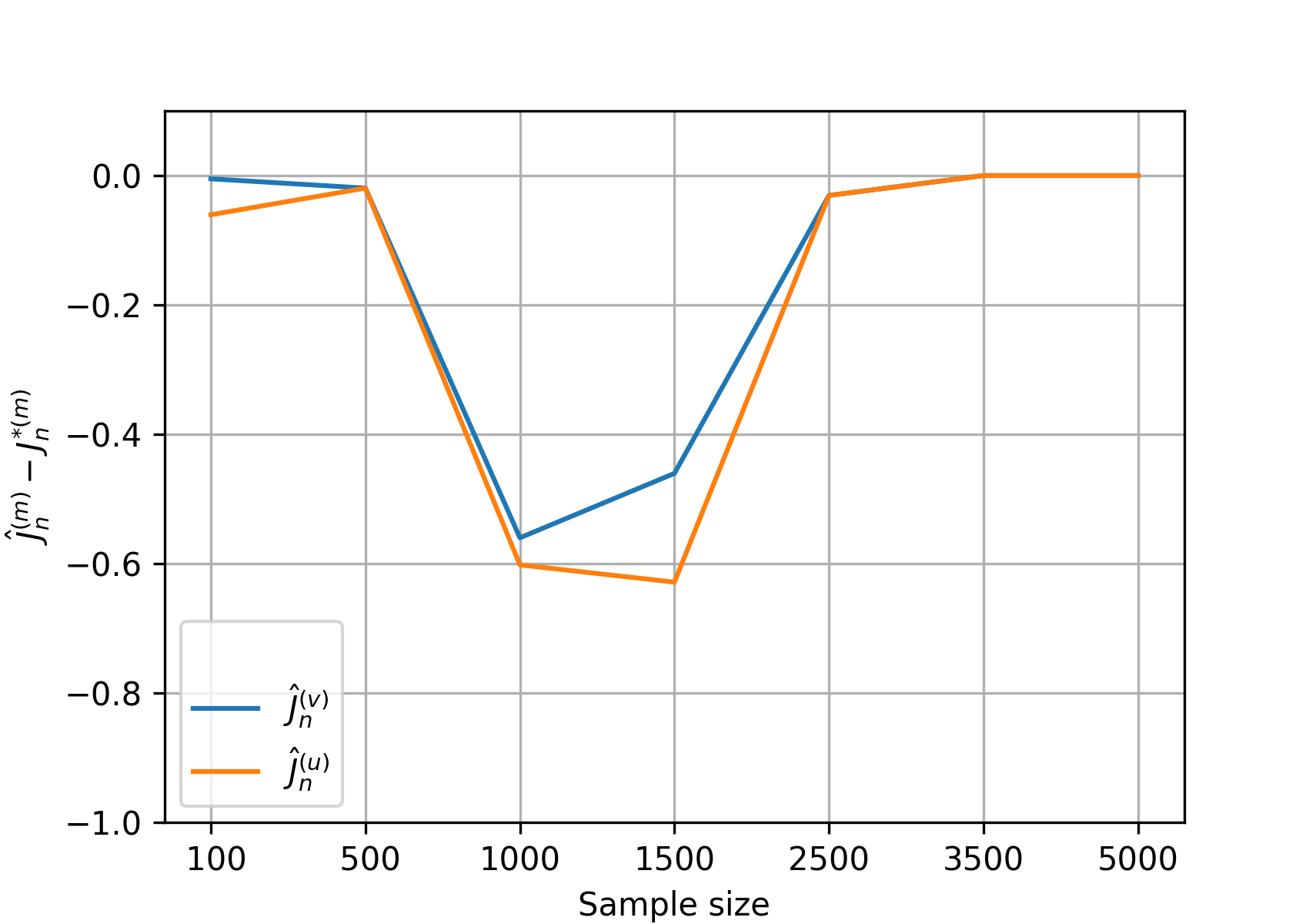}
		\caption{Average difference $\widehat{J}-J^*$ and $\widehat{J}_\circ-J_\circ^*$}
	\end{subfigure}
	\begin{subfigure}[t]{0.5\textwidth}
		\includegraphics[width=\textwidth]{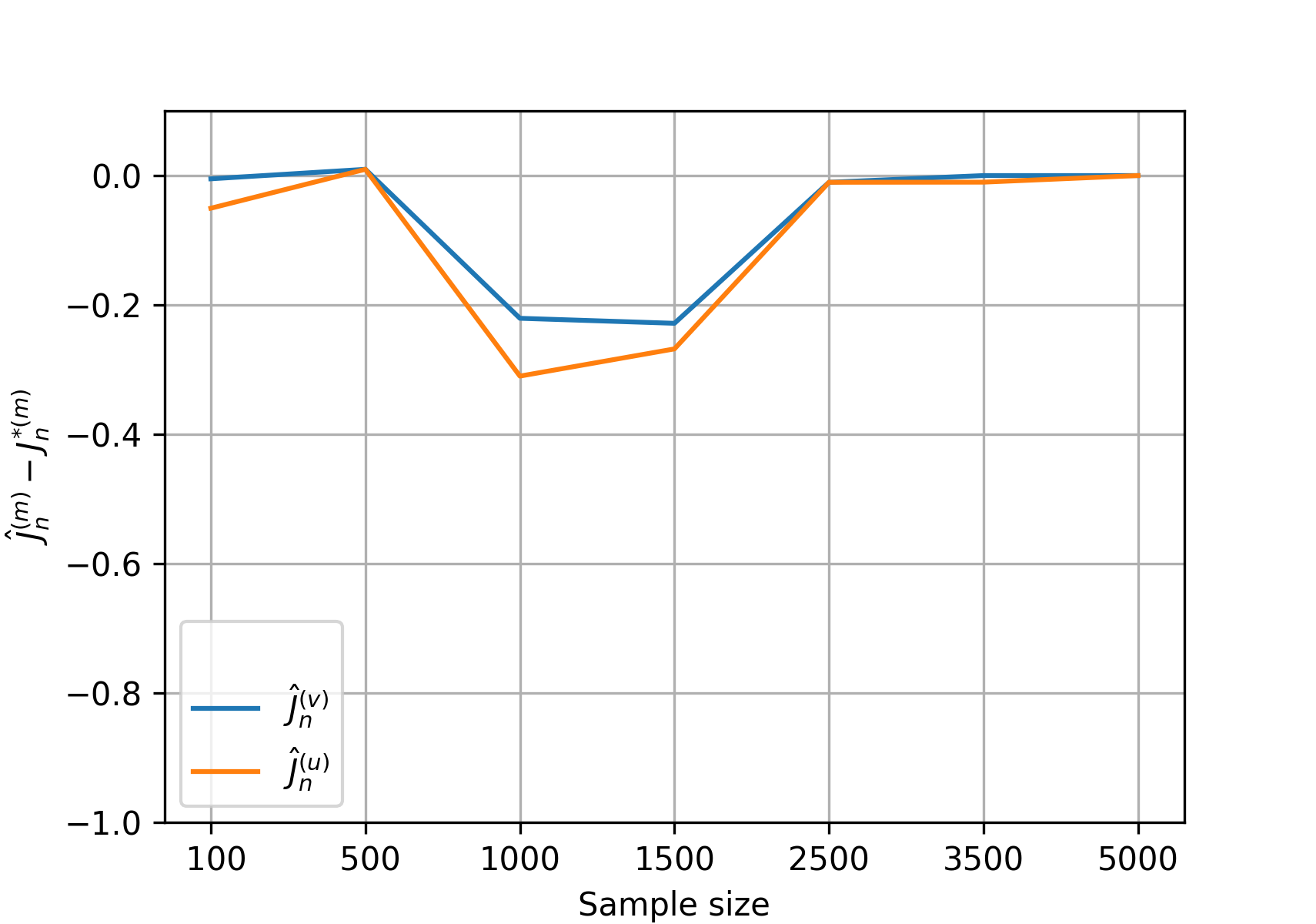}
		\caption{Average difference $\widehat{J}-J^*$ and $\widehat{J}_\circ-J_\circ^*$}
	\end{subfigure}
	\begin{subfigure}[t]{0.5\textwidth}
		\includegraphics[width=\textwidth]{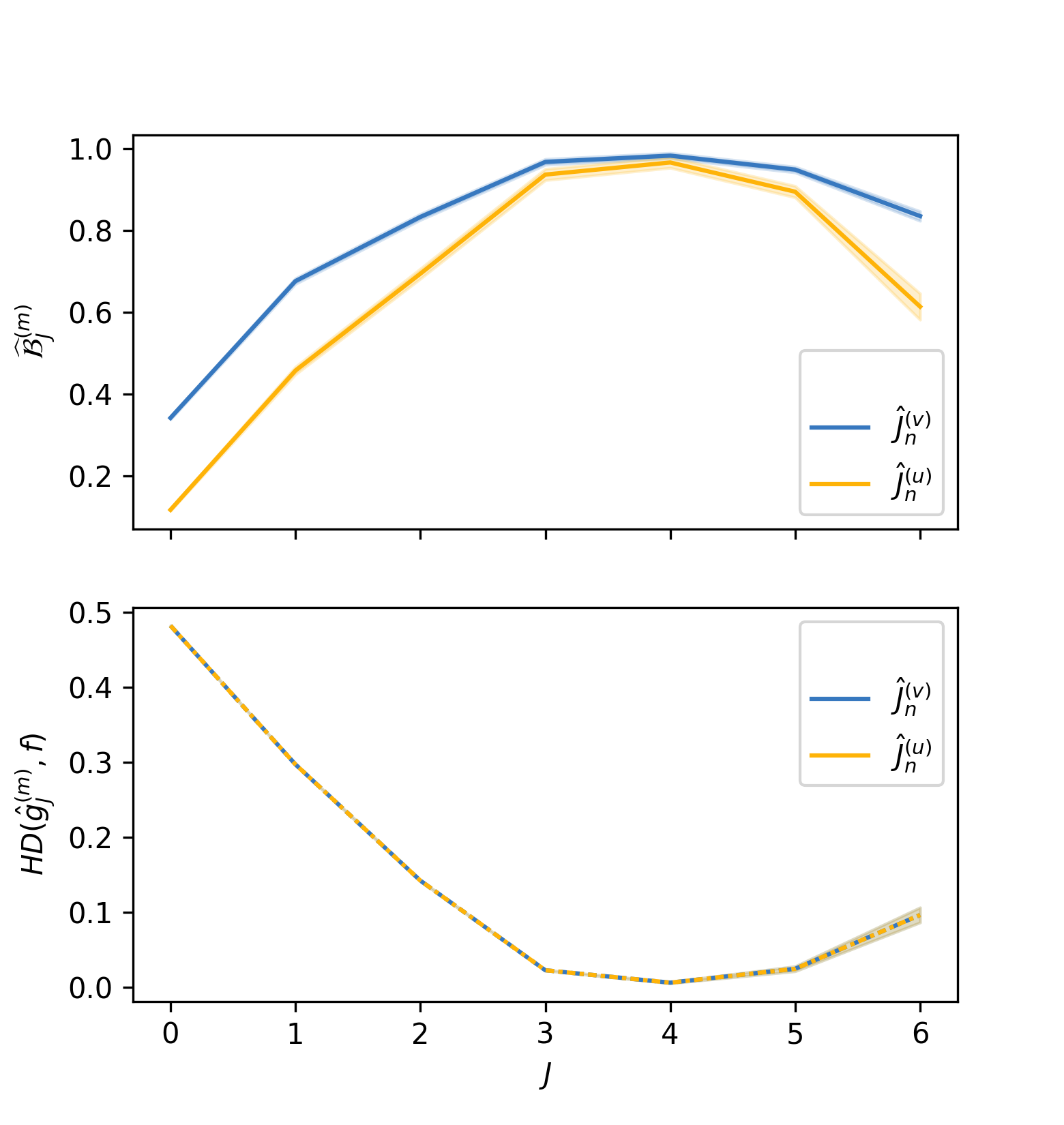}
		\caption{$\widehat{B}(J)$ and $\widehat{B}_\circ(J)$ (top) and $\Hs^2(J)$ (bottom); $n=5000$}
	\end{subfigure}
	\begin{subfigure}[t]{0.5\textwidth}
		\includegraphics[width=\textwidth]{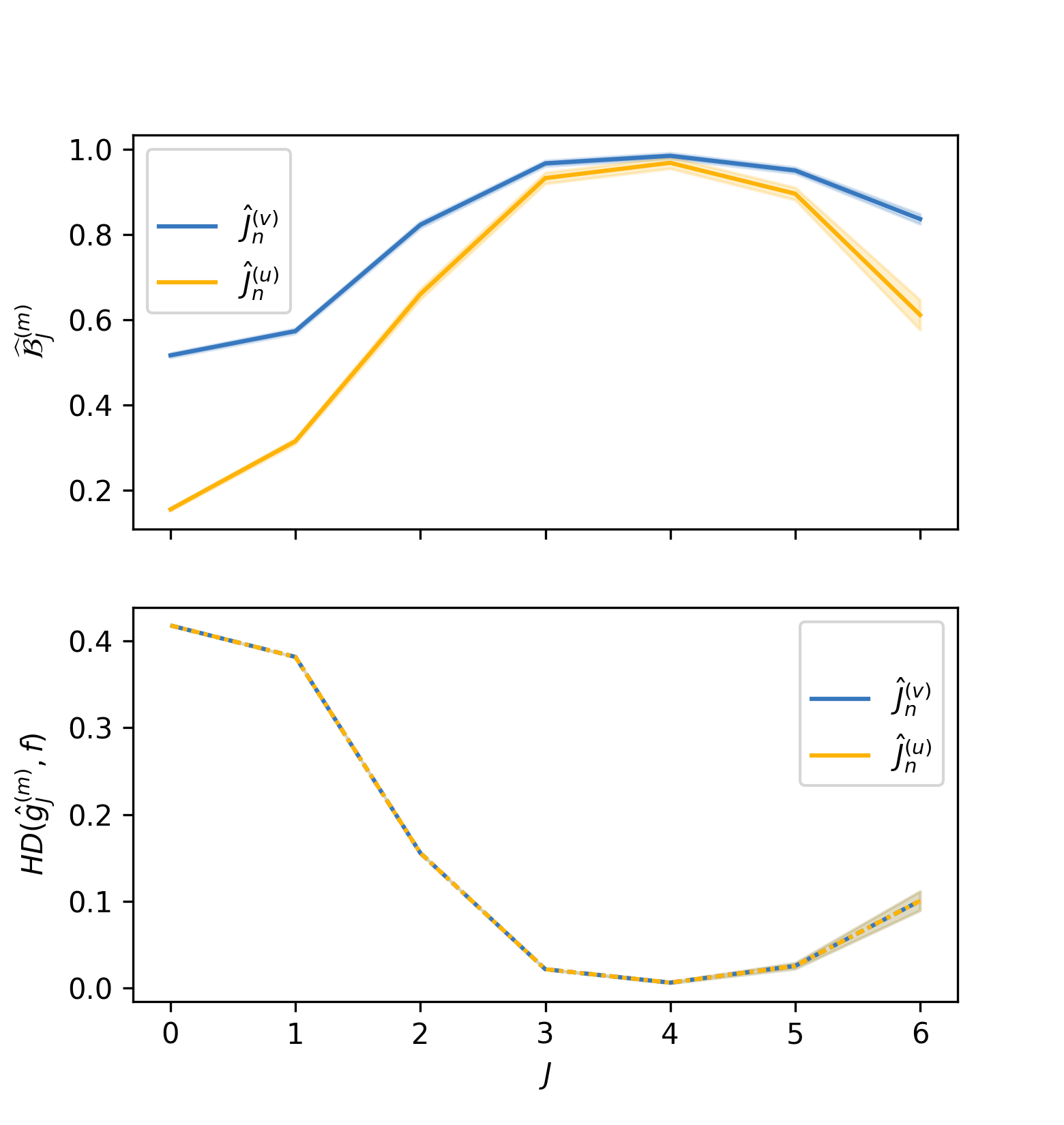}
		\caption{$\widehat{B}(J)$ and $\widehat{B}_\circ(J)$ (top) and $\Hs^2(J)$ (bottom); $n=5000$}
	\end{subfigure}
	\caption{Results for resolution selection rule; density \ref{fig:truedensj}(d) (Left: symlet 6, Right: biorthogonal spline 3.9).} \label{fig:bestjmix6}
\end{figure}

Finally, the `2D Gaussian mix 2' density (Density \ref{fig:truedensj}(d)) is a typical anisotropic density. Figure \ref{fig:bestjmix6} exhibit the same behaviour as above, whence the same conclusions may be drawn.

\medskip In conclusion, it appears that the novel Hellinger-\Bhat cross-validation criterion identifies the theoretically `best' resolution level -- or practically equivalent -- in a very reliable way, across a variety of simulation scenarios, sample sizes or wavelet bases.

\subsection{Thresholding} \label{subsec:thresholdingsimulation}

We proposed two threshold selection criteria (\ref{eqn:tauhatcirc}) and (\ref{eqn:tauhat}). Each may be used with any definition of the thresholding in \eqref{def:hardthreshold}: as benchmarks, we have considered here the `universal threshold' \cite{DJKP95} $\gamma_j \equiv 1$ (i.e., not level-dependent) and the level-dependent threshold setting $\gamma_j = \sqrt{j-j_0+1}$ \cite{Delyon96,DonohoJohnstone96}; which we compared to our novel approach involving the jackknife estimation of the variance of $\betahatqjz$ (\ref{eqn:lambda2}). The wavelet bases we used were symlets with 3 and 4 vanishing moments, and Daubechies with 4 vanishing movements. The value of the basis level $j_0$ was set to $\Delta J =$1, 2 or 3 levels {\it below} the best $J$ calculated by Hellinger-\Bhat cross-validation as described in previous sections. In all, we passed each density and sample size through a battery of $2 \times 3 \times 3 \times 3 = 54$ combinations, which we briefly report below.

\medskip

\begin{figure}[h]
	\begin{subfigure}[t]{0.5\textwidth}
		\includegraphics[width=\textwidth]{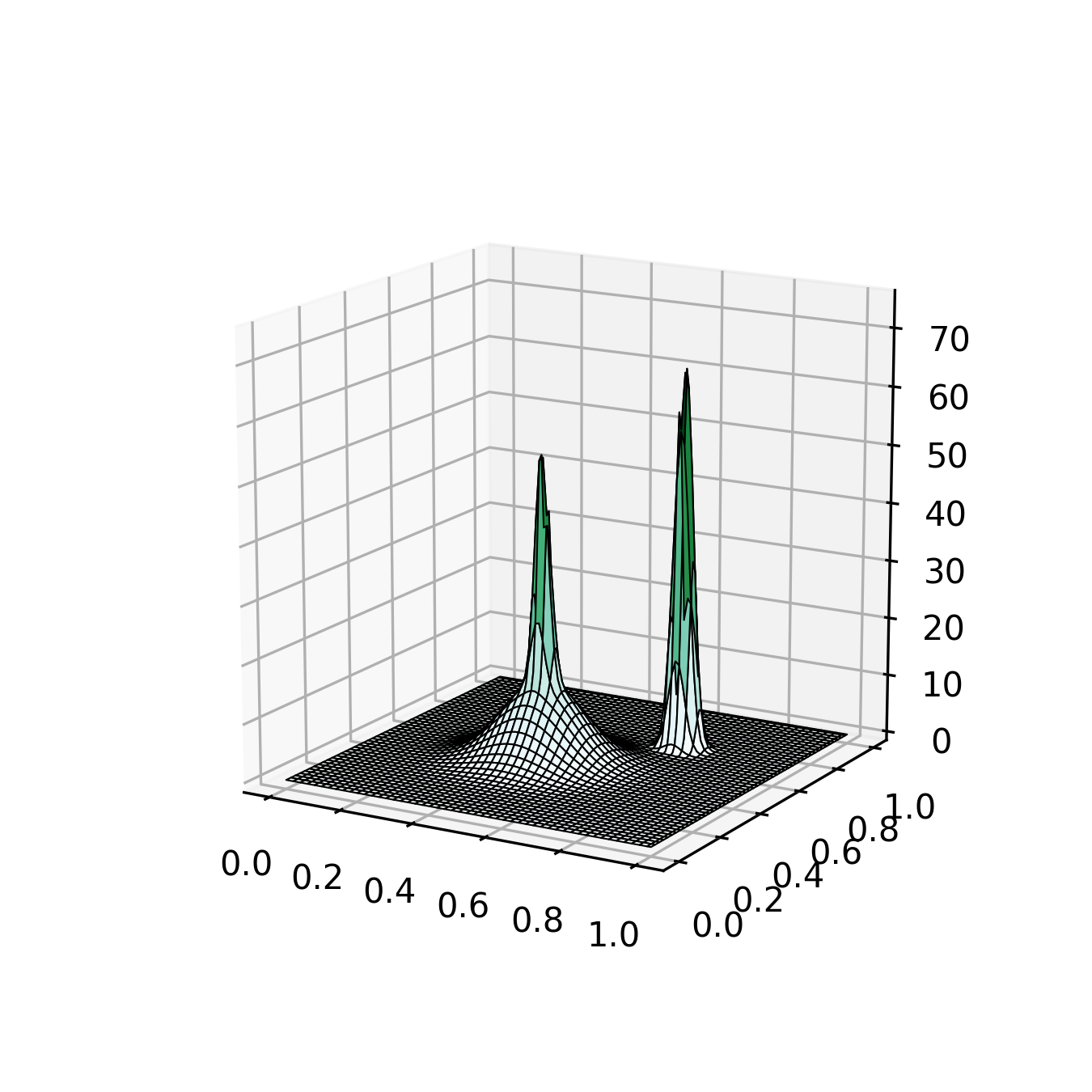}
		\caption{Kurtotic Mixture 1}
	\end{subfigure}
	\begin{subfigure}[t]{0.5\textwidth}
		\includegraphics[width=\textwidth]{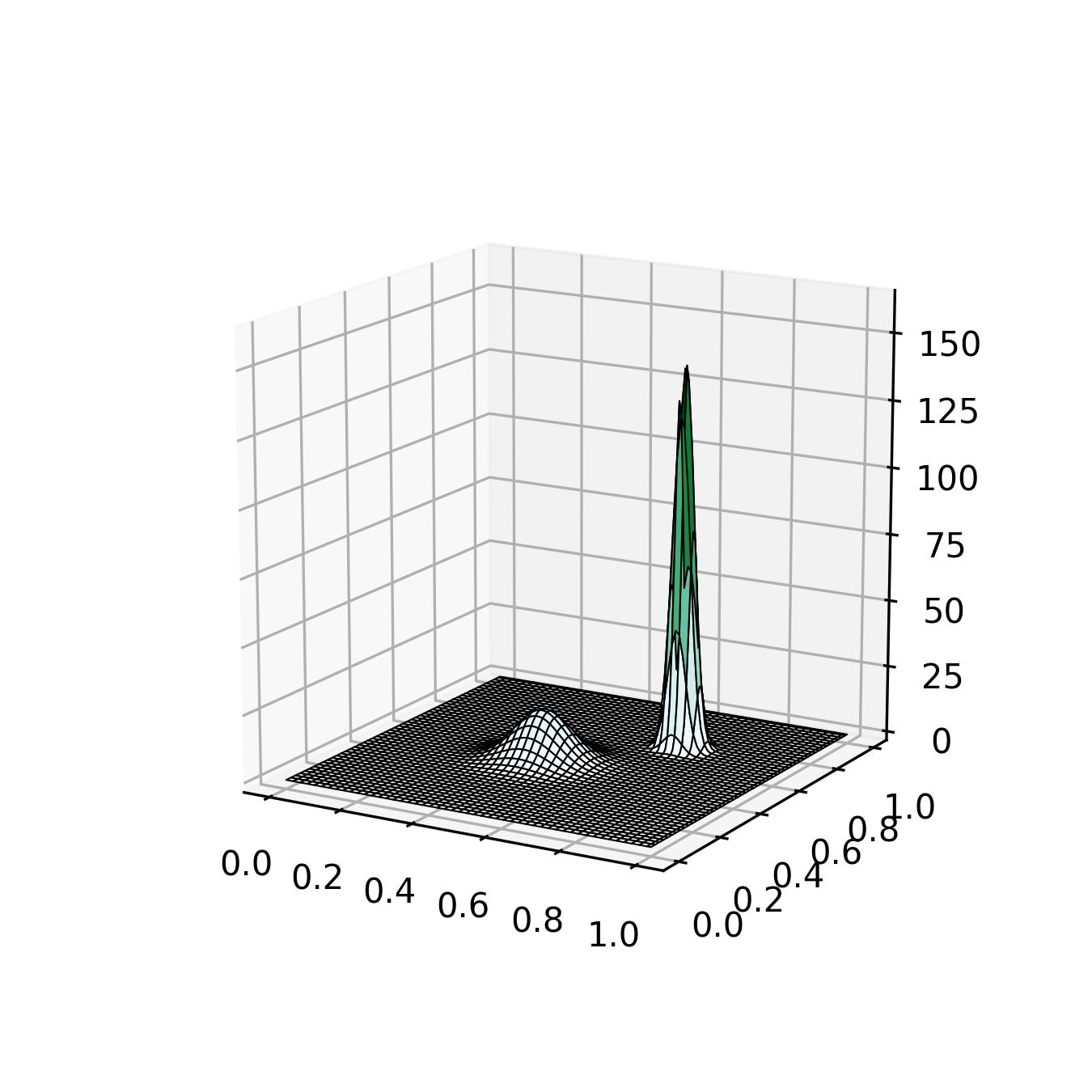}
		\caption{Mixture 2}
	\end{subfigure} \\
	\begin{subfigure}[t]{0.5\textwidth}
		\includegraphics[width=\textwidth]{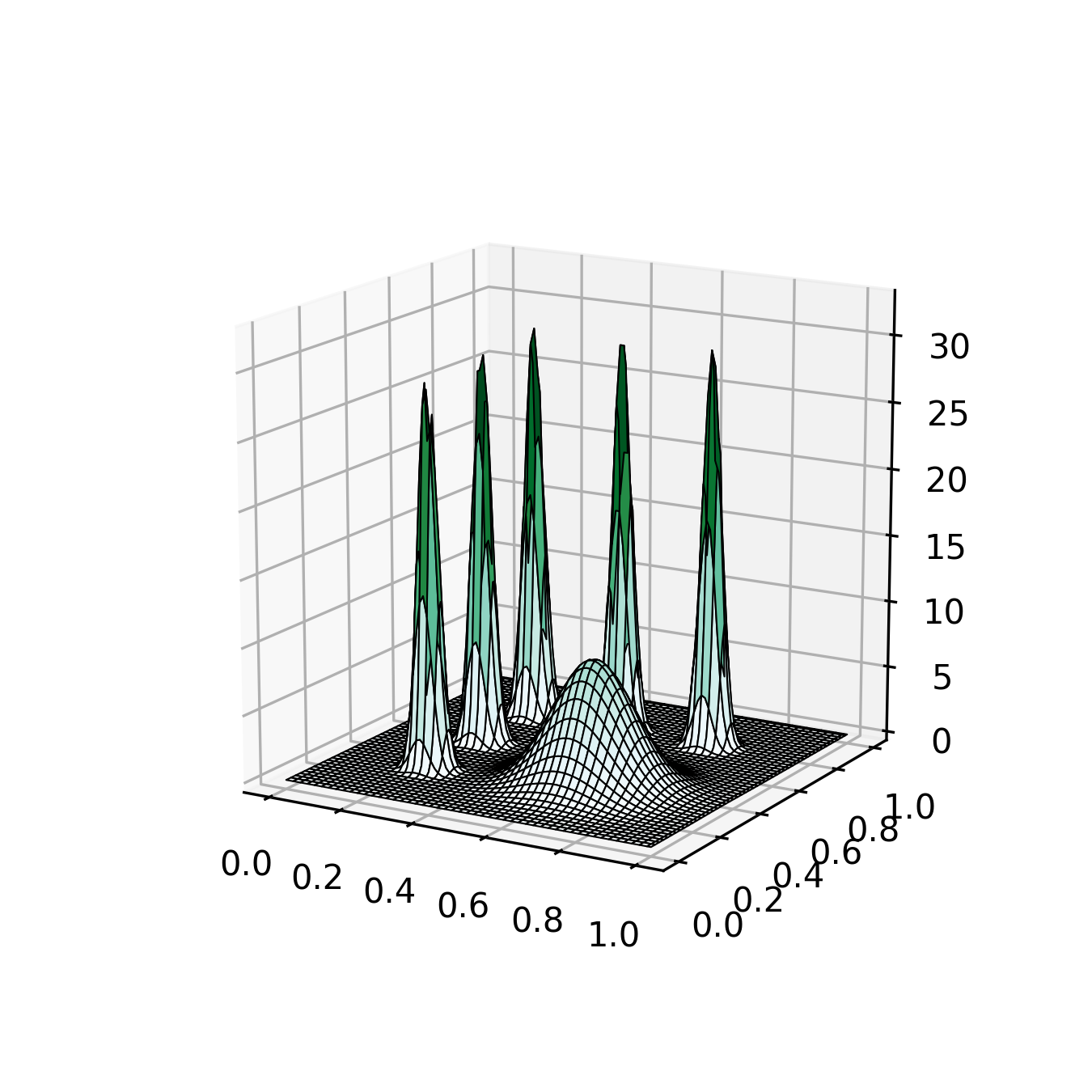}
		\caption{2D Comb 1 (claw)}
	\end{subfigure}
	\begin{subfigure}[t]{0.5\textwidth}
		\includegraphics[width=0.9\textwidth]{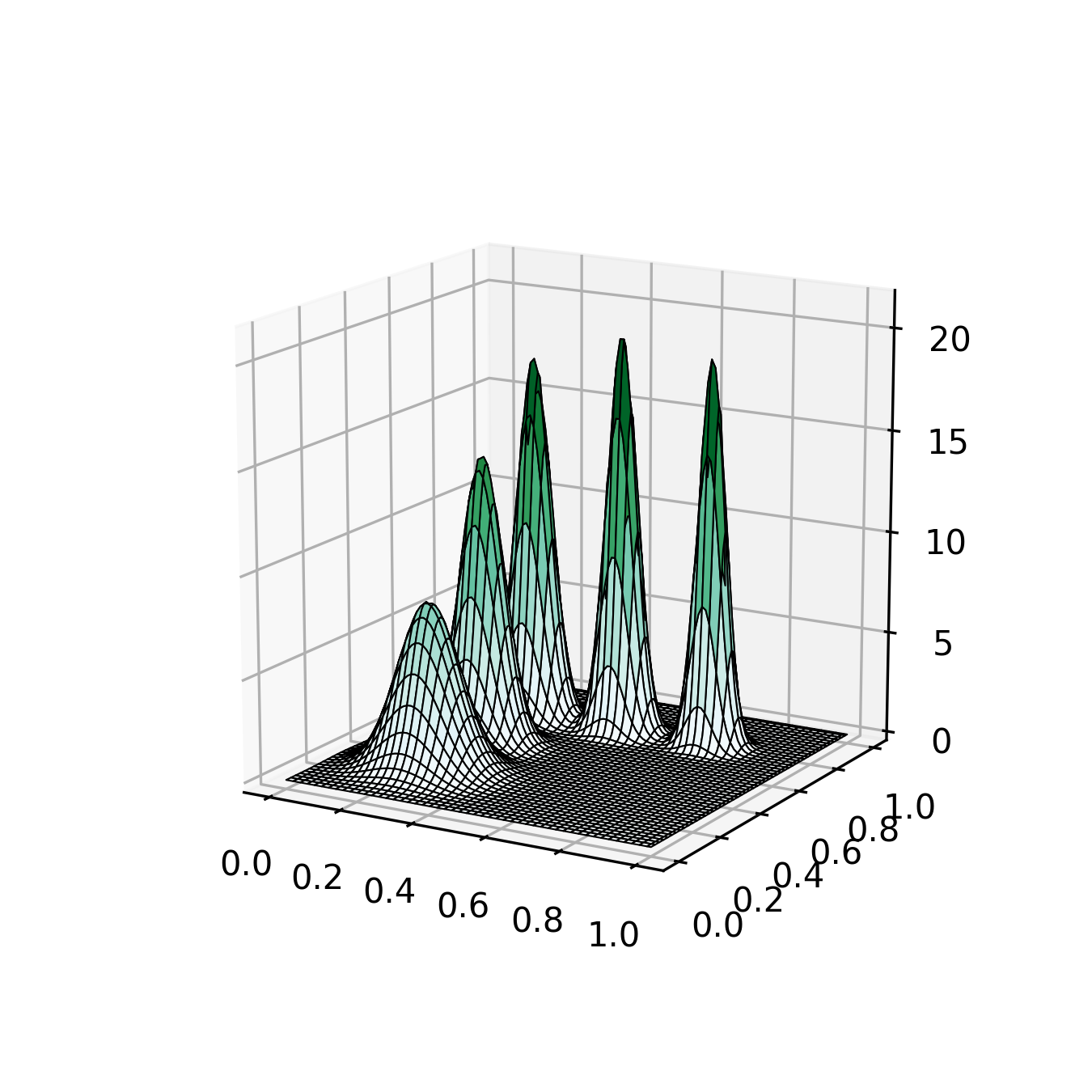}
		\caption{2D Smooth comb}
	\end{subfigure}
	\caption{Densities used in thresholding simulation study.} \label{fig:truedenshardt}
\end{figure}

We studied the 4 densities shown in Figure \ref{fig:truedenshardt}. Density (a) is a kurtotic, bimodal mixture made out of three Gaussians; (b) is a simple mixture of two Gaussians with different spread; (c) is similar to the claw density in \cite{Marron92}; and (d) is akin to the smooth comb there but in 2D. The analytic forms can be found in \cite{Aya20}.

\medskip 

We considered sample sizes from $n=250$ to $n=6,000$ (see Table \ref{tab:ex01} and following). For each sample size, we generated 100 samples. For each sample and each family of wavelets, we chose the `best' $J$ according to both criteria \eqref{eqn:Jhat} and \eqref{eqn:Jhatcirc}, followed by runs of the 9 combinations of the remaining parameters as explained above. For comparison, we also ran a multivariate Kernel Density Estimator (KDE) with a Gaussian kernel and a bandwidth matrix $H$; viz.
\begin{equation*}
	\hat{f}_H(x) = \frac{1}{n} \sum_{i=1}^{n} \Abs{H}^{-1/2} \left(2 \pi\right)^{-d/2} e^{-\frac{1}{2}\left(x - X_i\right)^T H^{-1} \left(x - X_i\right)}.
\end{equation*}
The matrix $H$ was determined by Maximum Likelihood Cross-Validation, using \cite{Seabold10}'s implementation.

\medskip 

The performance of each estimator was quantified by the squared Hellinger distance between the considered estimate and the true density. The results are shown in Tables \ref{tab:ex01} to \ref{tab:ex04}. For each scenario and each estimation scheme, we show the median, first and third quantiles of the 100 replicated squared Hellinger distances. We have used bold face to highlight the cases in which the median Hellinger distance for these estimators is less than or equal to the corresponding median for the KDE (last column). Note that we are comparing the kernel estimator, which does not drop any observation and in essence has $n$ free parameters, against our method that aims to reduce the number of parameters using thresholding. Reducing the number of free parameters without compromising performance is a key requirement in big data applications. So, the fact that our approach achieves similar performance is noteworthy.

\medskip

A first observation is that there is little difference in performance when the resolution level is chosen {\it after} or {\it before} normalisation (i.e., according to (\ref{eqn:Jhat}) or (\ref{eqn:Jhatcirc})). The automatically normalised criterion (\ref{eqn:Jhat}) seems to have a slight edge in smaller samples, though. Another important general observation drawn from Tables \ref{tab:ex01}-\ref{tab:ex04} is that the novel jackknife hard thresholding \eqref{eqn:lambda2} performs better than the other well-known thresholding strategies across all scenarios and sample sizes. This motivates further theoretical study of this idea in a follow-up paper. We see that, when this `jackknife thresholding' is used, the value of $\Delta J$ (that is, essentially the value $j_0$) is mostly irrelevant in terms of the accuracy of estimation -- this is not systematically the case for the other thresholding options (`universal' or `level-dependent'). 

\medskip 

For small sample sizes, the kernel estimator is usually (but not always, see Table \ref{tab:ex02}) doing better, but the proposed shape-preserving wavelet estimator catches up from $n$ around 1,000 and becomes superior for higher sample sizes -- the fact that wavelet methods are showing their advantages at larger sample sizes and high signal-to-noise ratios is well documented in the literature (see, e.g., \cite{Hall18}). In fact, it is quite remarkable that at $n=6000$ one can get better performance than the KDE by using typically less than $500$ coefficients (see supplementary results in \cite{Aya20}). An exception is here the `2D Smooth comb' density (Figure \ref{fig:truedenshardt}(d)), for which one can see our estimator performing better than KDE for small sample sizes (Table \ref{tab:ex04}) but no more as the sample size increases. Nonetheless, we note that one achieves a remarkably low error using as little as 163 coefficients for $n=6,000$ (see supplementary results in ibid.). We discuss more on this, and potential improvements, in Section \ref{sec:conclusions}.

\input{tables/ch4-th-table-ex01}
\input{tables/ch4-th-table-ex02}
\input{tables/ch4-th-table-ex03}
\input{tables/ch4-th-table-ex04}

\subsection{Real data: Old Faithful geyser} \label{subsec:oldfaithfulnonlinear}

We close this section by revisiting the Old Faithful geyser dataset \cite[Sec. 5.2]{Aya18}, and providing insights on practical usage of the estimator. This well-known data set contains $n=272$ pairs of observations (eruption time, waiting time since previous eruption) of Old Faithful, a very active geyser in the Yellowstone National Park, USA. Given this rather small sample size, lead by the observations made in Section \ref{subsec:thresholdingsimulation}, we opt for the criteria (\ref{eqn:Jhat}) and (\ref{eqn:tauhat}), based on automatic normalisation of all estimators, for choosing the resolution $J$ and applying the thresholding. We also enforce the novel jackknife thresholding \eqref{eqn:lambda2}, as this does not require any arbitrary choice of level-dependent $\gamma_j$ and was shown to be the best option in Section \ref{subsec:thresholdingsimulation}. 

\medskip 

\begin{figure}[h]
	\begin{subfigure}[t]{0.5\textwidth}
		\includegraphics[width=\textwidth]{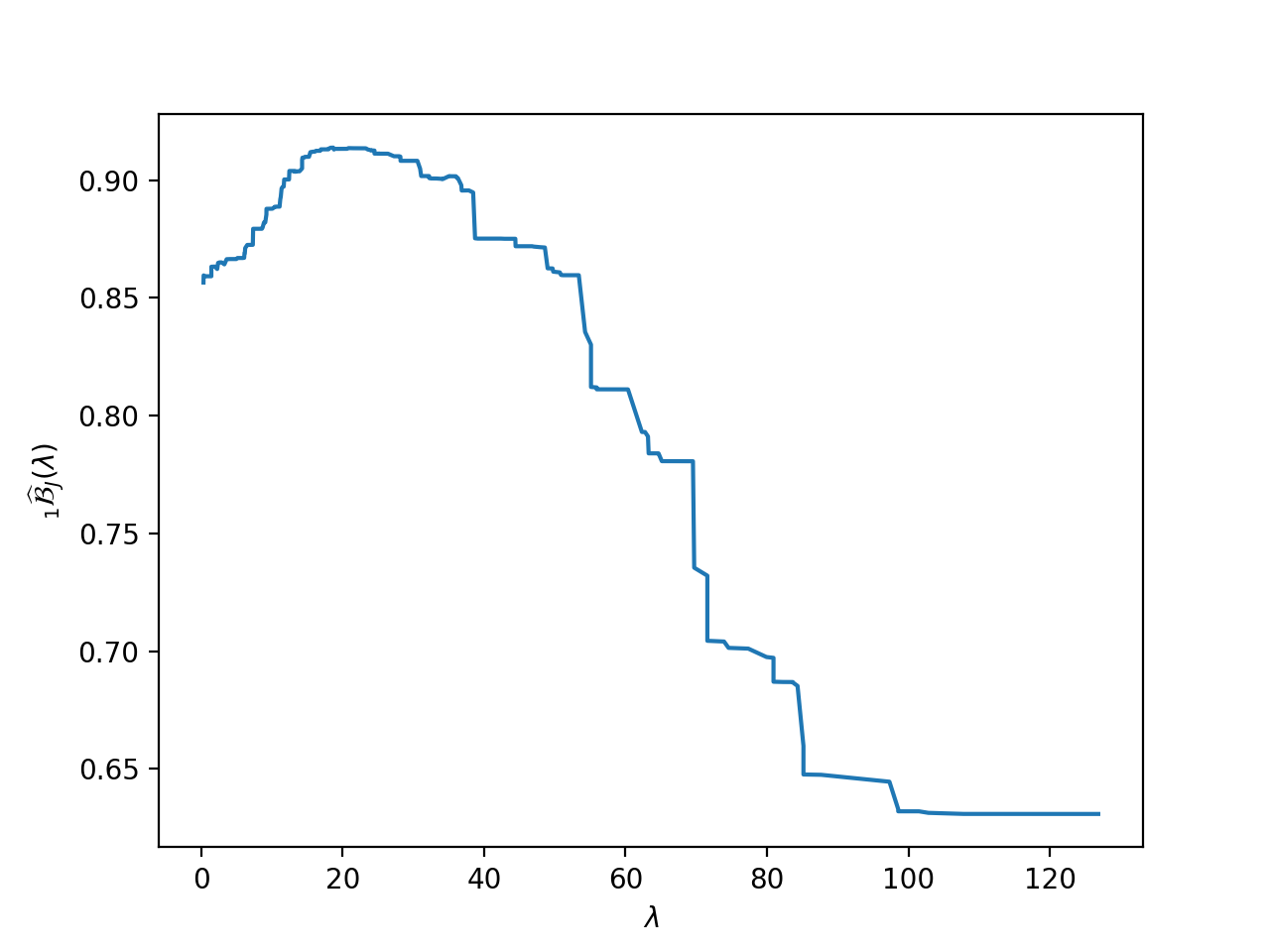}
	\end{subfigure}
	\begin{subfigure}[t]{0.5\textwidth}
		\includegraphics[width=1.2\textwidth]{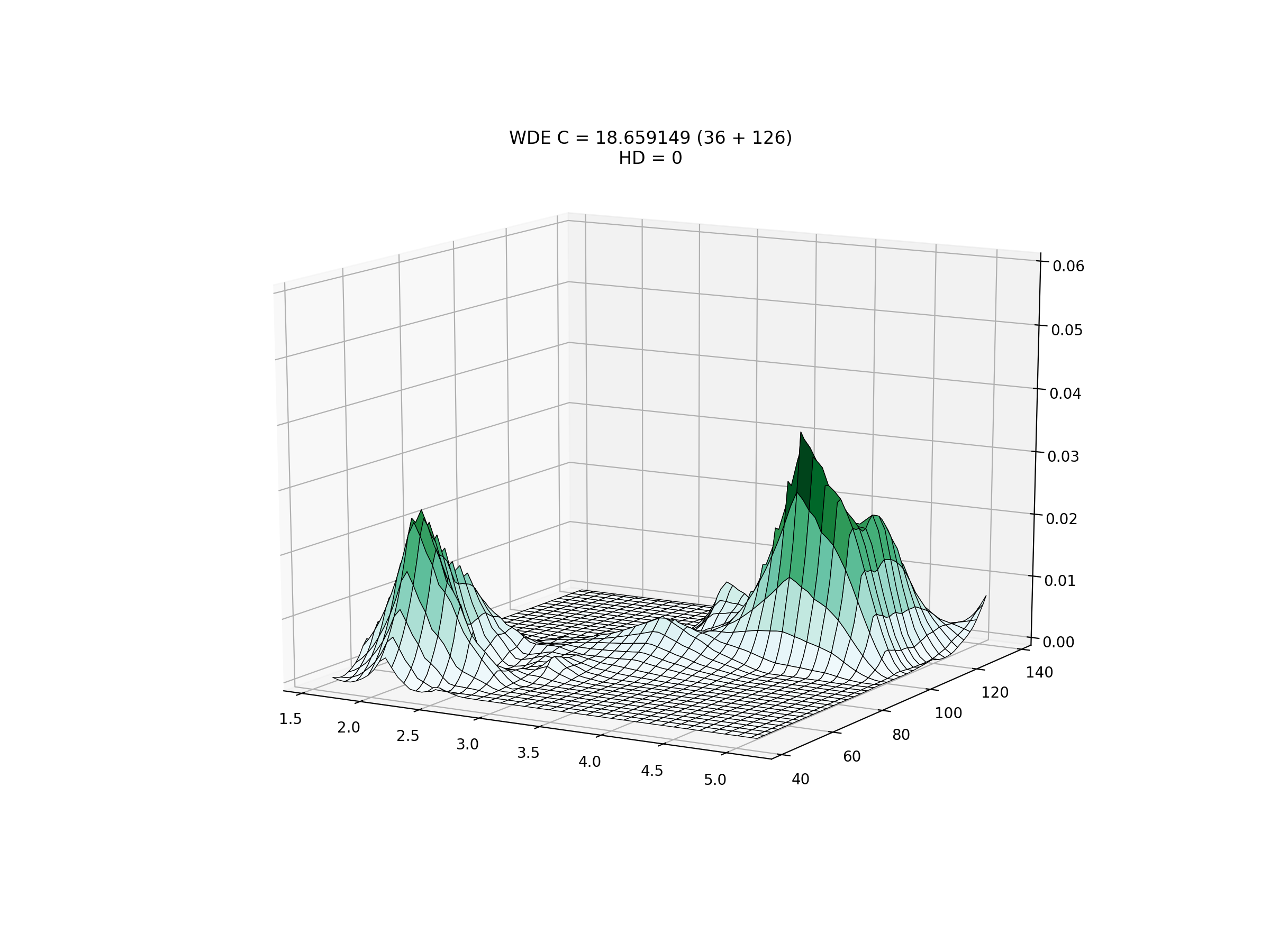}
	\end{subfigure}
	\caption{Optimisation curve, left, and density, right, for the Old Faithful geyser dataset using Daubechies 3 and $\Delta J = 2$.} \label{fig:geyserdb3j2}
\end{figure}

\begin{figure}[h]
	\begin{subfigure}[t]{0.5\textwidth}
		\includegraphics[width=\textwidth]{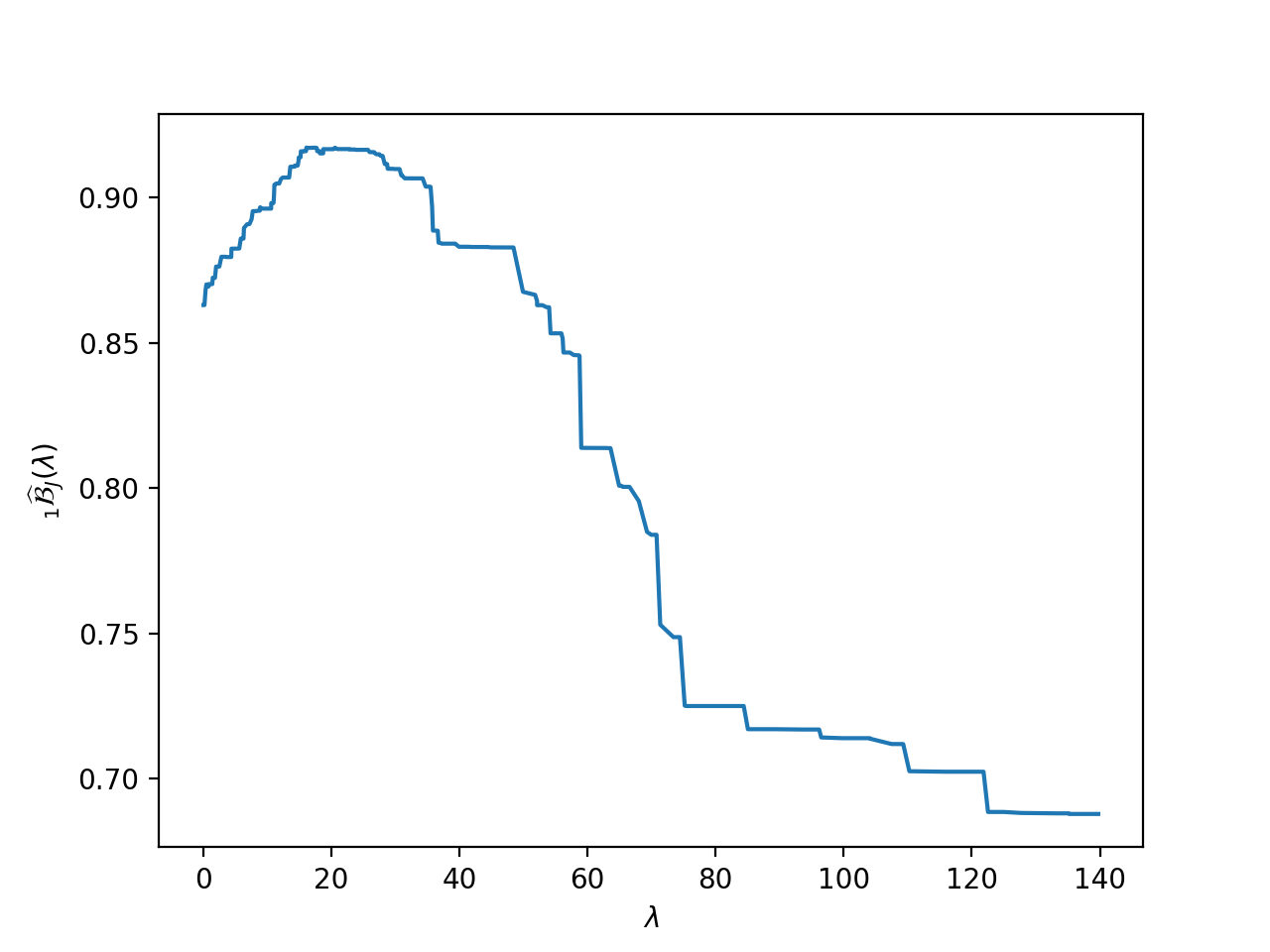}
	\end{subfigure}
	\begin{subfigure}[t]{0.5\textwidth}
		\includegraphics[width=1.2\textwidth]{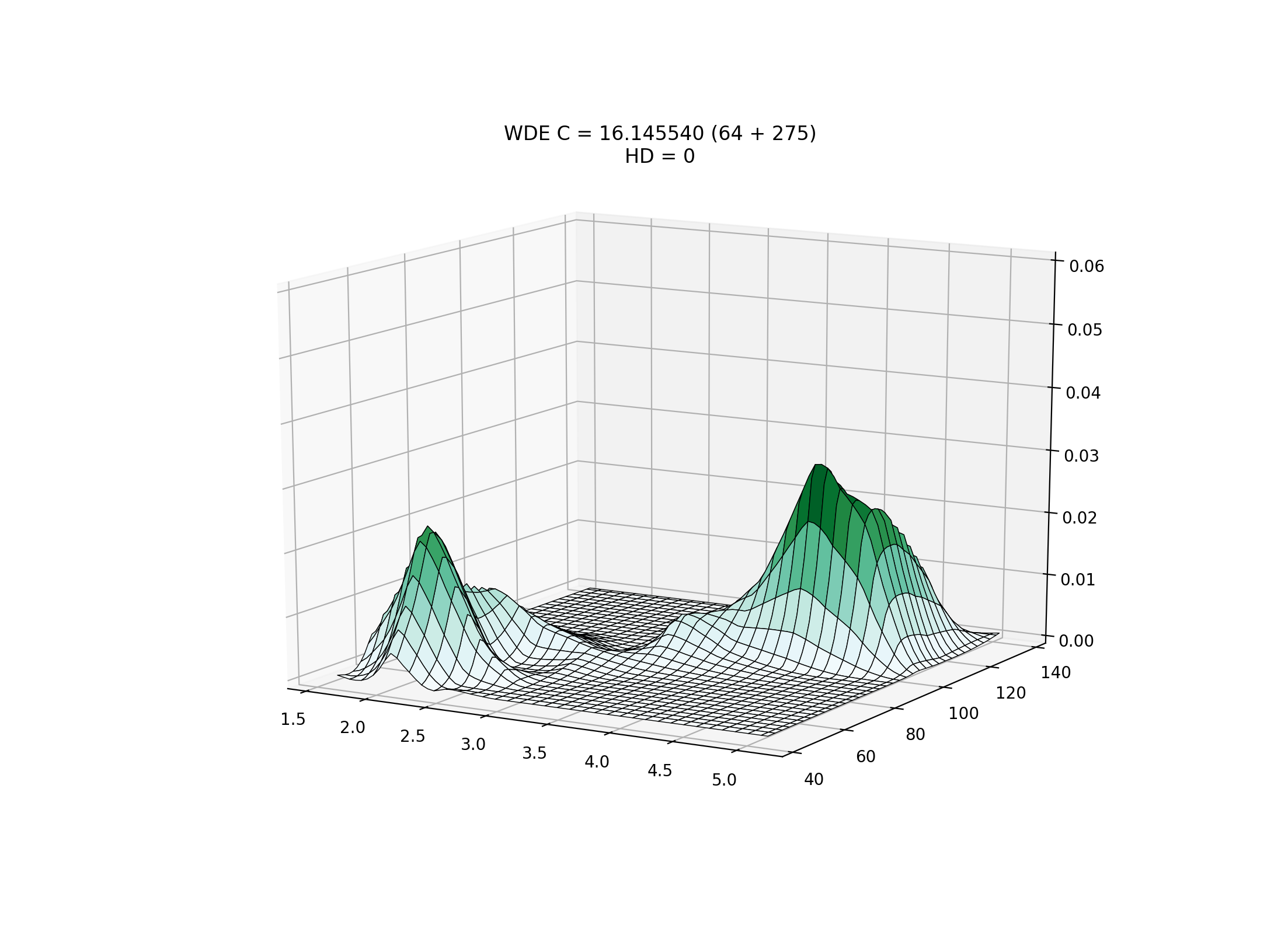}
	\end{subfigure}
	\caption{Optimisation curve, left, and density, right, for the Old Faithful geyser dataset using Symlet $4$ and $\Delta J = 2$.} \label{fig:geysersym4j2}
\end{figure}

In addition, we illustrate here how the Hellinger-\Bhat criterion can also offer some insight into which wavelet basis to work with. Consider, for example, Figures \ref{fig:geyserdb3j2} and \ref{fig:geysersym4j2} which show the appearance of criterion (\ref{eqn:Bhattau}) when based on `jackknife thresholding' \eqref{eqn:lambda2} for the Daubechies $3$ and Symlet $4$ wavelet bases using $\Delta J = 2$ levels of beta coefficients. Symlet $4$ produces a smoother density, of course, keeping $321$ coefficients alphas and betas. Daubechies $3$ requires fewer coefficents (162). The corresponding maximum values of $\widehat{\mathcal{B}}(\tau)$ are 0.91383 (Daubechies $3$) and 0.91719 (Symlet $4$) which, if not by the inclination for a smoother density, indicates that Symlet $4$ might be preferred in this case.


\begin{figure}[h]
	\begin{subfigure}[t]{0.5\textwidth}
		\includegraphics[width=\textwidth]{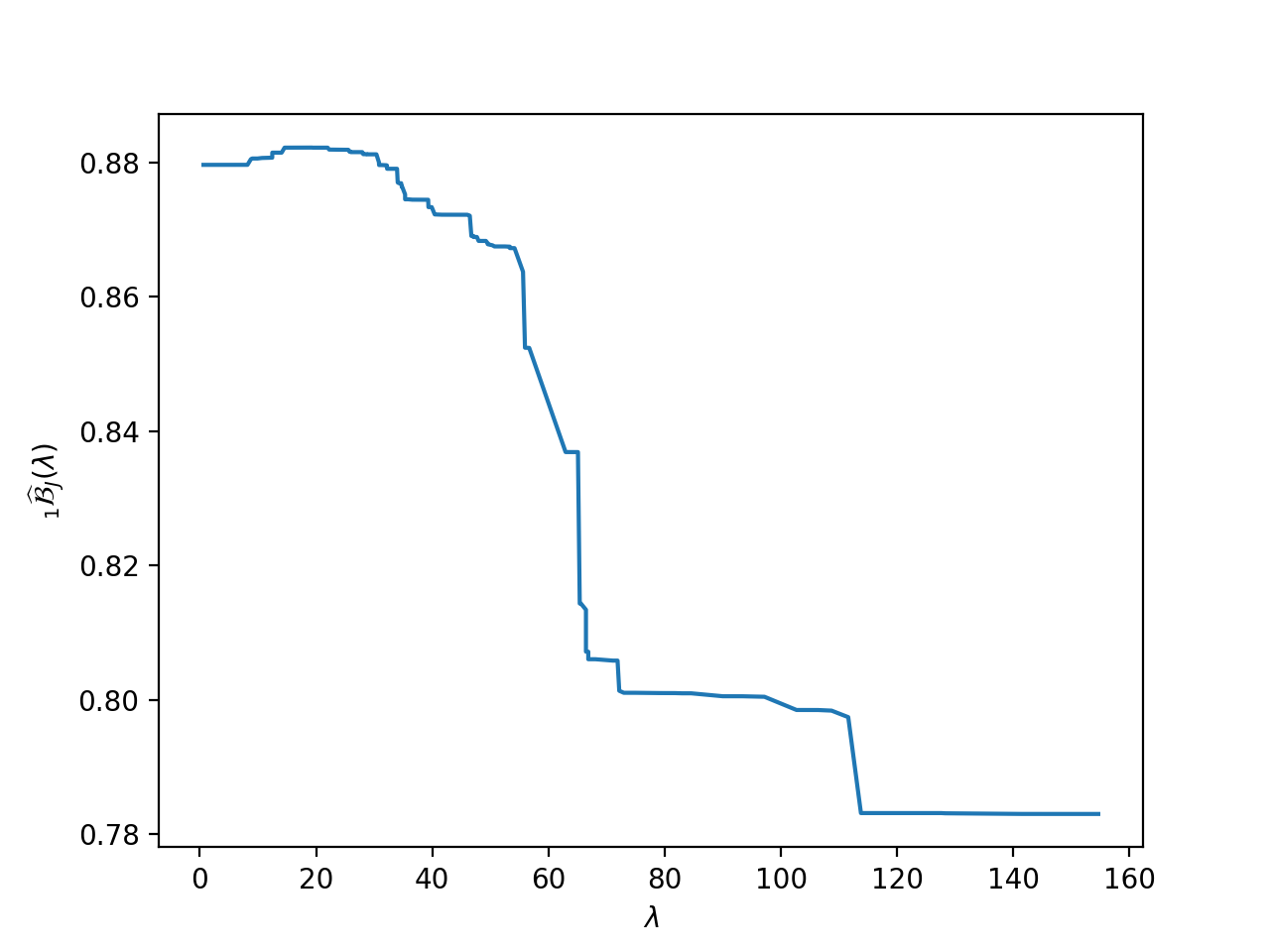}
	\end{subfigure}
	\begin{subfigure}[t]{0.5\textwidth}
		\includegraphics[width=1.2\textwidth]{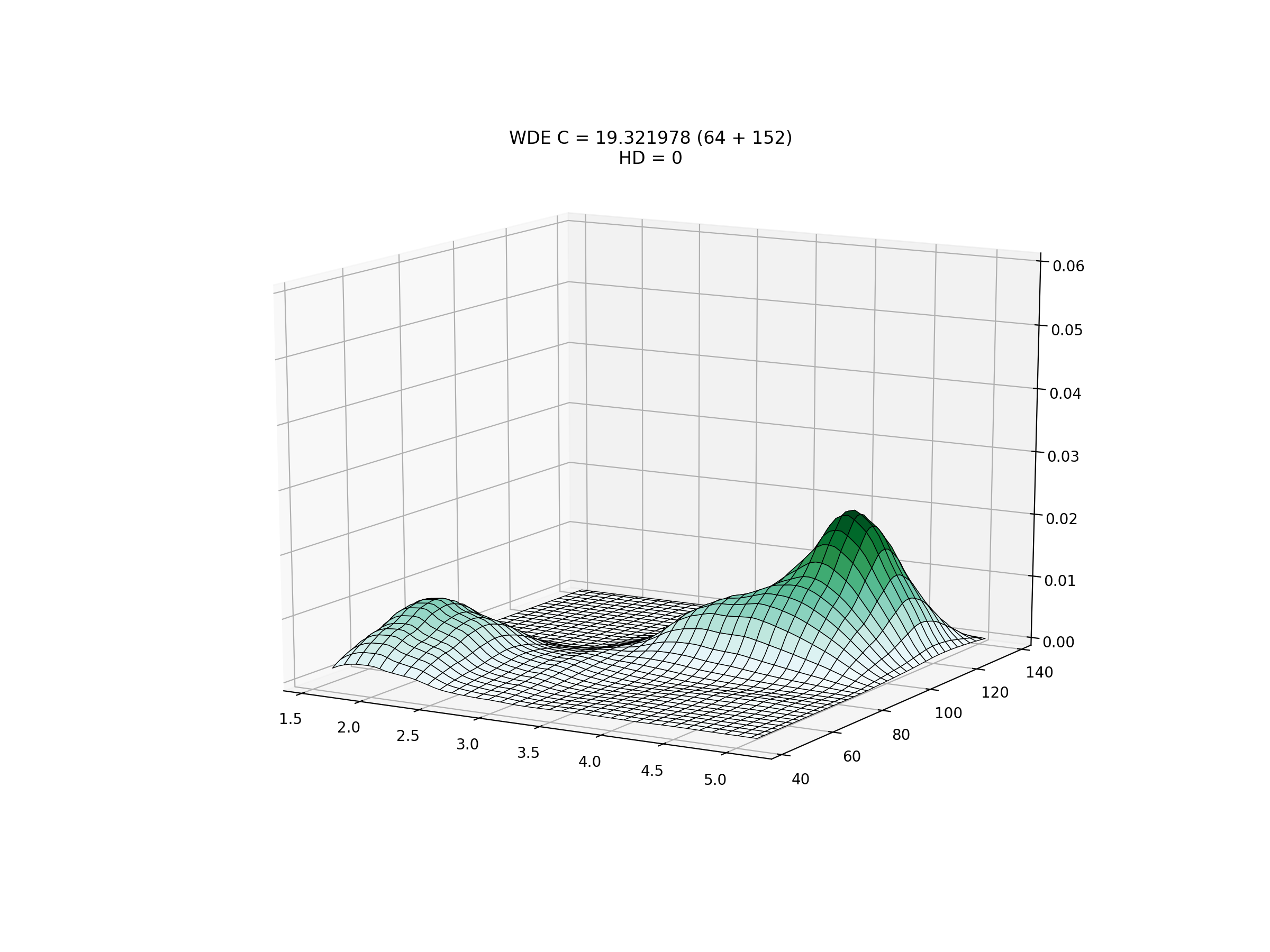}
	\end{subfigure}
	\caption{Optimisation curve, left, and density, right, for the Old Faithful geyser dataset using Daubechies $4$ and $\Delta J = 1$.} \label{fig:geyserbaddb4level1}
\end{figure}
\begin{figure}[h]
	\begin{subfigure}[t]{0.5\textwidth}
		\includegraphics[width=\textwidth]{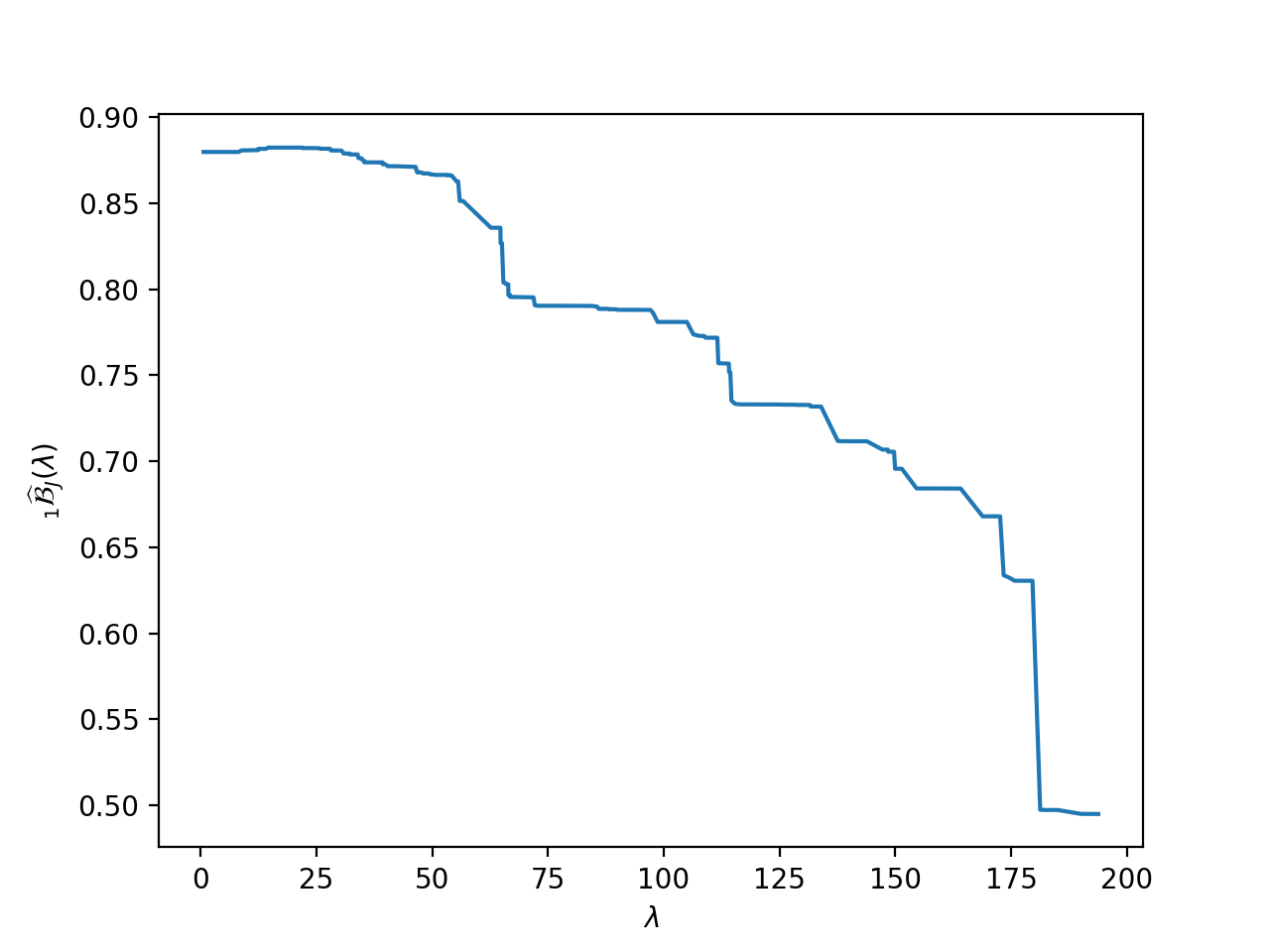}
	\end{subfigure}
	\begin{subfigure}[t]{0.5\textwidth}
		\includegraphics[width=1.2\textwidth]{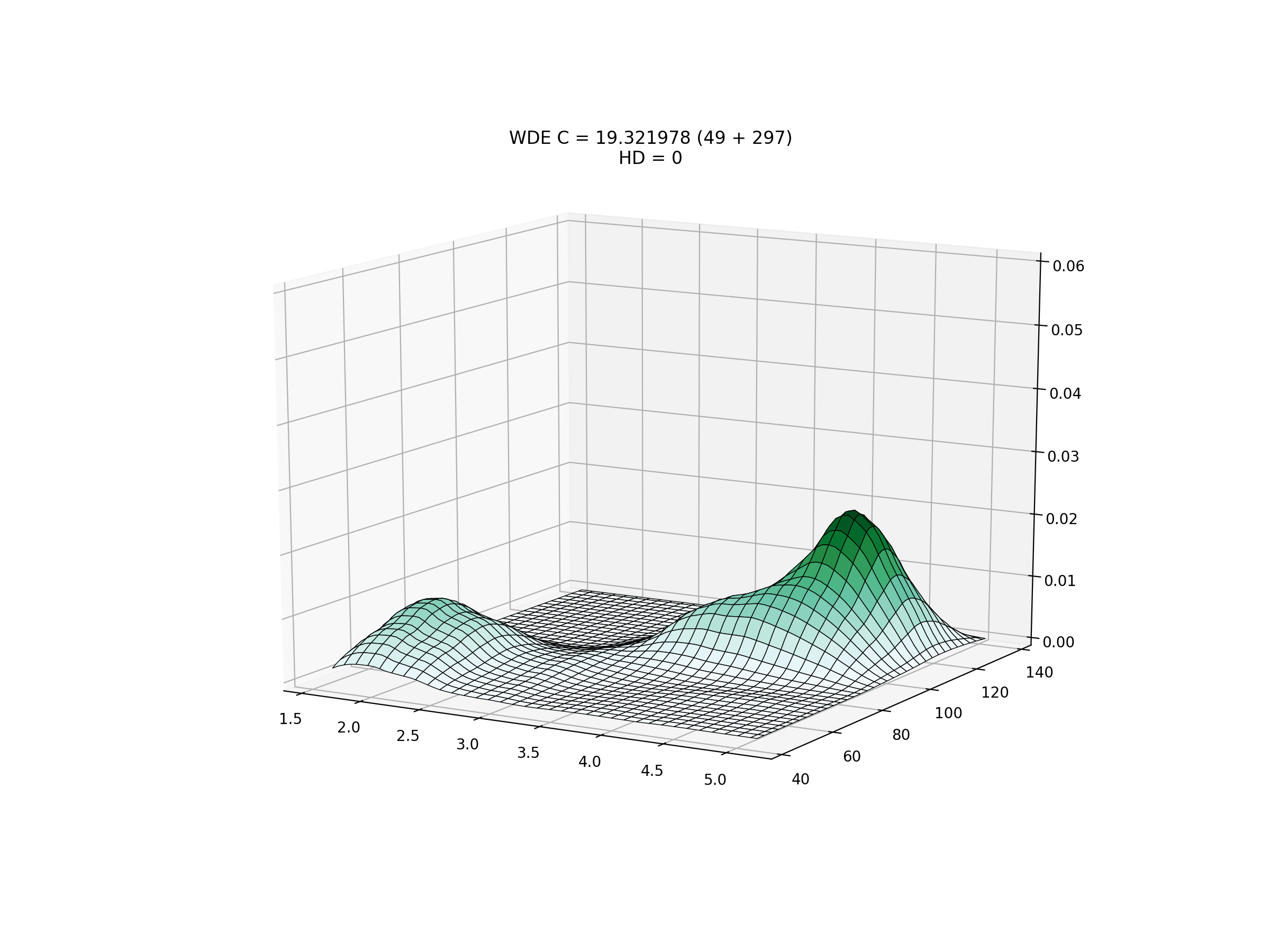}
	\end{subfigure}
	\caption{Optimisation curve, left, and density, right, for the Old Faithful geyser dataset using Daubechies $4$ and $\Delta J = 2$.} \label{fig:geyserbaddb4level2}
\end{figure}
\begin{figure}[h]
	\begin{subfigure}[t]{0.5\textwidth}
		\includegraphics[width=1.2\textwidth]{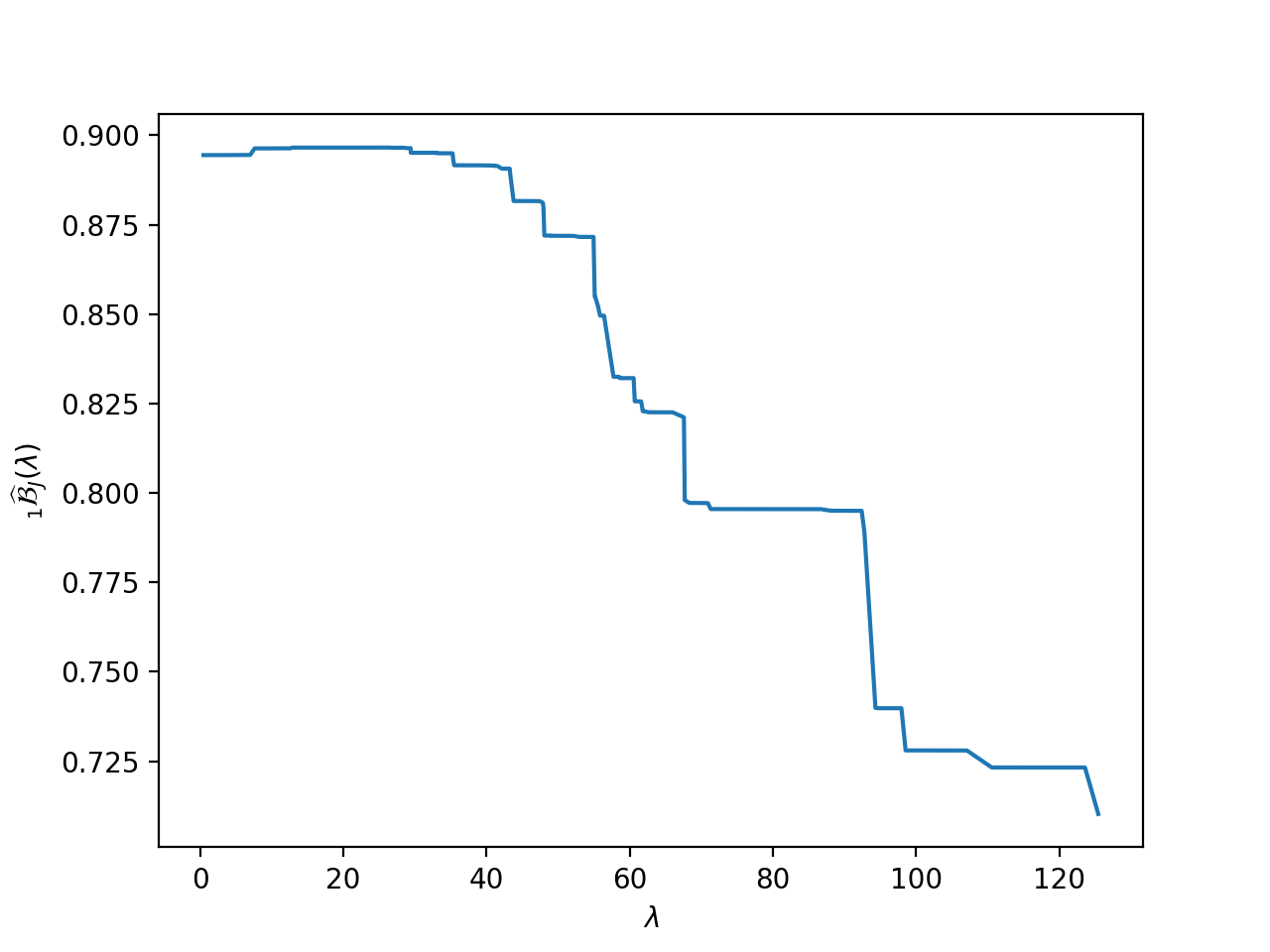}
	\end{subfigure}
	\begin{subfigure}[t]{0.5\textwidth}
		\includegraphics[width=1.2\textwidth]{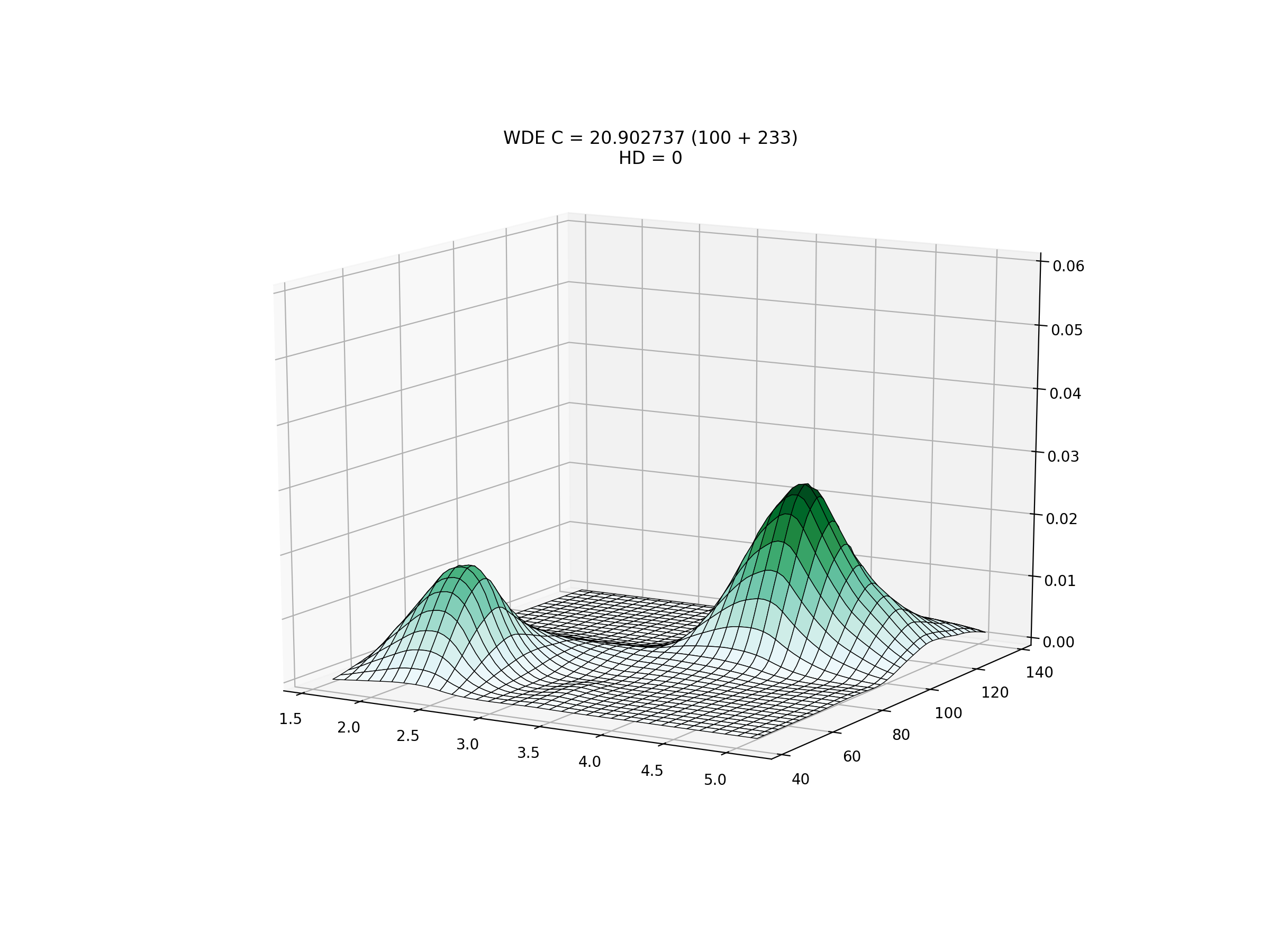}
	\end{subfigure}
	\caption{Optimisation curve, left, and density, right, for the Old Faithful geyser dataset using Symlet $5$ and $\Delta J = 2$.} \label{fig:geysersym5level1}
\end{figure}
\begin{figure}[h]
	\begin{subfigure}[t]{0.5\textwidth}
		\includegraphics[width=1.2\textwidth]{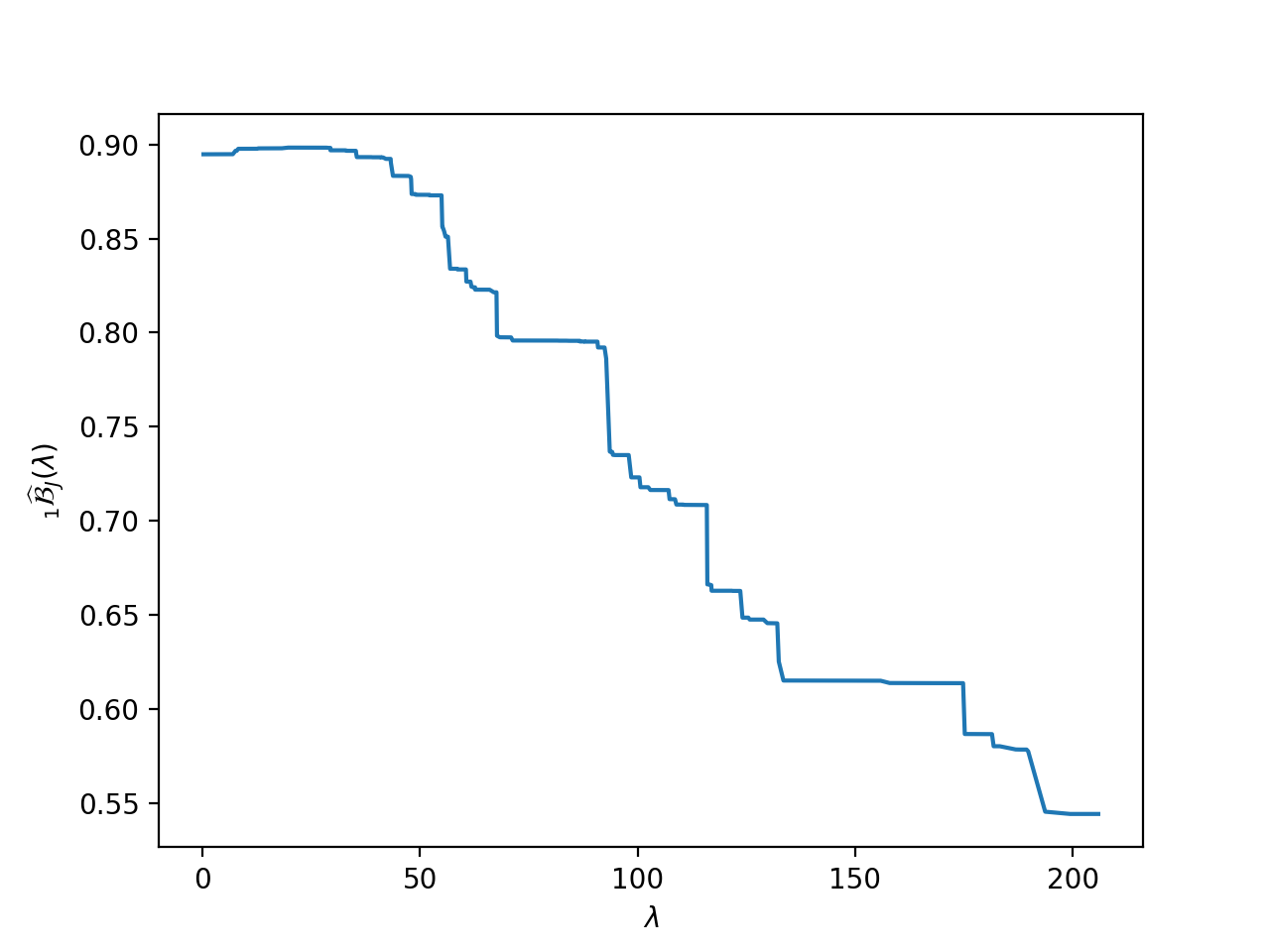}
	\end{subfigure}
	\begin{subfigure}[t]{0.5\textwidth}
		\includegraphics[width=1.2\textwidth]{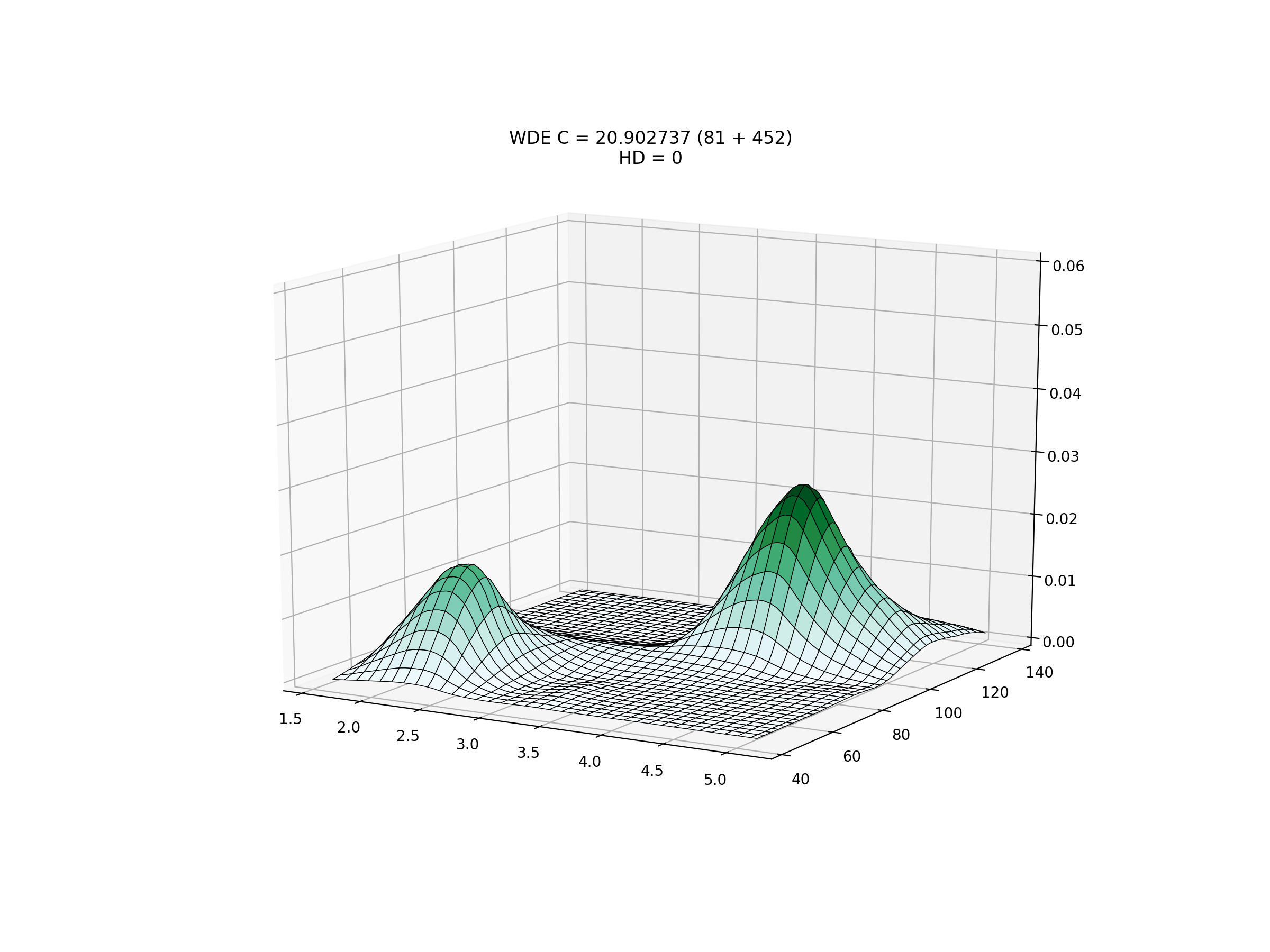}
	\end{subfigure}
	\caption{Optimisation curve, left, and density, right, for the Old Faithful geyser dataset using Symlet $5$ and $\Delta J = 2$.} \label{fig:geysersym5level2}
\end{figure}

We may contrast the above curves with those obtained if one increases the number of vanishing moments in both cases. Using Daubechies $4$ wavelets with $\Delta J = 1$ (Figure \ref{fig:geyserbaddb4level1}), one can see that most of the curve lies on the right hand side of its maximum, indicating that the threshold is going to be too drastic. Indeed, the resulting density seems over-smoothed compared to above. The situation is not much different for $\Delta J = 2$ as shown in Figure \ref{fig:geyserbaddb4level2}, or for Symlet $5$ with $\Delta J = 1$ (Figure \ref{fig:geysersym5level1}) and $\Delta J = 2$ (Figure \ref{fig:geysersym5level2}). We note that the maximum value of $\widehat{\mathcal{B}}(\tau)$ is, in all these cases, smaller than what it was above for Symlet 4 and $\Delta J = 2$.

\medskip

We finalise this chapter by considering biorthogonal bases, the most general wavelet construction, which has lately become very popular in the literature on multidimensional, sparse wavelet expansions. Here we constructed our multivariate wavelet expansions via tensor product (e.g. \cite[Chapter 10]{Daubechies92}) as opposed to the anisotropic methods of curvelets \cite{Candes04,Candes05}, shearlets \cite{Labate05} and $\alpha$-molecules \cite{Grohs13}. Although the results below are promising, theoretical and practical work is required to bring those novel algorithms to our construction. Indeed, in the general anisotropic case, the number of non-zero beta coefficients using the tensor product is $O(n)$, i.e. not optimal for a general sparse representation \cite{Starck10}.

\medskip

\begin{figure}[h]
	\begin{subfigure}[t]{0.5\textwidth}
		\includegraphics[width=\textwidth]{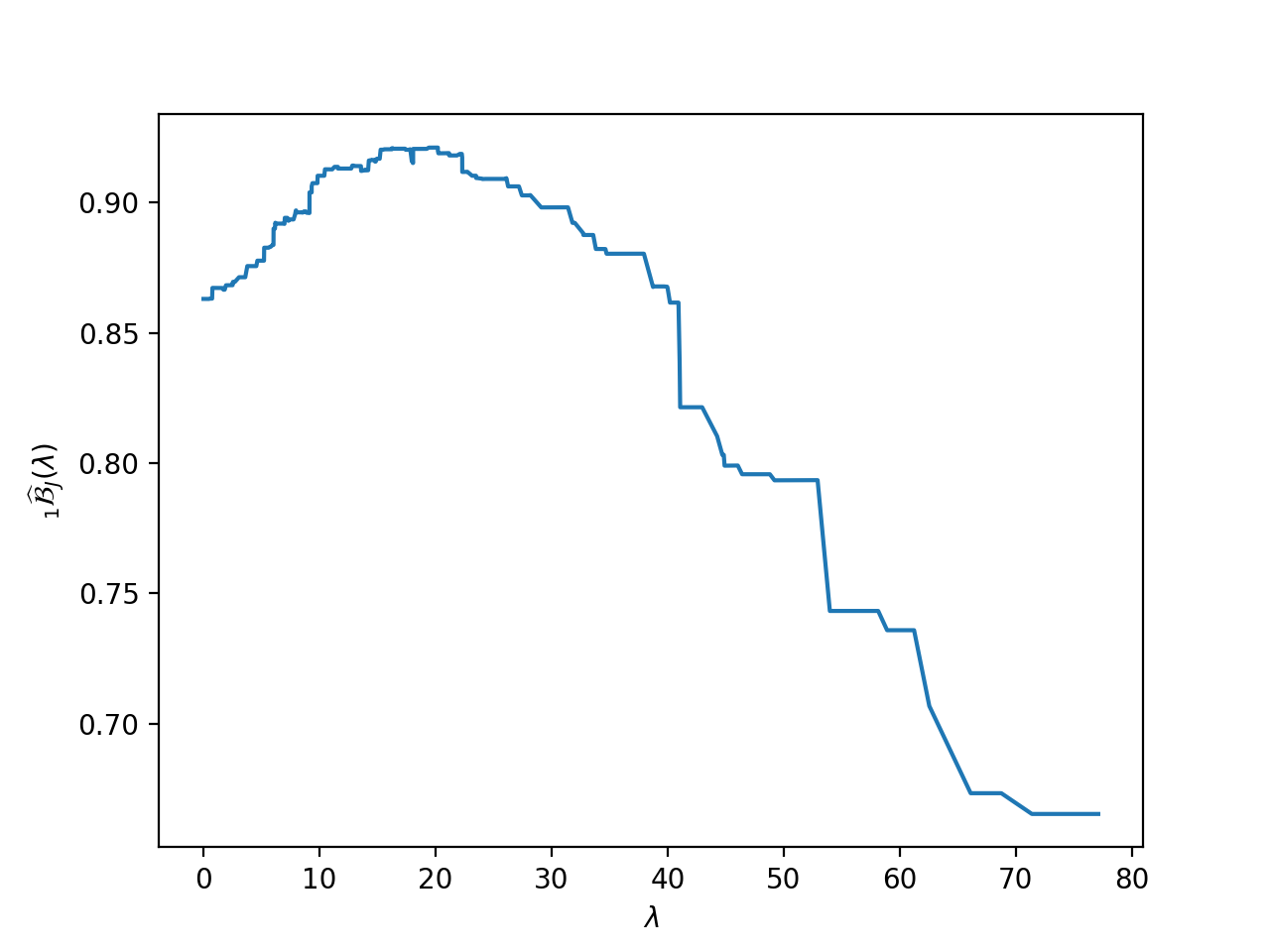}
	\end{subfigure}
	\begin{subfigure}[t]{0.5\textwidth}
		\includegraphics[width=\textwidth]{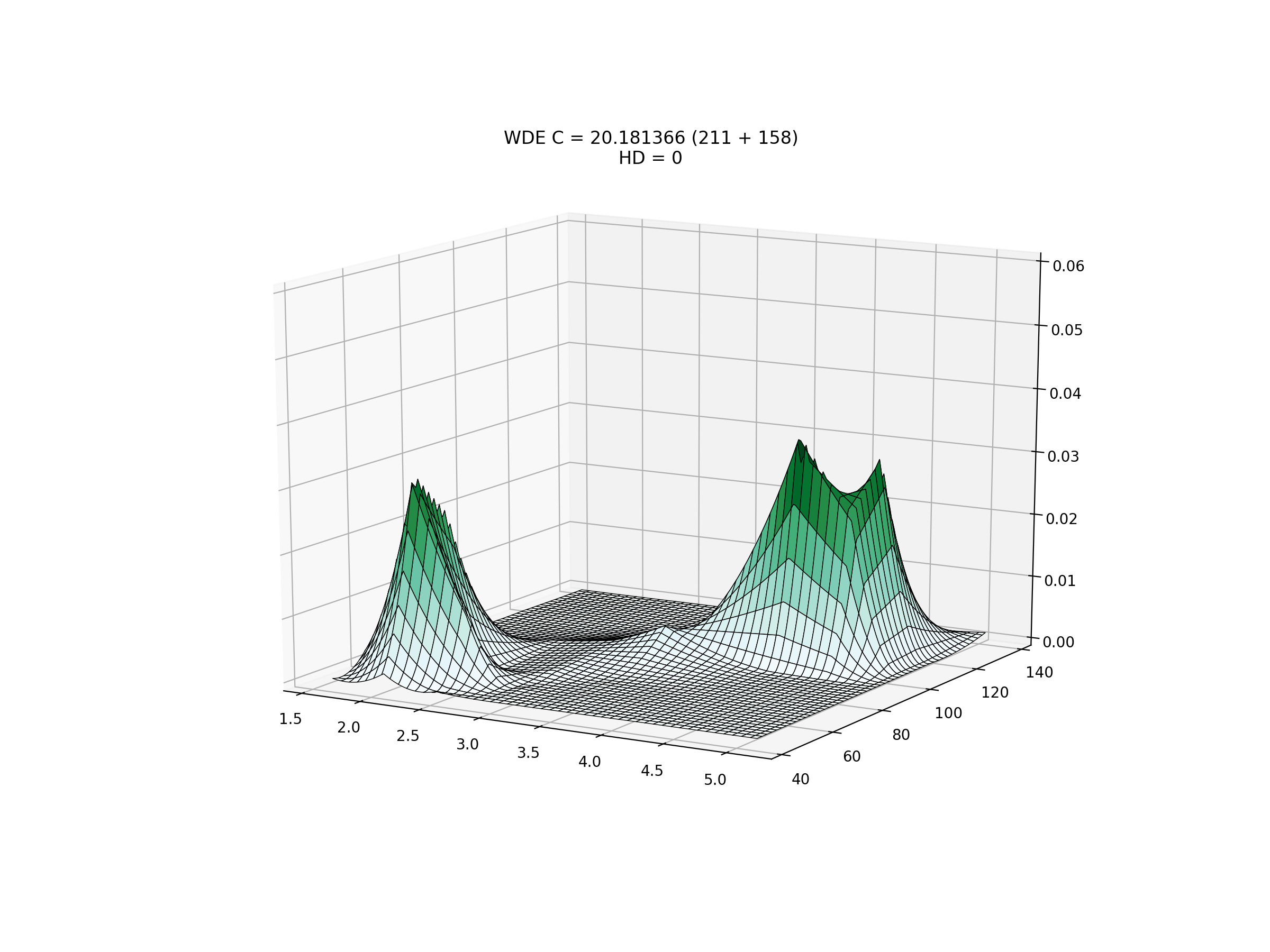}
	\end{subfigure}
	\caption{Optimisation curve, left, and density, right, for the Old Faithful geyser dataset using biorthogonal spline wavelets $2.6$ and $\Delta J = 2$.} \label{fig:geyserbior26level2}
\end{figure}

\begin{figure}[h]
	\begin{subfigure}[t]{0.5\textwidth}
		\includegraphics[width=\textwidth]{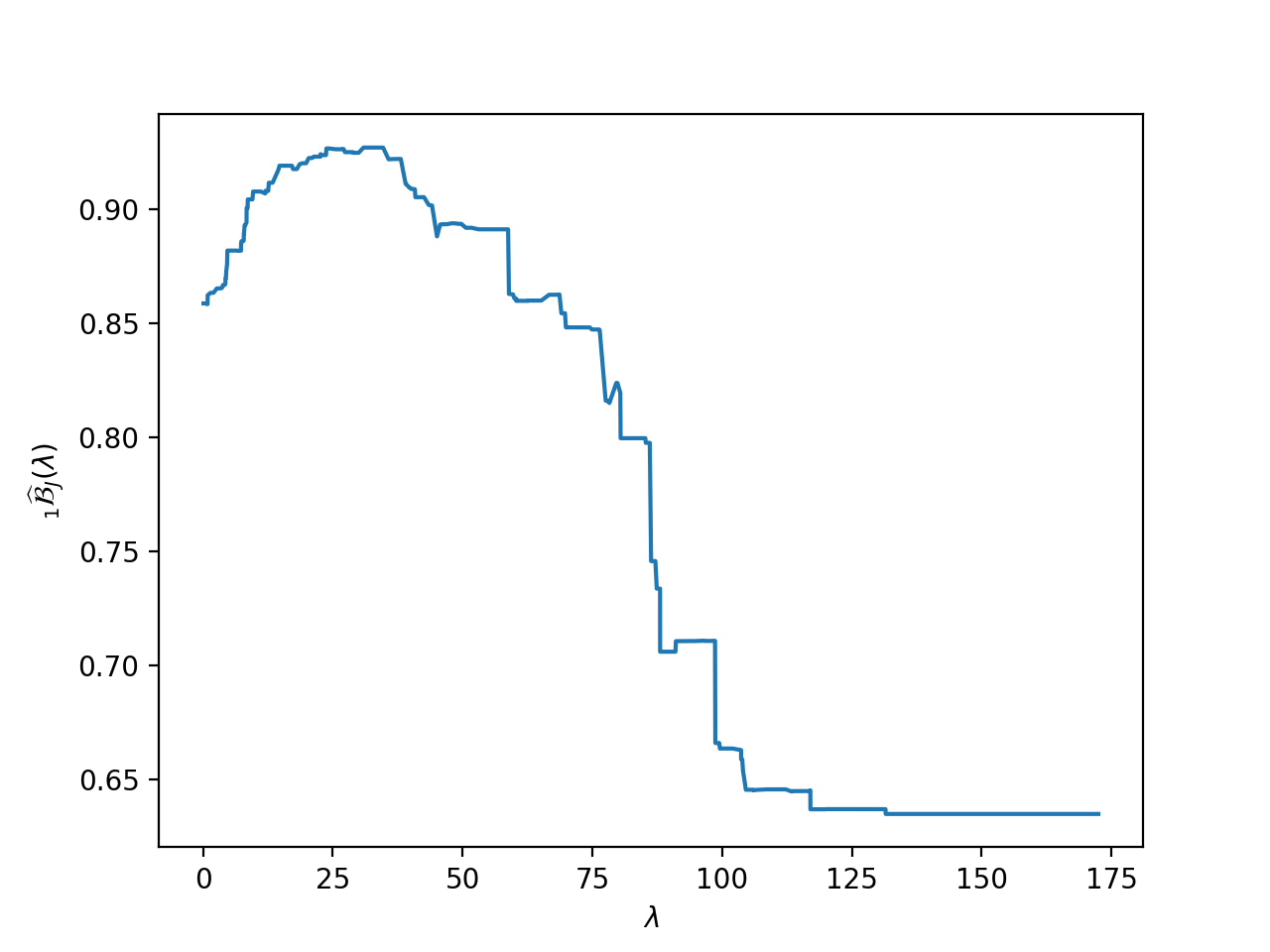}
	\end{subfigure}
	\begin{subfigure}[t]{0.5\textwidth}
		\includegraphics[width=\textwidth]{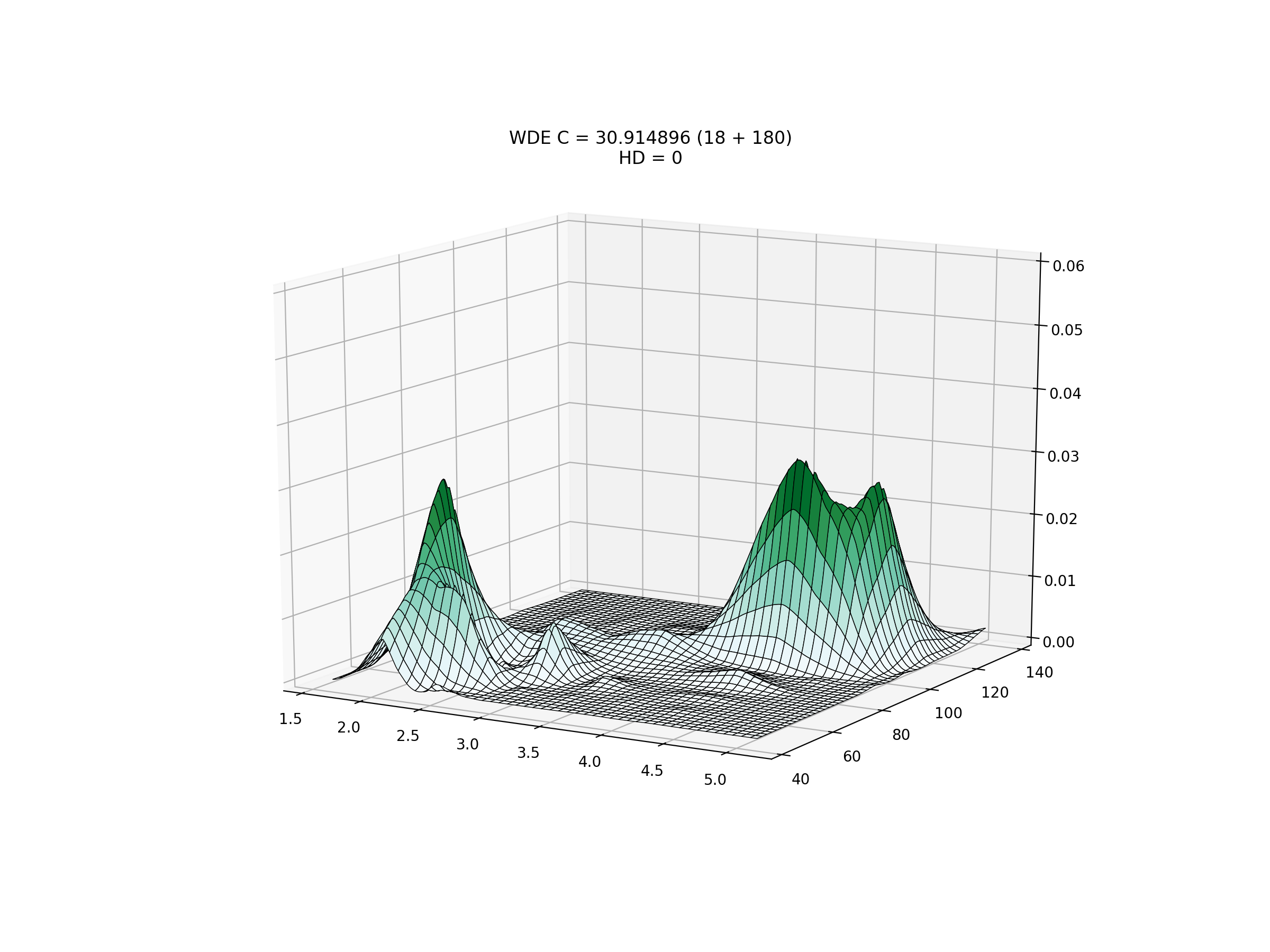}
	\end{subfigure}
	\caption{Optimisation curve, left, and density, right, for the Old Faithful geyser dataset using biorthogonal spline wavelets $6.2$ and $\Delta J = 2$.} \label{fig:geyserrbio26level2}
\end{figure}

Figure \ref{fig:geyserbior26level2} shows the optimisation curve and the density estimate using the biorthogonal, spline wavelets of \cite{Cohen92} for the case $2.6$ - this is, $6$ vanishing moments on the deconstruction filters and $2$ on the synthesis phase. We used $\Delta J = 2$ levels of betas. The result is not smooth: this uses a linear spline ($2$ vanishing moments) to reconstruct the signal but the optimisation curve seems balanced as observed in the Daubechies $3$ and Symlet $4$ results above. Reversing the role of these bases using now $6.2$, one gets a smoother density and the optimisation curve is still well-behaved, as shown in Figure \ref{fig:geyserrbio26level2}. The maximum value of $\widehat{\mathcal{B}}(\tau)$ shows an improvement from 0.92118 for 2.6 to 0.92689 for 6.2 respectively.

\medskip 

\section{Conclusion} \label{sec:conclusions}

In this paper we have addressed several gaps in the theory and practice of the `shape-preserving' wavelet density estimator recently proposed in \cite{Aya18}. That estimator, which always produces a {\it bona fide} density as output, is based on a primary estimation of the square-root of the density of interest. The natural statistical divergence involving square-root of densities is the {Hellinger distance}, itself strongly related to the \Bhat affinity coefficient. Hence this study is articulated around a novel cross-validation-like criterion which approximates the Hellinger-\Bhat distance between the estimator and the true density being estimated. User-defined parameters, such as the resolution level or the thresholding cut-off point, may be practically determined by optimising that Hellinger-\Bhat cross-validation criterion. We proved theoretically the optimality of that procedure, while the simulations evidence very strong performance in practice as well. Along the way, we proposed a novel thresholding approach, based on a jackknife estimation of the variance of each estimated wavelet coefficient -- this again fits perfectly within the considered framework of Hellinger-\Bhat cross-validation. This thresholding scheme has proved superior to other, more classical thresholding options in the simulations, and definitely deserves further research in a follow-up paper.

\appendix

\section*{Appendix}

Let $A_{n}$ the search space for $\lambda$ above in the `weighted delta sequence' $\eqref{eqn:deltarootalpha}$ for a sample of size $n$. We introduce the following additional assumptions on $A_{n}$:
\begin{assumption} \label{ass:optionsbounded}
	The cardinality of $A_{n}$ is bounded by a power of the sample size, i.e. $\#\left(A_{n}\right) \leqslant \mathcal{C} n^{\rho}$ for some $\mathcal{C}, \rho>0$.
\end{assumption}
\begin{assumption} \label{ass:valuebounded}
	There are two positive constants $\mathcal{C}', \epsilon > 0$, such that for any possible parameter choice $\lambda \in A_n$, $\mathcal{C}'^{-1} n^{\epsilon} \leq \lambda \leq \mathcal{C}' n^{1-\epsilon}$.
\end{assumption}
Assumption \ref{ass:optionsbounded} first appears in \cite{Stone84} and Assumption \ref{ass:valuebounded} is motivated in \cite{Hall83a} for the case of kernel estimators where the bandwidth $h$ is of order $n^{-1/5}$.

\printbibliography[title={References}]
\end{document}

%% file: tables/ch4-th-table-ex01.tex
\begin{table}[ht!]
	\fontsize{7}{10}\selectfont
	\centering
	\begin{minipage}[t]{0.95\textwidth}
		\centering
		\begin{tabular}{|r|r|r|r|r|r|r|r|r|r|r|r|r|}
			\hline
			\multirow{2}{*}{$n$} & \multirow{2}{*}{} & \multirow{2}{*}{$\Delta J$} & \multicolumn{3}{c|}{\textbf{universal}} & \multicolumn{3}{c|}{\textbf{level-dependent}} & \multicolumn{3}{c|}{\textbf{jackknife}} & \multirow{2}{*}{KDE} \\
			\cline{4-12}
			& & & $Q_1$ & $Med$ & $Q_3$ & $Q_1$ & $Med$ & $Q_3$ & $Q_1$ & $Med$ & $Q_3$ & \\
			\hline
			\multirow{6}{*}{250} &
			\multirow{3}{*}{$\widehat{\Bs}(J)$} &
			1 &
			{\tiny 0.0805} &
			0.0881 &
			{\tiny 0.0952} &
			{\tiny 0.0805} &
			0.0881 &
			{\tiny 0.0952} &
			{\tiny 0.0810} &
			0.0889 &
			{\tiny 0.0962} &
			\multirow{6}{*}{0.0410} \\
			\cline{3-12}
			&
			&
			2 &
			{\tiny 0.0812} &
			0.0871 &
			{\tiny 0.0948} &
			{\tiny 0.0828} &
			0.0885 &
			{\tiny 0.0960} &
			{\tiny 0.0820} &
			0.0898 &
			{\tiny 0.0983} &
			\\
			\cline{3-12}
			&
			&
			3 &
			{\tiny 0.0816} &
			0.0871 &
			{\tiny 0.0944} &
			{\tiny 0.0828} &
			0.0879 &
			{\tiny 0.0942} &
			{\tiny 0.0820} &
			0.0898 &
			{\tiny 0.0983} &
			\\
			\cline{2-12}
			&
			\multirow{3}{*}{$\widehat{\Bs}_\circ(J)$} &
			1 &
			{\tiny 0.0845} &
			0.0904 &
			{\tiny 0.0966} &
			{\tiny 0.0845} &
			0.0904 &
			{\tiny 0.0966} &
			{\tiny 0.0859} &
			0.0924 &
			{\tiny 0.0963} &
			\\
			\cline{3-12}
			&
			&
			2 &
			{\tiny 0.0838} &
			0.0882 &
			{\tiny 0.0972} &
			{\tiny 0.0842} &
			0.0891 &
			{\tiny 0.0971} &
			{\tiny 0.0868} &
			0.0927 &
			{\tiny 0.0989} &
			\\
			\cline{3-12}
			&
			&
			3 &
			{\tiny 0.0838} &
			0.0879 &
			{\tiny 0.0952} &
			{\tiny 0.0836} &
			0.0879 &
			{\tiny 0.0940} &
			{\tiny 0.0868} &
			0.0927 &
			{\tiny 0.0990} &
			\\
			\hline
			\multirow{6}{*}{500} &
			\multirow{3}{*}{$\widehat{\Bs}(J)$} &
			1 &
			{\tiny 0.0374} &
			0.0447 &
			{\tiny 0.0781} &
			{\tiny 0.0374} &
			0.0447 &
			{\tiny 0.0781} &
			{\tiny 0.0315} &
			0.0396 &
			{\tiny 0.0785} &
			\multirow{6}{*}{0.0294} \\
			\cline{3-12}
			&
			&
			2 &
			{\tiny 0.0427} &
			0.0514 &
			{\tiny 0.0775} &
			{\tiny 0.0464} &
			0.0583 &
			{\tiny 0.0775} &
			{\tiny 0.0340} &
			0.0407 &
			{\tiny 0.0785} &
			\\
			\cline{3-12}
			&
			&
			3 &
			{\tiny 0.0438} &
			0.0523 &
			{\tiny 0.0776} &
			{\tiny 0.0482} &
			0.0591 &
			{\tiny 0.0776} &
			{\tiny 0.0341} &
			0.0407 &
			{\tiny 0.0786} &
			\\
			\cline{2-12}
			&
			\multirow{3}{*}{$\widehat{\Bs}_\circ(J)$} &
			1 &
			{\tiny 0.0387} &
			0.0772 &
			{\tiny 0.0810} &
			{\tiny 0.0387} &
			0.0772 &
			{\tiny 0.0810} &
			{\tiny 0.0340} &
			0.0778 &
			{\tiny 0.0807} &
			\\
			\cline{3-12}
			&
			&
			2 &
			{\tiny 0.0467} &
			0.0767 &
			{\tiny 0.0809} &
			{\tiny 0.0512} &
			0.0771 &
			{\tiny 0.0810} &
			{\tiny 0.0367} &
			0.0781 &
			{\tiny 0.0810} &
			\\
			\cline{3-12}
			&
			&
			3 &
			{\tiny 0.0481} &
			0.0768 &
			{\tiny 0.0809} &
			{\tiny 0.0518} &
			0.0768 &
			{\tiny 0.0810} &
			{\tiny 0.0367} &
			0.0781 &
			{\tiny 0.0810} &
			\\
			\hline
			\multirow{6}{*}{1000} &
			\multirow{3}{*}{$\widehat{\Bs}(J)$} &
			1 &
			{\tiny 0.0215} &
			0.0239 &
			{\tiny 0.0275} &
			{\tiny 0.0215} &
			0.0239 &
			{\tiny 0.0275} &
			{\tiny 0.0186} &
			0.0203 &
			{\tiny 0.0229} &
			\multirow{6}{*}{0.0199} \\
			\cline{3-12}
			&
			&
			2 &
			{\tiny 0.0226} &
			0.0255 &
			{\tiny 0.0299} &
			{\tiny 0.0241} &
			0.0282 &
			{\tiny 0.0326} &
			{\tiny 0.0192} &
			0.0214 &
			{\tiny 0.0241} &
			\\
			\cline{3-12}
			&
			&
			3 &
			{\tiny 0.0242} &
			0.0264 &
			{\tiny 0.0309} &
			{\tiny 0.0261} &
			0.0295 &
			{\tiny 0.0336} &
			{\tiny 0.0192} &
			0.0215 &
			{\tiny 0.0242} &
			\\
			\cline{2-12}
			&
			\multirow{3}{*}{$\widehat{\Bs}_\circ(J)$} &
			1 &
			{\tiny 0.0218} &
			0.0239 &
			{\tiny 0.0271} &
			{\tiny 0.0218} &
			0.0239 &
			{\tiny 0.0271} &
			{\tiny 0.0187} &
			0.0205 &
			{\tiny 0.0230} &
			\\
			\cline{3-12}
			&
			&
			2 &
			{\tiny 0.0229} &
			0.0258 &
			{\tiny 0.0302} &
			{\tiny 0.0248} &
			0.0287 &
			{\tiny 0.0331} &
			{\tiny 0.0192} &
			0.0216 &
			{\tiny 0.0241} &
			\\
			\cline{3-12}
			&
			&
			3 &
			{\tiny 0.0243} &
			0.0274 &
			{\tiny 0.0310} &
			{\tiny 0.0265} &
			0.0303 &
			{\tiny 0.0339} &
			{\tiny 0.0193} &
			0.0217 &
			{\tiny 0.0242} &
			\\
			\hline
			\multirow{6}{*}{1500} &
			\multirow{3}{*}{$\widehat{\Bs}(J)$} &
			1 &
			{\tiny 0.0154} &
			0.0174 &
			{\tiny 0.0198} &
			{\tiny 0.0154} &
			0.0174 &
			{\tiny 0.0198} &
			{\tiny 0.0139} &
			\textbf{0.0156} &
			{\tiny 0.0171} &
			\multirow{6}{*}{0.0157} \\
			\cline{3-12}
			&
			&
			2 &
			{\tiny 0.0166} &
			0.0186 &
			{\tiny 0.0211} &
			{\tiny 0.0176} &
			0.0200 &
			{\tiny 0.0226} &
			{\tiny 0.0144} &
			0.0164 &
			{\tiny 0.0178} &
			\\
			\cline{3-12}
			&
			&
			3 &
			{\tiny 0.0172} &
			0.0192 &
			{\tiny 0.0220} &
			{\tiny 0.0179} &
			0.0205 &
			{\tiny 0.0231} &
			{\tiny 0.0145} &
			0.0164 &
			{\tiny 0.0179} &
			\\
			\cline{2-12}
			&
			\multirow{3}{*}{$\widehat{\Bs}_\circ(J)$} &
			1 &
			{\tiny 0.0152} &
			0.0174 &
			{\tiny 0.0198} &
			{\tiny 0.0152} &
			0.0174 &
			{\tiny 0.0198} &
			{\tiny 0.0139} &
			\textbf{0.0153} &
			{\tiny 0.0169} &
			\\
			\cline{3-12}
			&
			&
			2 &
			{\tiny 0.0165} &
			0.0186 &
			{\tiny 0.0211} &
			{\tiny 0.0174} &
			0.0200 &
			{\tiny 0.0226} &
			{\tiny 0.0143} &
			0.0163 &
			{\tiny 0.0176} &
			\\
			\cline{3-12}
			&
			&
			3 &
			{\tiny 0.0171} &
			0.0192 &
			{\tiny 0.0220} &
			{\tiny 0.0179} &
			0.0205 &
			{\tiny 0.0231} &
			{\tiny 0.0144} &
			0.0163 &
			{\tiny 0.0176} &
			\\
			\hline
			\multirow{6}{*}{2000} &
			\multirow{3}{*}{$\widehat{\Bs}(J)$} &
			1 &
			{\tiny 0.0133} &
			0.0145 &
			{\tiny 0.0165} &
			{\tiny 0.0133} &
			0.0145 &
			{\tiny 0.0165} &
			{\tiny 0.0119} &
			\textbf{0.0132} &
			{\tiny 0.0145} &
			\multirow{6}{*}{0.0136} \\
			\cline{3-12}
			&
			&
			2 &
			{\tiny 0.0139} &
			0.0151 &
			{\tiny 0.0169} &
			{\tiny 0.0141} &
			0.0163 &
			{\tiny 0.0175} &
			{\tiny 0.0121} &
			\textbf{0.0133} &
			{\tiny 0.0151} &
			\\
			\cline{3-12}
			&
			&
			3 &
			{\tiny 0.0143} &
			0.0156 &
			{\tiny 0.0177} &
			{\tiny 0.0148} &
			0.0164 &
			{\tiny 0.0180} &
			{\tiny 0.0122} &
			\textbf{0.0134} &
			{\tiny 0.0151} &
			\\
			\cline{2-12}
			&
			\multirow{3}{*}{$\widehat{\Bs}_\circ(J)$} &
			1 &
			{\tiny 0.0133} &
			0.0145 &
			{\tiny 0.0164} &
			{\tiny 0.0133} &
			0.0145 &
			{\tiny 0.0164} &
			{\tiny 0.0119} &
			\textbf{0.0131} &
			{\tiny 0.0144} &
			\\
			\cline{3-12}
			&
			&
			2 &
			{\tiny 0.0140} &
			0.0151 &
			{\tiny 0.0169} &
			{\tiny 0.0144} &
			0.0163 &
			{\tiny 0.0175} &
			{\tiny 0.0121} &
			\textbf{0.0133} &
			{\tiny 0.0150} &
			\\
			\cline{3-12}
			&
			&
			3 &
			{\tiny 0.0143} &
			0.0156 &
			{\tiny 0.0174} &
			{\tiny 0.0148} &
			0.0164 &
			{\tiny 0.0179} &
			{\tiny 0.0122} &
			\textbf{0.0134} &
			{\tiny 0.0150} &
			\\
			\hline
			\multirow{6}{*}{3000} &
			\multirow{3}{*}{$\widehat{\Bs}(J)$} &
			1 &
			{\tiny 0.0105} &
			0.0110 &
			{\tiny 0.0120} &
			{\tiny 0.0105} &
			0.0110 &
			{\tiny 0.0120} &
			{\tiny 0.0095} &
			\textbf{0.0100} &
			{\tiny 0.0107} &
			\multirow{6}{*}{0.0109} \\
			\cline{3-12}
			&
			&
			2 &
			{\tiny 0.0107} &
			0.0116 &
			{\tiny 0.0124} &
			{\tiny 0.0109} &
			0.0121 &
			{\tiny 0.0130} &
			{\tiny 0.0097} &
			\textbf{0.0103} &
			{\tiny 0.0110} &
			\\
			\cline{3-12}
			&
			&
			3 &
			{\tiny 0.0108} &
			0.0116 &
			{\tiny 0.0127} &
			{\tiny 0.0111} &
			0.0122 &
			{\tiny 0.0132} &
			{\tiny 0.0097} &
			\textbf{0.0103} &
			{\tiny 0.0110} &
			\\
			\cline{2-12}
			&
			\multirow{3}{*}{$\widehat{\Bs}_\circ(J)$} &
			1 &
			{\tiny 0.0105} &
			0.0110 &
			{\tiny 0.0120} &
			{\tiny 0.0105} &
			0.0110 &
			{\tiny 0.0120} &
			{\tiny 0.0094} &
			\textbf{0.0100} &
			{\tiny 0.0107} &
			\\
			\cline{3-12}
			&
			&
			2 &
			{\tiny 0.0107} &
			0.0116 &
			{\tiny 0.0124} &
			{\tiny 0.0110} &
			0.0121 &
			{\tiny 0.0129} &
			{\tiny 0.0097} &
			\textbf{0.0103} &
			{\tiny 0.0111} &
			\\
			\cline{3-12}
			&
			&
			3 &
			{\tiny 0.0108} &
			0.0117 &
			{\tiny 0.0126} &
			{\tiny 0.0111} &
			0.0122 &
			{\tiny 0.0133} &
			{\tiny 0.0097} &
			\textbf{0.0104} &
			{\tiny 0.0110} &
			\\
			\hline
			\multirow{6}{*}{4000} &
			\multirow{3}{*}{$\widehat{\Bs}(J)$} &
			1 &
			{\tiny 0.0086} &
			\textbf{0.0091} &
			{\tiny 0.0099} &
			{\tiny 0.0086} &
			\textbf{0.0091} &
			{\tiny 0.0099} &
			{\tiny 0.0081} &
			\textbf{0.0087} &
			{\tiny 0.0092} &
			\multirow{6}{*}{0.0094} \\
			\cline{3-12}
			&
			&
			2 &
			{\tiny 0.0091} &
			0.0095 &
			{\tiny 0.0105} &
			{\tiny 0.0094} &
			0.0097 &
			{\tiny 0.0107} &
			{\tiny 0.0082} &
			\textbf{0.0087} &
			{\tiny 0.0094} &
			\\
			\cline{3-12}
			&
			&
			3 &
			{\tiny 0.0091} &
			0.0096 &
			{\tiny 0.0106} &
			{\tiny 0.0094} &
			0.0100 &
			{\tiny 0.0109} &
			{\tiny 0.0082} &
			\textbf{0.0087} &
			{\tiny 0.0094} &
			\\
			\cline{2-12}
			&
			\multirow{3}{*}{$\widehat{\Bs}_\circ(J)$} &
			1 &
			{\tiny 0.0087} &
			\textbf{0.0092} &
			{\tiny 0.0101} &
			{\tiny 0.0087} &
			\textbf{0.0092} &
			{\tiny 0.0101} &
			{\tiny 0.0081} &
			\textbf{0.0087} &
			{\tiny 0.0092} &
			\\
			\cline{3-12}
			&
			&
			2 &
			{\tiny 0.0091} &
			0.0096 &
			{\tiny 0.0106} &
			{\tiny 0.0094} &
			0.0098 &
			{\tiny 0.0108} &
			{\tiny 0.0081} &
			\textbf{0.0088} &
			{\tiny 0.0093} &
			\\
			\cline{3-12}
			&
			&
			3 &
			{\tiny 0.0091} &
			0.0097 &
			{\tiny 0.0107} &
			{\tiny 0.0095} &
			0.0100 &
			{\tiny 0.0112} &
			{\tiny 0.0082} &
			\textbf{0.0088} &
			{\tiny 0.0093} &
			\\
			\hline
			\multirow{6}{*}{6000} &
			\multirow{3}{*}{$\widehat{\Bs}(J)$} &
			1 &
			{\tiny 0.0068} &
			0.0075 &
			{\tiny 0.0081} &
			{\tiny 0.0068} &
			0.0075 &
			{\tiny 0.0081} &
			{\tiny 0.0065} &
			\textbf{0.0070} &
			{\tiny 0.0077} &
			\multirow{6}{*}{0.0072} \\
			\cline{3-12}
			&
			&
			2 &
			{\tiny 0.0070} &
			0.0076 &
			{\tiny 0.0082} &
			{\tiny 0.0074} &
			0.0079 &
			{\tiny 0.0085} &
			{\tiny 0.0067} &
			\textbf{0.0072} &
			{\tiny 0.0077} &
			\\
			\cline{3-12}
			&
			&
			3 &
			{\tiny 0.0073} &
			0.0077 &
			{\tiny 0.0080} &
			{\tiny 0.0077} &
			0.0081 &
			{\tiny 0.0088} &
			{\tiny 0.0067} &
			\textbf{0.0072} &
			{\tiny 0.0077} &
			\\
			\cline{2-12}
			&
			\multirow{3}{*}{$\widehat{\Bs}_\circ(J)$} &
			1 &
			{\tiny 0.0067} &
			0.0073 &
			{\tiny 0.0078} &
			{\tiny 0.0067} &
			0.0073 &
			{\tiny 0.0078} &
			{\tiny 0.0064} &
			\textbf{0.0069} &
			{\tiny 0.0073} &
			\\
			\cline{3-12}
			&
			&
			2 &
			{\tiny 0.0068} &
			0.0074 &
			{\tiny 0.0080} &
			{\tiny 0.0071} &
			0.0076 &
			{\tiny 0.0084} &
			{\tiny 0.0065} &
			\textbf{0.0070} &
			{\tiny 0.0074} &
			\\
			\cline{3-12}
			&
			&
			3 &
			{\tiny 0.0069} &
			0.0074 &
			{\tiny 0.0080} &
			{\tiny 0.0072} &
			0.0077 &
			{\tiny 0.0085} &
			{\tiny 0.0065} &
			\textbf{0.0070} &
			{\tiny 0.0074} &
			\\
			\hline
		\end{tabular}
	\end{minipage}
    \caption{Hellinger Distances between the true density `Kurtotic Mixture 1' (Figure \ref{fig:truedenshardt}(a)) and estimators obtained through various estimation schemes. The wavelet family is Daubechies 4. See text for column descriptions.}
	\label{tab:ex01}
\end{table}

%% file: tables/ch4-th-table-ex02.tex
\begin{table}[ht!]
	\fontsize{7}{10}\selectfont
	\centering
	\begin{minipage}[t]{0.95\textwidth}
		\centering
		\begin{tabular}{|r|r|r|r|r|r|r|r|r|r|r|r|r|}
			\hline
			\multirow{2}{*}{$n$} & \multirow{2}{*}{} & \multirow{2}{*}{$\Delta J$} & \multicolumn{3}{c|}{\textbf{universal}} & \multicolumn{3}{c|}{\textbf{level-dependent}} & \multicolumn{3}{c|}{\textbf{jackknife}} & \multirow{2}{*}{KDE} \\
			\cline{4-12}
			& & & $Q_1$ & $Med$ & $Q_3$ & $Q_1$ & $Med$ & $Q_3$ & $Q_1$ & $Med$ & $Q_3$ & \\
			\hline
			\multirow{6}{*}{250} &
			\multirow{3}{*}{$\widehat{\Bs}(J)$} &
			1 &
			{\tiny 0.0365} &
			0.0401 &
			{\tiny 0.0479} &
			{\tiny 0.0365} &
			0.0401 &
			{\tiny 0.0479} &
			{\tiny 0.0250} &
			\textbf{0.0300} &
			{\tiny 0.0344} &
			\multirow{6}{*}{0.0351} \\
			\cline{3-12}
			&
			&
			2 &
			{\tiny 0.0426} &
			0.0483 &
			{\tiny 0.0538} &
			{\tiny 0.0456} &
			0.0510 &
			{\tiny 0.0596} &
			{\tiny 0.0289} &
			\textbf{0.0326} &
			{\tiny 0.0377} &
			\\
			\cline{3-12}
			&
			&
			3 &
			{\tiny 0.0441} &
			0.0496 &
			{\tiny 0.0562} &
			{\tiny 0.0460} &
			0.0527 &
			{\tiny 0.0621} &
			{\tiny 0.0295} &
			\textbf{0.0336} &
			{\tiny 0.0386} &
			\\
			\cline{2-12}
			&
			\multirow{3}{*}{$\widehat{\Bs}_\circ(J)$} &
			1 &
			{\tiny 0.0365} &
			0.0399 &
			{\tiny 0.0463} &
			{\tiny 0.0365} &
			0.0399 &
			{\tiny 0.0463} &
			{\tiny 0.0246} &
			\textbf{0.0289} &
			{\tiny 0.0340} &
			\\
			\cline{3-12}
			&
			&
			2 &
			{\tiny 0.0421} &
			0.0483 &
			{\tiny 0.0545} &
			{\tiny 0.0456} &
			0.0510 &
			{\tiny 0.0597} &
			{\tiny 0.0290} &
			\textbf{0.0326} &
			{\tiny 0.0372} &
			\\
			\cline{3-12}
			&
			&
			3 &
			{\tiny 0.0447} &
			0.0502 &
			{\tiny 0.0562} &
			{\tiny 0.0468} &
			0.0527 &
			{\tiny 0.0625} &
			{\tiny 0.0292} &
			\textbf{0.0348} &
			{\tiny 0.0380} &
			\\
			\hline
			\multirow{6}{*}{500} &
			\multirow{3}{*}{$\widehat{\Bs}(J)$} &
			1 &
			{\tiny 0.0231} &
			0.0267 &
			{\tiny 0.0329} &
			{\tiny 0.0231} &
			0.0267 &
			{\tiny 0.0329} &
			{\tiny 0.0176} &
			\textbf{0.0206} &
			{\tiny 0.0238} &
			\multirow{6}{*}{0.0247} \\
			\cline{3-12}
			&
			&
			2 &
			{\tiny 0.0254} &
			0.0298 &
			{\tiny 0.0346} &
			{\tiny 0.0259} &
			0.0320 &
			{\tiny 0.0370} &
			{\tiny 0.0185} &
			\textbf{0.0223} &
			{\tiny 0.0265} &
			\\
			\cline{3-12}
			&
			&
			3 &
			{\tiny 0.0253} &
			0.0302 &
			{\tiny 0.0355} &
			{\tiny 0.0263} &
			0.0306 &
			{\tiny 0.0377} &
			{\tiny 0.0188} &
			\textbf{0.0223} &
			{\tiny 0.0269} &
			\\
			\cline{2-12}
			&
			\multirow{3}{*}{$\widehat{\Bs}_\circ(J)$} &
			1 &
			{\tiny 0.0230} &
			0.0267 &
			{\tiny 0.0326} &
			{\tiny 0.0230} &
			0.0267 &
			{\tiny 0.0326} &
			{\tiny 0.0176} &
			\textbf{0.0202} &
			{\tiny 0.0235} &
			\\
			\cline{3-12}
			&
			&
			2 &
			{\tiny 0.0251} &
			0.0298 &
			{\tiny 0.0353} &
			{\tiny 0.0266} &
			0.0320 &
			{\tiny 0.0372} &
			{\tiny 0.0185} &
			\textbf{0.0223} &
			{\tiny 0.0262} &
			\\
			\cline{3-12}
			&
			&
			3 &
			{\tiny 0.0264} &
			0.0302 &
			{\tiny 0.0355} &
			{\tiny 0.0263} &
			0.0318 &
			{\tiny 0.0379} &
			{\tiny 0.0190} &
			\textbf{0.0223} &
			{\tiny 0.0269} &
			\\
			\hline
			\multirow{6}{*}{1000} &
			\multirow{3}{*}{$\widehat{\Bs}(J)$} &
			1 &
			{\tiny 0.0149} &
			\textbf{0.0159} &
			{\tiny 0.0183} &
			{\tiny 0.0149} &
			\textbf{0.0159} &
			{\tiny 0.0183} &
			{\tiny 0.0128} &
			\textbf{0.0140} &
			{\tiny 0.0155} &
			\multirow{6}{*}{0.0163} \\
			\cline{3-12}
			&
			&
			2 &
			{\tiny 0.0153} &
			0.0169 &
			{\tiny 0.0195} &
			{\tiny 0.0156} &
			0.0177 &
			{\tiny 0.0198} &
			{\tiny 0.0132} &
			\textbf{0.0146} &
			{\tiny 0.0162} &
			\\
			\cline{3-12}
			&
			&
			3 &
			{\tiny 0.0156} &
			0.0171 &
			{\tiny 0.0198} &
			{\tiny 0.0157} &
			0.0179 &
			{\tiny 0.0207} &
			{\tiny 0.0132} &
			\textbf{0.0146} &
			{\tiny 0.0162} &
			\\
			\cline{2-12}
			&
			\multirow{3}{*}{$\widehat{\Bs}_\circ(J)$} &
			1 &
			{\tiny 0.0149} &
			\textbf{0.0159} &
			{\tiny 0.0184} &
			{\tiny 0.0149} &
			\textbf{0.0159} &
			{\tiny 0.0184} &
			{\tiny 0.0128} &
			\textbf{0.0140} &
			{\tiny 0.0155} &
			\\
			\cline{3-12}
			&
			&
			2 &
			{\tiny 0.0154} &
			0.0169 &
			{\tiny 0.0196} &
			{\tiny 0.0158} &
			0.0179 &
			{\tiny 0.0198} &
			{\tiny 0.0132} &
			\textbf{0.0146} &
			{\tiny 0.0162} &
			\\
			\cline{3-12}
			&
			&
			3 &
			{\tiny 0.0156} &
			0.0171 &
			{\tiny 0.0199} &
			{\tiny 0.0158} &
			0.0180 &
			{\tiny 0.0207} &
			{\tiny 0.0132} &
			\textbf{0.0146} &
			{\tiny 0.0162} &
			\\
			\hline
			\multirow{6}{*}{1500} &
			\multirow{3}{*}{$\widehat{\Bs}(J)$} &
			1 &
			{\tiny 0.0114} &
			\textbf{0.0124} &
			{\tiny 0.0141} &
			{\tiny 0.0114} &
			\textbf{0.0124} &
			{\tiny 0.0141} &
			{\tiny 0.0102} &
			\textbf{0.0114} &
			{\tiny 0.0126} &
			\multirow{6}{*}{0.0128} \\
			\cline{3-12}
			&
			&
			2 &
			{\tiny 0.0119} &
			0.0130 &
			{\tiny 0.0147} &
			{\tiny 0.0123} &
			0.0136 &
			{\tiny 0.0155} &
			{\tiny 0.0105} &
			\textbf{0.0115} &
			{\tiny 0.0129} &
			\\
			\cline{3-12}
			&
			&
			3 &
			{\tiny 0.0118} &
			0.0130 &
			{\tiny 0.0150} &
			{\tiny 0.0123} &
			0.0136 &
			{\tiny 0.0157} &
			{\tiny 0.0105} &
			\textbf{0.0115} &
			{\tiny 0.0129} &
			\\
			\cline{2-12}
			&
			\multirow{3}{*}{$\widehat{\Bs}_\circ(J)$} &
			1 &
			{\tiny 0.0114} &
			\textbf{0.0124} &
			{\tiny 0.0141} &
			{\tiny 0.0114} &
			\textbf{0.0124} &
			{\tiny 0.0141} &
			{\tiny 0.0102} &
			\textbf{0.0114} &
			{\tiny 0.0126} &
			\\
			\cline{3-12}
			&
			&
			2 &
			{\tiny 0.0119} &
			0.0130 &
			{\tiny 0.0147} &
			{\tiny 0.0123} &
			0.0136 &
			{\tiny 0.0155} &
			{\tiny 0.0105} &
			\textbf{0.0115} &
			{\tiny 0.0129} &
			\\
			\cline{3-12}
			&
			&
			3 &
			{\tiny 0.0118} &
			0.0130 &
			{\tiny 0.0150} &
			{\tiny 0.0123} &
			0.0133 &
			{\tiny 0.0157} &
			{\tiny 0.0105} &
			\textbf{0.0115} &
			{\tiny 0.0129} &
			\\
			\hline
			\multirow{6}{*}{2000} &
			\multirow{3}{*}{$\widehat{\Bs}(J)$} &
			1 &
			{\tiny 0.0098} &
			\textbf{0.0104} &
			{\tiny 0.0113} &
			{\tiny 0.0098} &
			\textbf{0.0104} &
			{\tiny 0.0113} &
			{\tiny 0.0090} &
			\textbf{0.0097} &
			{\tiny 0.0107} &
			\multirow{6}{*}{0.0110} \\
			\cline{3-12}
			&
			&
			2 &
			{\tiny 0.0103} &
			0.0111 &
			{\tiny 0.0120} &
			{\tiny 0.0105} &
			0.0113 &
			{\tiny 0.0124} &
			{\tiny 0.0090} &
			\textbf{0.0099} &
			{\tiny 0.0108} &
			\\
			\cline{3-12}
			&
			&
			3 &
			{\tiny 0.0101} &
			0.0111 &
			{\tiny 0.0121} &
			{\tiny 0.0103} &
			0.0112 &
			{\tiny 0.0125} &
			{\tiny 0.0090} &
			\textbf{0.0099} &
			{\tiny 0.0108} &
			\\
			\cline{2-12}
			&
			\multirow{3}{*}{$\widehat{\Bs}_\circ(J)$} &
			1 &
			{\tiny 0.0099} &
			\textbf{0.0104} &
			{\tiny 0.0113} &
			{\tiny 0.0099} &
			\textbf{0.0104} &
			{\tiny 0.0113} &
			{\tiny 0.0090} &
			\textbf{0.0097} &
			{\tiny 0.0107} &
			\\
			\cline{3-12}
			&
			&
			2 &
			{\tiny 0.0103} &
			0.0111 &
			{\tiny 0.0120} &
			{\tiny 0.0105} &
			0.0113 &
			{\tiny 0.0124} &
			{\tiny 0.0090} &
			\textbf{0.0099} &
			{\tiny 0.0108} &
			\\
			\cline{3-12}
			&
			&
			3 &
			{\tiny 0.0101} &
			0.0111 &
			{\tiny 0.0121} &
			{\tiny 0.0104} &
			0.0112 &
			{\tiny 0.0125} &
			{\tiny 0.0090} &
			\textbf{0.0099} &
			{\tiny 0.0108} &
			\\
			\hline
			\multirow{6}{*}{3000} &
			\multirow{3}{*}{$\widehat{\Bs}(J)$} &
			1 &
			{\tiny 0.0078} &
			\textbf{0.0087} &
			{\tiny 0.0096} &
			{\tiny 0.0078} &
			\textbf{0.0087} &
			{\tiny 0.0096} &
			{\tiny 0.0075} &
			\textbf{0.0084} &
			{\tiny 0.0091} &
			\multirow{6}{*}{0.0087} \\
			\cline{3-12}
			&
			&
			2 &
			{\tiny 0.0082} &
			0.0090 &
			{\tiny 0.0099} &
			{\tiny 0.0083} &
			0.0090 &
			{\tiny 0.0100} &
			{\tiny 0.0075} &
			\textbf{0.0085} &
			{\tiny 0.0091} &
			\\
			\cline{3-12}
			&
			&
			3 &
			{\tiny 0.0081} &
			0.0089 &
			{\tiny 0.0097} &
			{\tiny 0.0082} &
			0.0091 &
			{\tiny 0.0098} &
			{\tiny 0.0075} &
			\textbf{0.0085} &
			{\tiny 0.0091} &
			\\
			\cline{2-12}
			&
			\multirow{3}{*}{$\widehat{\Bs}_\circ(J)$} &
			1 &
			{\tiny 0.0078} &
			\textbf{0.0087} &
			{\tiny 0.0096} &
			{\tiny 0.0078} &
			\textbf{0.0087} &
			{\tiny 0.0096} &
			{\tiny 0.0075} &
			\textbf{0.0084} &
			{\tiny 0.0090} &
			\\
			\cline{3-12}
			&
			&
			2 &
			{\tiny 0.0082} &
			0.0090 &
			{\tiny 0.0099} &
			{\tiny 0.0083} &
			0.0090 &
			{\tiny 0.0100} &
			{\tiny 0.0075} &
			\textbf{0.0085} &
			{\tiny 0.0091} &
			\\
			\cline{3-12}
			&
			&
			3 &
			{\tiny 0.0081} &
			0.0089 &
			{\tiny 0.0097} &
			{\tiny 0.0082} &
			0.0091 &
			{\tiny 0.0098} &
			{\tiny 0.0075} &
			\textbf{0.0085} &
			{\tiny 0.0091} &
			\\
			\hline
			\multirow{6}{*}{4000} &
			\multirow{3}{*}{$\widehat{\Bs}(J)$} &
			1 &
			{\tiny 0.0069} &
			\textbf{0.0072} &
			{\tiny 0.0080} &
			{\tiny 0.0069} &
			\textbf{0.0072} &
			{\tiny 0.0080} &
			{\tiny 0.0065} &
			\textbf{0.0070} &
			{\tiny 0.0076} &
			\multirow{6}{*}{0.0073} \\
			\cline{3-12}
			&
			&
			2 &
			{\tiny 0.0070} &
			0.0074 &
			{\tiny 0.0080} &
			{\tiny 0.0071} &
			0.0075 &
			{\tiny 0.0082} &
			{\tiny 0.0066} &
			\textbf{0.0071} &
			{\tiny 0.0076} &
			\\
			\cline{3-12}
			&
			&
			3 &
			{\tiny 0.0070} &
			\textbf{0.0073} &
			{\tiny 0.0080} &
			{\tiny 0.0071} &
			0.0074 &
			{\tiny 0.0081} &
			{\tiny 0.0066} &
			\textbf{0.0071} &
			{\tiny 0.0076} &
			\\
			\cline{2-12}
			&
			\multirow{3}{*}{$\widehat{\Bs}_\circ(J)$} &
			1 &
			{\tiny 0.0069} &
			\textbf{0.0072} &
			{\tiny 0.0080} &
			{\tiny 0.0069} &
			\textbf{0.0072} &
			{\tiny 0.0080} &
			{\tiny 0.0065} &
			\textbf{0.0070} &
			{\tiny 0.0076} &
			\\
			\cline{3-12}
			&
			&
			2 &
			{\tiny 0.0070} &
			0.0074 &
			{\tiny 0.0080} &
			{\tiny 0.0072} &
			0.0075 &
			{\tiny 0.0082} &
			{\tiny 0.0066} &
			\textbf{0.0071} &
			{\tiny 0.0076} &
			\\
			\cline{3-12}
			&
			&
			3 &
			{\tiny 0.0070} &
			0.0074 &
			{\tiny 0.0081} &
			{\tiny 0.0071} &
			0.0075 &
			{\tiny 0.0082} &
			{\tiny 0.0066} &
			\textbf{0.0071} &
			{\tiny 0.0076} &
			\\
			\hline
			\multirow{6}{*}{6000} &
			\multirow{3}{*}{$\widehat{\Bs}(J)$} &
			1 &
			{\tiny 0.0059} &
			0.0062 &
			{\tiny 0.0065} &
			{\tiny 0.0059} &
			0.0062 &
			{\tiny 0.0065} &
			{\tiny 0.0057} &
			0.0061 &
			{\tiny 0.0064} &
			\multirow{6}{*}{0.0057} \\
			\cline{3-12}
			&
			&
			2 &
			{\tiny 0.0060} &
			0.0063 &
			{\tiny 0.0066} &
			{\tiny 0.0061} &
			0.0064 &
			{\tiny 0.0068} &
			{\tiny 0.0057} &
			0.0061 &
			{\tiny 0.0064} &
			\\
			\cline{3-12}
			&
			&
			3 &
			{\tiny 0.0060} &
			0.0064 &
			{\tiny 0.0066} &
			{\tiny 0.0061} &
			0.0065 &
			{\tiny 0.0067} &
			{\tiny 0.0057} &
			0.0061 &
			{\tiny 0.0064} &
			\\
			\cline{2-12}
			&
			\multirow{3}{*}{$\widehat{\Bs}_\circ(J)$} &
			1 &
			{\tiny 0.0059} &
			0.0062 &
			{\tiny 0.0065} &
			{\tiny 0.0059} &
			0.0062 &
			{\tiny 0.0065} &
			{\tiny 0.0057} &
			0.0061 &
			{\tiny 0.0064} &
			\\
			\cline{3-12}
			&
			&
			2 &
			{\tiny 0.0060} &
			0.0063 &
			{\tiny 0.0066} &
			{\tiny 0.0061} &
			0.0064 &
			{\tiny 0.0068} &
			{\tiny 0.0057} &
			0.0061 &
			{\tiny 0.0064} &
			\\
			\cline{3-12}
			&
			&
			3 &
			{\tiny 0.0060} &
			0.0064 &
			{\tiny 0.0066} &
			{\tiny 0.0061} &
			0.0065 &
			{\tiny 0.0067} &
			{\tiny 0.0057} &
			0.0061 &
			{\tiny 0.0064} &
			\\
			\hline
		\end{tabular}
	\end{minipage}
    \caption{Hellinger Distances between the true density `Mixture 2' (Figure \ref{fig:truedenshardt}(b)) and estimators obtained through various estimation schemes. The wavelet family is Daubechies 4. See text for column descriptions.}
	\label{tab:ex02}
\end{table}

%% file: tables/ch4-th-table-ex03.tex
\begin{table}[ht!]
	\fontsize{7}{10}\selectfont
	\centering
	\begin{minipage}[t]{0.95\textwidth}
		\centering
		\begin{tabular}{|r|r|r|r|r|r|r|r|r|r|r|r|r|}
			\hline
			\multirow{2}{*}{$n$} & \multirow{2}{*}{} & \multirow{2}{*}{$\Delta J$} & \multicolumn{3}{c|}{\textbf{universal}} & \multicolumn{3}{c|}{\textbf{level-dependent}} & \multicolumn{3}{c|}{\textbf{jackknife}} & \multirow{2}{*}{KDE} \\
			\cline{4-12}
			& & & $Q_1$ & $Med$ & $Q_3$ & $Q_1$ & $Med$ & $Q_3$ & $Q_1$ & $Med$ & $Q_3$ & \\
			\hline
			\multirow{6}{*}{250} &
			\multirow{3}{*}{$\widehat{\Bs}_\circ(J)$} &
			1 &
			{\tiny 0.0609} &
			0.0664 &
			{\tiny 0.0714} &
			{\tiny 0.0609} &
			0.0664 &
			{\tiny 0.0714} &
			{\tiny 0.0629} &
			0.0671 &
			{\tiny 0.0749} &
			\multirow{6}{*}{0.0574} \\
			\cline{3-12}
			&
			&
			2 &
			{\tiny 0.0622} &
			0.0658 &
			{\tiny 0.0720} &
			{\tiny 0.0623} &
			0.0651 &
			{\tiny 0.0718} &
			{\tiny 0.0634} &
			0.0693 &
			{\tiny 0.0754} &
			\\
			\cline{3-12}
			&
			&
			3 &
			{\tiny 0.0610} &
			0.0656 &
			{\tiny 0.0708} &
			{\tiny 0.0603} &
			0.0652 &
			{\tiny 0.0691} &
			{\tiny 0.0634} &
			0.0693 &
			{\tiny 0.0753} &
			\\
			\cline{2-12}
			&
			\multirow{3}{*}{$\widehat{\mathcal{B}}_J^{(u)}$} &
			1 &
			{\tiny 0.0609} &
			0.0664 &
			{\tiny 0.0718} &
			{\tiny 0.0609} &
			0.0664 &
			{\tiny 0.0718} &
			{\tiny 0.0630} &
			0.0676 &
			{\tiny 0.0752} &
			\\
			\cline{3-12}
			&
			&
			2 &
			{\tiny 0.0620} &
			0.0658 &
			{\tiny 0.0725} &
			{\tiny 0.0623} &
			0.0655 &
			{\tiny 0.0718} &
			{\tiny 0.0644} &
			0.0695 &
			{\tiny 0.0754} &
			\\
			\cline{3-12}
			&
			&
			3 &
			{\tiny 0.0606} &
			0.0651 &
			{\tiny 0.0708} &
			{\tiny 0.0603} &
			0.0656 &
			{\tiny 0.0695} &
			{\tiny 0.0646} &
			0.0695 &
			{\tiny 0.0754} &
			\\
			\hline
			\multirow{6}{*}{500} &
			\multirow{3}{*}{$\widehat{\Bs}_\circ(J)$} &
			1 &
			{\tiny 0.0497} &
			0.0523 &
			{\tiny 0.0564} &
			{\tiny 0.0497} &
			0.0523 &
			{\tiny 0.0564} &
			{\tiny 0.0498} &
			0.0530 &
			{\tiny 0.0571} &
			\multirow{6}{*}{0.0386} \\
			\cline{3-12}
			&
			&
			2 &
			{\tiny 0.0501} &
			0.0531 &
			{\tiny 0.0559} &
			{\tiny 0.0503} &
			0.0533 &
			{\tiny 0.0564} &
			{\tiny 0.0504} &
			0.0533 &
			{\tiny 0.0571} &
			\\
			\cline{3-12}
			&
			&
			3 &
			{\tiny 0.0499} &
			0.0528 &
			{\tiny 0.0553} &
			{\tiny 0.0499} &
			0.0524 &
			{\tiny 0.0554} &
			{\tiny 0.0506} &
			0.0534 &
			{\tiny 0.0571} &
			\\
			\cline{2-12}
			&
			\multirow{3}{*}{$\widehat{\mathcal{B}}_J^{(u)}$} &
			1 &
			{\tiny 0.0501} &
			0.0528 &
			{\tiny 0.0567} &
			{\tiny 0.0501} &
			0.0528 &
			{\tiny 0.0567} &
			{\tiny 0.0504} &
			0.0537 &
			{\tiny 0.0572} &
			\\
			\cline{3-12}
			&
			&
			2 &
			{\tiny 0.0503} &
			0.0532 &
			{\tiny 0.0560} &
			{\tiny 0.0504} &
			0.0533 &
			{\tiny 0.0559} &
			{\tiny 0.0509} &
			0.0537 &
			{\tiny 0.0571} &
			\\
			\cline{3-12}
			&
			&
			3 &
			{\tiny 0.0499} &
			0.0528 &
			{\tiny 0.0557} &
			{\tiny 0.0499} &
			0.0523 &
			{\tiny 0.0551} &
			{\tiny 0.0509} &
			0.0537 &
			{\tiny 0.0571} &
			\\
			\hline
			\multirow{6}{*}{1000} &
			\multirow{3}{*}{$\widehat{\Bs}_\circ(J)$} &
			1 &
			{\tiny 0.0233} &
			0.0266 &
			{\tiny 0.0334} &
			{\tiny 0.0233} &
			0.0266 &
			{\tiny 0.0334} &
			{\tiny 0.0210} &
			\textbf{0.0249} &
			{\tiny 0.0300} &
			\multirow{6}{*}{0.0257} \\
			\cline{3-12}
			&
			&
			2 &
			{\tiny 0.0259} &
			0.0299 &
			{\tiny 0.0361} &
			{\tiny 0.0275} &
			0.0322 &
			{\tiny 0.0398} &
			{\tiny 0.0222} &
			0.0262 &
			{\tiny 0.0314} &
			\\
			\cline{3-12}
			&
			&
			3 &
			{\tiny 0.0269} &
			0.0307 &
			{\tiny 0.0388} &
			{\tiny 0.0301} &
			0.0344 &
			{\tiny 0.0421} &
			{\tiny 0.0224} &
			0.0263 &
			{\tiny 0.0315} &
			\\
			\cline{2-12}
			&
			\multirow{3}{*}{$\widehat{\mathcal{B}}_J^{(u)}$} &
			1 &
			{\tiny 0.0234} &
			0.0281 &
			{\tiny 0.0451} &
			{\tiny 0.0234} &
			0.0281 &
			{\tiny 0.0451} &
			{\tiny 0.0216} &
			0.0261 &
			{\tiny 0.0452} &
			\\
			\cline{3-12}
			&
			&
			2 &
			{\tiny 0.0261} &
			0.0305 &
			{\tiny 0.0450} &
			{\tiny 0.0276} &
			0.0330 &
			{\tiny 0.0449} &
			{\tiny 0.0227} &
			0.0270 &
			{\tiny 0.0454} &
			\\
			\cline{3-12}
			&
			&
			3 &
			{\tiny 0.0272} &
			0.0315 &
			{\tiny 0.0450} &
			{\tiny 0.0306} &
			0.0349 &
			{\tiny 0.0449} &
			{\tiny 0.0228} &
			0.0270 &
			{\tiny 0.0454} &
			\\
			\hline
			\multirow{6}{*}{1500} &
			\multirow{3}{*}{$\widehat{\Bs}_\circ(J)$} &
			1 &
			{\tiny 0.0170} &
			\textbf{0.0190} &
			{\tiny 0.0214} &
			{\tiny 0.0170} &
			\textbf{0.0190} &
			{\tiny 0.0214} &
			{\tiny 0.0161} &
			\textbf{0.0177} &
			{\tiny 0.0197} &
			\multirow{6}{*}{0.0205} \\
			\cline{3-12}
			&
			&
			2 &
			{\tiny 0.0185} &
			0.0207 &
			{\tiny 0.0224} &
			{\tiny 0.0202} &
			0.0222 &
			{\tiny 0.0246} &
			{\tiny 0.0168} &
			\textbf{0.0183} &
			{\tiny 0.0199} &
			\\
			\cline{3-12}
			&
			&
			3 &
			{\tiny 0.0197} &
			0.0215 &
			{\tiny 0.0240} &
			{\tiny 0.0216} &
			0.0239 &
			{\tiny 0.0262} &
			{\tiny 0.0167} &
			\textbf{0.0183} &
			{\tiny 0.0202} &
			\\
			\cline{2-12}
			&
			\multirow{3}{*}{$\widehat{\mathcal{B}}_J^{(u)}$} &
			1 &
			{\tiny 0.0170} &
			\textbf{0.0189} &
			{\tiny 0.0214} &
			{\tiny 0.0170} &
			\textbf{0.0189} &
			{\tiny 0.0214} &
			{\tiny 0.0160} &
			\textbf{0.0177} &
			{\tiny 0.0200} &
			\\
			\cline{3-12}
			&
			&
			2 &
			{\tiny 0.0185} &
			0.0207 &
			{\tiny 0.0226} &
			{\tiny 0.0201} &
			0.0222 &
			{\tiny 0.0246} &
			{\tiny 0.0167} &
			\textbf{0.0183} &
			{\tiny 0.0203} &
			\\
			\cline{3-12}
			&
			&
			3 &
			{\tiny 0.0197} &
			0.0216 &
			{\tiny 0.0240} &
			{\tiny 0.0218} &
			0.0239 &
			{\tiny 0.0266} &
			{\tiny 0.0167} &
			\textbf{0.0182} &
			{\tiny 0.0202} &
			\\
			\hline
			\multirow{6}{*}{2000} &
			\multirow{3}{*}{$\widehat{\Bs}_\circ(J)$} &
			1 &
			{\tiny 0.0145} &
			\textbf{0.0161} &
			{\tiny 0.0177} &
			{\tiny 0.0145} &
			\textbf{0.0161} &
			{\tiny 0.0177} &
			{\tiny 0.0135} &
			\textbf{0.0147} &
			{\tiny 0.0159} &
			\multirow{6}{*}{0.0174} \\
			\cline{3-12}
			&
			&
			2 &
			{\tiny 0.0158} &
			0.0178 &
			{\tiny 0.0190} &
			{\tiny 0.0170} &
			0.0189 &
			{\tiny 0.0204} &
			{\tiny 0.0140} &
			\textbf{0.0149} &
			{\tiny 0.0163} &
			\\
			\cline{3-12}
			&
			&
			3 &
			{\tiny 0.0169} &
			0.0185 &
			{\tiny 0.0197} &
			{\tiny 0.0182} &
			0.0198 &
			{\tiny 0.0220} &
			{\tiny 0.0140} &
			\textbf{0.0149} &
			{\tiny 0.0163} &
			\\
			\cline{2-12}
			&
			\multirow{3}{*}{$\widehat{\mathcal{B}}_J^{(u)}$} &
			1 &
			{\tiny 0.0145} &
			\textbf{0.0161} &
			{\tiny 0.0176} &
			{\tiny 0.0145} &
			\textbf{0.0161} &
			{\tiny 0.0176} &
			{\tiny 0.0135} &
			\textbf{0.0147} &
			{\tiny 0.0157} &
			\\
			\cline{3-12}
			&
			&
			2 &
			{\tiny 0.0157} &
			0.0175 &
			{\tiny 0.0188} &
			{\tiny 0.0170} &
			0.0189 &
			{\tiny 0.0205} &
			{\tiny 0.0140} &
			\textbf{0.0149} &
			{\tiny 0.0163} &
			\\
			\cline{3-12}
			&
			&
			3 &
			{\tiny 0.0169} &
			0.0184 &
			{\tiny 0.0198} &
			{\tiny 0.0182} &
			0.0200 &
			{\tiny 0.0221} &
			{\tiny 0.0139} &
			\textbf{0.0149} &
			{\tiny 0.0164} &
			\\
			\hline
			\multirow{6}{*}{3000} &
			\multirow{3}{*}{$\widehat{\Bs}_\circ(J)$} &
			1 &
			{\tiny 0.0119} &
			\textbf{0.0126} &
			{\tiny 0.0138} &
			{\tiny 0.0119} &
			\textbf{0.0126} &
			{\tiny 0.0138} &
			{\tiny 0.0112} &
			\textbf{0.0120} &
			{\tiny 0.0129} &
			\multirow{6}{*}{0.0138} \\
			\cline{3-12}
			&
			&
			2 &
			{\tiny 0.0127} &
			\textbf{0.0135} &
			{\tiny 0.0143} &
			{\tiny 0.0133} &
			0.0142 &
			{\tiny 0.0150} &
			{\tiny 0.0114} &
			\textbf{0.0123} &
			{\tiny 0.0130} &
			\\
			\cline{3-12}
			&
			&
			3 &
			{\tiny 0.0133} &
			0.0139 &
			{\tiny 0.0150} &
			{\tiny 0.0138} &
			0.0146 &
			{\tiny 0.0158} &
			{\tiny 0.0115} &
			\textbf{0.0123} &
			{\tiny 0.0130} &
			\\
			\cline{2-12}
			&
			\multirow{3}{*}{$\widehat{\mathcal{B}}_J^{(u)}$} &
			1 &
			{\tiny 0.0119} &
			\textbf{0.0126} &
			{\tiny 0.0136} &
			{\tiny 0.0119} &
			\textbf{0.0126} &
			{\tiny 0.0136} &
			{\tiny 0.0111} &
			\textbf{0.0119} &
			{\tiny 0.0128} &
			\\
			\cline{3-12}
			&
			&
			2 &
			{\tiny 0.0127} &
			\textbf{0.0133} &
			{\tiny 0.0143} &
			{\tiny 0.0133} &
			0.0142 &
			{\tiny 0.0149} &
			{\tiny 0.0112} &
			\textbf{0.0122} &
			{\tiny 0.0129} &
			\\
			\cline{3-12}
			&
			&
			3 &
			{\tiny 0.0133} &
			0.0139 &
			{\tiny 0.0149} &
			{\tiny 0.0138} &
			0.0146 &
			{\tiny 0.0158} &
			{\tiny 0.0113} &
			\textbf{0.0123} &
			{\tiny 0.0130} &
			\\
			\hline
			\multirow{6}{*}{4000} &
			\multirow{3}{*}{$\widehat{\Bs}_\circ(J)$} &
			1 &
			{\tiny 0.0105} &
			\textbf{0.0113} &
			{\tiny 0.0119} &
			{\tiny 0.0105} &
			\textbf{0.0113} &
			{\tiny 0.0119} &
			{\tiny 0.0098} &
			\textbf{0.0105} &
			{\tiny 0.0110} &
			\multirow{6}{*}{0.0115} \\
			\cline{3-12}
			&
			&
			2 &
			{\tiny 0.0112} &
			0.0118 &
			{\tiny 0.0127} &
			{\tiny 0.0115} &
			0.0121 &
			{\tiny 0.0131} &
			{\tiny 0.0101} &
			\textbf{0.0107} &
			{\tiny 0.0114} &
			\\
			\cline{3-12}
			&
			&
			3 &
			{\tiny 0.0115} &
			0.0120 &
			{\tiny 0.0129} &
			{\tiny 0.0117} &
			0.0125 &
			{\tiny 0.0132} &
			{\tiny 0.0101} &
			\textbf{0.0107} &
			{\tiny 0.0114} &
			\\
			\cline{2-12}
			&
			\multirow{3}{*}{$\widehat{\mathcal{B}}_J^{(u)}$} &
			1 &
			{\tiny 0.0105} &
			\textbf{0.0113} &
			{\tiny 0.0119} &
			{\tiny 0.0105} &
			\textbf{0.0113} &
			{\tiny 0.0119} &
			{\tiny 0.0098} &
			\textbf{0.0105} &
			{\tiny 0.0110} &
			\\
			\cline{3-12}
			&
			&
			2 &
			{\tiny 0.0112} &
			0.0118 &
			{\tiny 0.0127} &
			{\tiny 0.0115} &
			0.0121 &
			{\tiny 0.0131} &
			{\tiny 0.0101} &
			\textbf{0.0107} &
			{\tiny 0.0114} &
			\\
			\cline{3-12}
			&
			&
			3 &
			{\tiny 0.0115} &
			0.0120 &
			{\tiny 0.0129} &
			{\tiny 0.0117} &
			0.0125 &
			{\tiny 0.0132} &
			{\tiny 0.0101} &
			\textbf{0.0107} &
			{\tiny 0.0114} &
			\\
			\hline
			\multirow{6}{*}{6000} &
			\multirow{3}{*}{$\widehat{\Bs}_\circ(J)$} &
			1 &
			{\tiny 0.0091} &
			0.0093 &
			{\tiny 0.0098} &
			{\tiny 0.0091} &
			0.0093 &
			{\tiny 0.0098} &
			{\tiny 0.0083} &
			\textbf{0.0088} &
			{\tiny 0.0091} &
			\multirow{6}{*}{0.0092} \\
			\cline{3-12}
			&
			&
			2 &
			{\tiny 0.0093} &
			0.0097 &
			{\tiny 0.0100} &
			{\tiny 0.0093} &
			0.0098 &
			{\tiny 0.0102} &
			{\tiny 0.0085} &
			\textbf{0.0088} &
			{\tiny 0.0092} &
			\\
			\cline{3-12}
			&
			&
			3 &
			{\tiny 0.0093} &
			0.0097 &
			{\tiny 0.0101} &
			{\tiny 0.0094} &
			0.0099 &
			{\tiny 0.0104} &
			{\tiny 0.0085} &
			\textbf{0.0088} &
			{\tiny 0.0092} &
			\\
			\cline{2-12}
			&
			\multirow{3}{*}{$\widehat{\mathcal{B}}_J^{(u)}$} &
			1 &
			{\tiny 0.0089} &
			0.0093 &
			{\tiny 0.0098} &
			{\tiny 0.0089} &
			0.0093 &
			{\tiny 0.0098} &
			{\tiny 0.0082} &
			\textbf{0.0087} &
			{\tiny 0.0090} &
			\\
			\cline{3-12}
			&
			&
			2 &
			{\tiny 0.0092} &
			0.0097 &
			{\tiny 0.0100} &
			{\tiny 0.0093} &
			0.0098 &
			{\tiny 0.0103} &
			{\tiny 0.0083} &
			\textbf{0.0088} &
			{\tiny 0.0092} &
			\\
			\cline{3-12}
			&
			&
			3 &
			{\tiny 0.0092} &
			0.0097 &
			{\tiny 0.0103} &
			{\tiny 0.0093} &
			0.0099 &
			{\tiny 0.0104} &
			{\tiny 0.0084} &
			\textbf{0.0088} &
			{\tiny 0.0091} &
			\\
			\hline
		\end{tabular}
	\end{minipage}
    \caption{Hellinger Distances between the true density `2D Comb 1' (Figure \ref{fig:truedenshardt}(c)) and estimators obtained through various estimation schemes. The wavelet family is Symlet 3. See text for column descriptions.}
	\label{tab:ex03}
\end{table}

%% file: tables/ch4-th-table-ex04.tex
\begin{table}[ht!]
	\fontsize{7}{10}\selectfont
	\centering
	\begin{minipage}[t]{0.95\textwidth}
		\centering
		\begin{tabular}{|r|r|r|r|r|r|r|r|r|r|r|r|r|}
			\hline
			\multirow{2}{*}{$n$} & \multirow{2}{*}{} & \multirow{2}{*}{$\Delta J$} & \multicolumn{3}{c|}{\textbf{universal}} & \multicolumn{3}{c|}{\textbf{level-dependent}} & \multicolumn{3}{c|}{\textbf{jackknife}} & \multirow{2}{*}{KDE} \\
			\cline{4-12}
			& & & $Q_1$ & $Med$ & $Q_3$ & $Q_1$ & $Med$ & $Q_3$ & $Q_1$ & $Med$ & $Q_3$ & \\
			\hline
			\multirow{6}{*}{250} &
			\multirow{3}{*}{$\widehat{\Bs}(J)$} &
			1 &
			{\tiny 0.0241} &
			\textbf{0.0282} &
			{\tiny 0.0347} &
			{\tiny 0.0241} &
			\textbf{0.0282} &
			{\tiny 0.0347} &
			{\tiny 0.0221} &
			\textbf{0.0265} &
			{\tiny 0.0323} &
			\multirow{6}{*}{0.0312} \\
			\cline{3-12}
			&
			&
			2 &
			{\tiny 0.0244} &
			\textbf{0.0283} &
			{\tiny 0.0348} &
			{\tiny 0.0249} &
			\textbf{0.0296} &
			{\tiny 0.0351} &
			{\tiny 0.0224} &
			\textbf{0.0277} &
			{\tiny 0.0336} &
			\\
			\cline{3-12}
			&
			&
			3 &
			{\tiny 0.0245} &
			\textbf{0.0281} &
			{\tiny 0.0336} &
			{\tiny 0.0244} &
			\textbf{0.0282} &
			{\tiny 0.0324} &
			{\tiny 0.0224} &
			\textbf{0.0277} &
			{\tiny 0.0336} &
			\\
			\cline{2-12}
			&
			\multirow{3}{*}{$\widehat{\Bs}_\circ(J)$} &
			1 &
			{\tiny 0.0241} &
			\textbf{0.0283} &
			{\tiny 0.0351} &
			{\tiny 0.0241} &
			\textbf{0.0283} &
			{\tiny 0.0351} &
			{\tiny 0.0217} &
			\textbf{0.0265} &
			{\tiny 0.0322} &
			\\
			\cline{3-12}
			&
			&
			2 &
			{\tiny 0.0248} &
			\textbf{0.0286} &
			{\tiny 0.0349} &
			{\tiny 0.0248} &
			\textbf{0.0294} &
			{\tiny 0.0354} &
			{\tiny 0.0223} &
			\textbf{0.0277} &
			{\tiny 0.0335} &
			\\
			\cline{3-12}
			&
			&
			3 &
			{\tiny 0.0246} &
			\textbf{0.0280} &
			{\tiny 0.0336} &
			{\tiny 0.0244} &
			\textbf{0.0278} &
			{\tiny 0.0325} &
			{\tiny 0.0223} &
			\textbf{0.0277} &
			{\tiny 0.0335} &
			\\
			\hline
			\multirow{6}{*}{500} &
			\multirow{3}{*}{$\widehat{\Bs}(J)$} &
			1 &
			{\tiny 0.0176} &
			\textbf{0.0196} &
			{\tiny 0.0229} &
			{\tiny 0.0176} &
			\textbf{0.0196} &
			{\tiny 0.0229} &
			{\tiny 0.0159} &
			\textbf{0.0186} &
			{\tiny 0.0206} &
			\multirow{6}{*}{0.0207} \\
			\cline{3-12}
			&
			&
			2 &
			{\tiny 0.0176} &
			\textbf{0.0199} &
			{\tiny 0.0229} &
			{\tiny 0.0177} &
			\textbf{0.0200} &
			{\tiny 0.0232} &
			{\tiny 0.0161} &
			\textbf{0.0190} &
			{\tiny 0.0213} &
			\\
			\cline{3-12}
			&
			&
			3 &
			{\tiny 0.0172} &
			\textbf{0.0194} &
			{\tiny 0.0230} &
			{\tiny 0.0170} &
			\textbf{0.0196} &
			{\tiny 0.0224} &
			{\tiny 0.0161} &
			\textbf{0.0190} &
			{\tiny 0.0213} &
			\\
			\cline{2-12}
			&
			\multirow{3}{*}{$\widehat{\Bs}_\circ(J)$} &
			1 &
			{\tiny 0.0176} &
			\textbf{0.0199} &
			{\tiny 0.0231} &
			{\tiny 0.0176} &
			\textbf{0.0199} &
			{\tiny 0.0231} &
			{\tiny 0.0159} &
			\textbf{0.0186} &
			{\tiny 0.0206} &
			\\
			\cline{3-12}
			&
			&
			2 &
			{\tiny 0.0177} &
			\textbf{0.0199} &
			{\tiny 0.0229} &
			{\tiny 0.0178} &
			\textbf{0.0200} &
			{\tiny 0.0232} &
			{\tiny 0.0161} &
			\textbf{0.0190} &
			{\tiny 0.0214} &
			\\
			\cline{3-12}
			&
			&
			3 &
			{\tiny 0.0172} &
			\textbf{0.0194} &
			{\tiny 0.0230} &
			{\tiny 0.0171} &
			\textbf{0.0196} &
			{\tiny 0.0224} &
			{\tiny 0.0161} &
			\textbf{0.0190} &
			{\tiny 0.0214} &
			\\
			\hline
			\multirow{6}{*}{1000} &
			\multirow{3}{*}{$\widehat{\Bs}(J)$} &
			1 &
			{\tiny 0.0126} &
			\textbf{0.0132} &
			{\tiny 0.0148} &
			{\tiny 0.0126} &
			\textbf{0.0132} &
			{\tiny 0.0148} &
			{\tiny 0.0122} &
			\textbf{0.0132} &
			{\tiny 0.0144} &
			\multirow{6}{*}{0.0138} \\
			\cline{3-12}
			&
			&
			2 &
			{\tiny 0.0125} &
			\textbf{0.0132} &
			{\tiny 0.0146} &
			{\tiny 0.0126} &
			\textbf{0.0133} &
			{\tiny 0.0146} &
			{\tiny 0.0125} &
			\textbf{0.0133} &
			{\tiny 0.0147} &
			\\
			\cline{3-12}
			&
			&
			3 &
			{\tiny 0.0124} &
			\textbf{0.0132} &
			{\tiny 0.0146} &
			{\tiny 0.0124} &
			\textbf{0.0131} &
			{\tiny 0.0144} &
			{\tiny 0.0125} &
			\textbf{0.0133} &
			{\tiny 0.0147} &
			\\
			\cline{2-12}
			&
			\multirow{3}{*}{$\widehat{\Bs}_\circ(J)$} &
			1 &
			{\tiny 0.0126} &
			\textbf{0.0132} &
			{\tiny 0.0148} &
			{\tiny 0.0126} &
			\textbf{0.0132} &
			{\tiny 0.0148} &
			{\tiny 0.0122} &
			\textbf{0.0132} &
			{\tiny 0.0144} &
			\\
			\cline{3-12}
			&
			&
			2 &
			{\tiny 0.0125} &
			\textbf{0.0132} &
			{\tiny 0.0146} &
			{\tiny 0.0126} &
			\textbf{0.0133} &
			{\tiny 0.0146} &
			{\tiny 0.0125} &
			\textbf{0.0133} &
			{\tiny 0.0147} &
			\\
			\cline{3-12}
			&
			&
			3 &
			{\tiny 0.0124} &
			\textbf{0.0132} &
			{\tiny 0.0146} &
			{\tiny 0.0124} &
			\textbf{0.0131} &
			{\tiny 0.0144} &
			{\tiny 0.0125} &
			\textbf{0.0133} &
			{\tiny 0.0147} &
			\\
			\hline
			\multirow{6}{*}{1500} &
			\multirow{3}{*}{$\widehat{\Bs}(J)$} &
			1 &
			{\tiny 0.0111} &
			0.0118 &
			{\tiny 0.0127} &
			{\tiny 0.0111} &
			0.0118 &
			{\tiny 0.0127} &
			{\tiny 0.0110} &
			0.0118 &
			{\tiny 0.0126} &
			\multirow{6}{*}{0.0109} \\
			\cline{3-12}
			&
			&
			2 &
			{\tiny 0.0110} &
			0.0117 &
			{\tiny 0.0126} &
			{\tiny 0.0110} &
			0.0118 &
			{\tiny 0.0126} &
			{\tiny 0.0110} &
			0.0118 &
			{\tiny 0.0127} &
			\\
			\cline{3-12}
			&
			&
			3 &
			{\tiny 0.0111} &
			0.0117 &
			{\tiny 0.0126} &
			{\tiny 0.0110} &
			0.0118 &
			{\tiny 0.0126} &
			{\tiny 0.0110} &
			0.0118 &
			{\tiny 0.0127} &
			\\
			\cline{2-12}
			&
			\multirow{3}{*}{$\widehat{\Bs}_\circ(J)$} &
			1 &
			{\tiny 0.0111} &
			0.0118 &
			{\tiny 0.0127} &
			{\tiny 0.0111} &
			0.0118 &
			{\tiny 0.0127} &
			{\tiny 0.0110} &
			0.0118 &
			{\tiny 0.0126} &
			\\
			\cline{3-12}
			&
			&
			2 &
			{\tiny 0.0110} &
			0.0117 &
			{\tiny 0.0126} &
			{\tiny 0.0110} &
			0.0118 &
			{\tiny 0.0126} &
			{\tiny 0.0111} &
			0.0118 &
			{\tiny 0.0127} &
			\\
			\cline{3-12}
			&
			&
			3 &
			{\tiny 0.0111} &
			0.0117 &
			{\tiny 0.0126} &
			{\tiny 0.0110} &
			0.0118 &
			{\tiny 0.0126} &
			{\tiny 0.0111} &
			0.0118 &
			{\tiny 0.0127} &
			\\
			\hline
			\multirow{6}{*}{2000} &
			\multirow{3}{*}{$\widehat{\Bs}(J)$} &
			1 &
			{\tiny 0.0105} &
			0.0110 &
			{\tiny 0.0118} &
			{\tiny 0.0105} &
			0.0110 &
			{\tiny 0.0118} &
			{\tiny 0.0105} &
			0.0111 &
			{\tiny 0.0118} &
			\multirow{6}{*}{0.0092} \\
			\cline{3-12}
			&
			&
			2 &
			{\tiny 0.0105} &
			0.0110 &
			{\tiny 0.0117} &
			{\tiny 0.0104} &
			0.0109 &
			{\tiny 0.0117} &
			{\tiny 0.0106} &
			0.0111 &
			{\tiny 0.0119} &
			\\
			\cline{3-12}
			&
			&
			3 &
			{\tiny 0.0105} &
			0.0109 &
			{\tiny 0.0116} &
			{\tiny 0.0104} &
			0.0109 &
			{\tiny 0.0117} &
			{\tiny 0.0106} &
			0.0111 &
			{\tiny 0.0119} &
			\\
			\cline{2-12}
			&
			\multirow{3}{*}{$\widehat{\Bs}_\circ(J)$} &
			1 &
			{\tiny 0.0105} &
			0.0110 &
			{\tiny 0.0118} &
			{\tiny 0.0105} &
			0.0110 &
			{\tiny 0.0118} &
			{\tiny 0.0105} &
			0.0111 &
			{\tiny 0.0118} &
			\\
			\cline{3-12}
			&
			&
			2 &
			{\tiny 0.0104} &
			0.0110 &
			{\tiny 0.0117} &
			{\tiny 0.0105} &
			0.0109 &
			{\tiny 0.0117} &
			{\tiny 0.0106} &
			0.0111 &
			{\tiny 0.0119} &
			\\
			\cline{3-12}
			&
			&
			3 &
			{\tiny 0.0105} &
			0.0109 &
			{\tiny 0.0116} &
			{\tiny 0.0104} &
			0.0109 &
			{\tiny 0.0117} &
			{\tiny 0.0106} &
			0.0111 &
			{\tiny 0.0119} &
			\\
			\hline
			\multirow{6}{*}{3000} &
			\multirow{3}{*}{$\widehat{\Bs}(J)$} &
			1 &
			{\tiny 0.0097} &
			0.0100 &
			{\tiny 0.0104} &
			{\tiny 0.0097} &
			0.0100 &
			{\tiny 0.0104} &
			{\tiny 0.0097} &
			0.0100 &
			{\tiny 0.0104} &
			\multirow{6}{*}{0.0073} \\
			\cline{3-12}
			&
			&
			2 &
			{\tiny 0.0096} &
			0.0100 &
			{\tiny 0.0103} &
			{\tiny 0.0096} &
			0.0100 &
			{\tiny 0.0104} &
			{\tiny 0.0097} &
			0.0101 &
			{\tiny 0.0105} &
			\\
			\cline{3-12}
			&
			&
			3 &
			{\tiny 0.0096} &
			0.0100 &
			{\tiny 0.0104} &
			{\tiny 0.0096} &
			0.0099 &
			{\tiny 0.0103} &
			{\tiny 0.0097} &
			0.0101 &
			{\tiny 0.0105} &
			\\
			\cline{2-12}
			&
			\multirow{3}{*}{$\widehat{\Bs}_\circ(J)$} &
			1 &
			{\tiny 0.0097} &
			0.0100 &
			{\tiny 0.0104} &
			{\tiny 0.0097} &
			0.0100 &
			{\tiny 0.0104} &
			{\tiny 0.0097} &
			0.0100 &
			{\tiny 0.0104} &
			\\
			\cline{3-12}
			&
			&
			2 &
			{\tiny 0.0096} &
			0.0100 &
			{\tiny 0.0103} &
			{\tiny 0.0096} &
			0.0100 &
			{\tiny 0.0104} &
			{\tiny 0.0097} &
			0.0101 &
			{\tiny 0.0105} &
			\\
			\cline{3-12}
			&
			&
			3 &
			{\tiny 0.0096} &
			0.0100 &
			{\tiny 0.0104} &
			{\tiny 0.0096} &
			0.0099 &
			{\tiny 0.0103} &
			{\tiny 0.0097} &
			0.0101 &
			{\tiny 0.0105} &
			\\
			\hline
			\multirow{6}{*}{4000} &
			\multirow{3}{*}{$\widehat{\Bs}(J)$} &
			1 &
			{\tiny 0.0093} &
			0.0095 &
			{\tiny 0.0098} &
			{\tiny 0.0093} &
			0.0095 &
			{\tiny 0.0098} &
			{\tiny 0.0093} &
			0.0097 &
			{\tiny 0.0099} &
			\multirow{6}{*}{0.0062} \\
			\cline{3-12}
			&
			&
			2 &
			{\tiny 0.0092} &
			0.0095 &
			{\tiny 0.0098} &
			{\tiny 0.0092} &
			0.0095 &
			{\tiny 0.0097} &
			{\tiny 0.0093} &
			0.0097 &
			{\tiny 0.0100} &
			\\
			\cline{3-12}
			&
			&
			3 &
			{\tiny 0.0092} &
			0.0095 &
			{\tiny 0.0098} &
			{\tiny 0.0092} &
			0.0095 &
			{\tiny 0.0097} &
			{\tiny 0.0093} &
			0.0097 &
			{\tiny 0.0100} &
			\\
			\cline{2-12}
			&
			\multirow{3}{*}{$\widehat{\Bs}_\circ(J)$} &
			1 &
			{\tiny 0.0093} &
			0.0095 &
			{\tiny 0.0098} &
			{\tiny 0.0093} &
			0.0095 &
			{\tiny 0.0098} &
			{\tiny 0.0093} &
			0.0097 &
			{\tiny 0.0100} &
			\\
			\cline{3-12}
			&
			&
			2 &
			{\tiny 0.0093} &
			0.0095 &
			{\tiny 0.0098} &
			{\tiny 0.0093} &
			0.0095 &
			{\tiny 0.0098} &
			{\tiny 0.0093} &
			0.0097 &
			{\tiny 0.0100} &
			\\
			\cline{3-12}
			&
			&
			3 &
			{\tiny 0.0092} &
			0.0095 &
			{\tiny 0.0098} &
			{\tiny 0.0092} &
			0.0095 &
			{\tiny 0.0098} &
			{\tiny 0.0093} &
			0.0097 &
			{\tiny 0.0100} &
			\\
			\hline
			\multirow{6}{*}{6000} &
			\multirow{3}{*}{$\widehat{\Bs}(J)$} &
			1 &
			{\tiny 0.0088} &
			0.0090 &
			{\tiny 0.0091} &
			{\tiny 0.0088} &
			0.0090 &
			{\tiny 0.0091} &
			{\tiny 0.0089} &
			0.0090 &
			{\tiny 0.0092} &
			\multirow{6}{*}{0.0048} \\
			\cline{3-12}
			&
			&
			2 &
			{\tiny 0.0088} &
			0.0090 &
			{\tiny 0.0092} &
			{\tiny 0.0088} &
			0.0089 &
			{\tiny 0.0091} &
			{\tiny 0.0089} &
			0.0090 &
			{\tiny 0.0092} &
			\\
			\cline{3-12}
			&
			&
			3 &
			{\tiny 0.0088} &
			0.0089 &
			{\tiny 0.0091} &
			{\tiny 0.0088} &
			0.0089 &
			{\tiny 0.0091} &
			{\tiny 0.0089} &
			0.0090 &
			{\tiny 0.0092} &
			\\
			\cline{2-12}
			&
			\multirow{3}{*}{$\widehat{\Bs}_\circ(J)$} &
			1 &
			{\tiny 0.0088} &
			0.0090 &
			{\tiny 0.0091} &
			{\tiny 0.0088} &
			0.0090 &
			{\tiny 0.0091} &
			{\tiny 0.0089} &
			0.0090 &
			{\tiny 0.0092} &
			\\
			\cline{3-12}
			&
			&
			2 &
			{\tiny 0.0088} &
			0.0089 &
			{\tiny 0.0091} &
			{\tiny 0.0088} &
			0.0089 &
			{\tiny 0.0091} &
			{\tiny 0.0089} &
			0.0090 &
			{\tiny 0.0092} &
			\\
			\cline{3-12}
			&
			&
			3 &
			{\tiny 0.0088} &
			0.0089 &
			{\tiny 0.0091} &
			{\tiny 0.0088} &
			0.0089 &
			{\tiny 0.0091} &
			{\tiny 0.0089} &
			0.0090 &
			{\tiny 0.0092} &
			\\
			\hline
		\end{tabular}
	\end{minipage}
    \caption{Hellinger Distances between the true density `2D Smooth Comb' (Figure \ref{fig:truedenshardt}(d)) and estimators obtained through various estimation schemes. The wavelet family is Symlet 3. See text for column descriptions.}
	\label{tab:ex04}
\end{table}